\documentclass{svmult}

\usepackage{makeidx}         
\usepackage{graphicx}        
\usepackage{multicol}        
\usepackage[bottom]{footmisc}
\usepackage{amssymb}
\makeindex             

\begin{document}
\title*{The Lyapunov Characteristic Exponents and their computation}
\titlerunning{Lyapunov Characteristic Exponents} \author{Charalampos
Skokos\inst{1} \inst{2}} \authorrunning{Ch.~Skokos} \institute{Astronomie et
Syst\`{e}mes Dynamiques, IMCCE, Observatoire de Paris, 77 avenue
Denfert--Rochereau, F-75014, Paris, France \and Max Planck Institute for the
Physics of Complex Systems, N\"othnitzer Strasse 38, D-01187, Dresden, Germany
\\ \texttt{hskokos@pks.mpg.de}}

\maketitle

\begin{flushright}
\begin{small}
\textit{For want of a nail the shoe was lost.\\
For want of a shoe the horse was lost.\\
For want of a horse the rider was lost.\\
For want of a rider the battle was lost.\\
For want of a battle the kingdom was lost.\\
And all for the want of a horseshoe nail.\\}
\textit{\textbf{For Want of a Nail}} (proverbial rhyme) \\
\end{small}
\end{flushright}


\begin{abstract}
We present a survey of the theory of the Lyapunov Characteristic Exponents
(LCEs) for dynamical systems, as well as of the numerical techniques
developed for the computation of the maximal, of few and of all of them. After
some historical notes on the first attempts for the numerical evaluation of
LCEs, we discuss in detail the multiplicative ergodic theorem of Oseledec
\cite{O_68}, which provides the theoretical basis for the computation of the
LCEs. Then, we analyze the algorithm for the computation of the maximal LCE,
whose value has been extensively used as an indicator of chaos, and the
algorithm of the so--called `standard method', developed by Benettin et
al.~\cite{BGGS_80b}, for the computation of many LCEs. We also consider
different discrete and continuous methods for computing the LCEs based on the
QR or the singular value decomposition techniques. Although, we are mainly
interested in finite--dimensional conservative systems, i.~e.~autonomous
Hamiltonian systems and symplectic maps, we also briefly refer to the
evaluation of LCEs of dissipative systems and time series. The relation of two
chaos detection techniques, namely the fast Lyapunov indicator (FLI) and the
generalized alignment index (GALI), to the computation of the LCEs is also
discussed.

\end{abstract} 

\setcounter{minitocdepth}{5}
\dominitoc

\section{Introduction}
\label{introduction}

One of the basic information in understanding the behavior of a dynamical
system is the knowledge of the spectrum of its \textit{Lyapunov Characteristic
Exponents (LCEs)}.  The LCEs are asymptotic measures characterizing the average
rate of growth (or shrinking) of small perturbations to the solutions of a
dynamical system. Their concept was introduced by Lyapunov when studying the
stability of nonstationary solutions of ordinary differential equations
\cite{Lyapunov_1892}, and has been widely employed in studying dynamical
systems since then. The value of the maximal LCE (mLCE) is an indicator of the
chaotic or regular nature of orbits, while the whole spectrum of LCEs is
related to entropy (Kolmogorov--Sinai entropy) and dimension--like (Lyapunov
dimension) quantities that characterize the underlying dynamics.

By \textit{dynamical system} we refer to a physical and/or mathematical system
consisting of a) a set of $l$ real state variables $x_1, x_2\, \ldots,x_l $,
whose current values define the state of the system, and b) a well--defined
rule from which the evolution of the state with respect to an independent real
variable (which is usually referred as the time $t$) can be derived. We refer
to the number $l$ of state variables as the \textit{dimension} of the system,
and denote a state using the vector $\vec{x}= (x_1, x_2\, \ldots,x_l)$, or the
matrix $\vec{x}= [ \begin{array}{cccc} x_1& x_2& \ldots & x_l
\end{array}]^{\mathrm{T}} $ notation, where $(^{\mathrm{T}})$ denotes the
transpose matrix. A particular state $\vec{x}$ corresponds to a point in an
$l$--dimensional space $\cal{S}$, the so--called \textit{phase space} of the
system, while a set of states $\vec{x}(t)$ parameterized by $t$ is referred as
an \textit{orbit} of the dynamical system.

Dynamical systems come in essentially two types:
\begin{enumerate}

\item \textit{Continuous dynamical systems} described by differential
  equations of the form
\[
\dot{\vec{x}}=\frac{d \vec{x}}{dt}=\vec{f}(\vec{x},t),
\]
with dot denoting derivative with respect to a continuous time $t$ and
$\vec{f}$ being a set of $l$ functions $f_1, f_2\, \ldots,f_l $ known
as the \textit{vector field}.

\item \textit{Discrete dynamical systems} or \textit{maps}, described
  by difference equations of the form
\[
\vec{x}_{n+1}=\vec{f}(\vec{x}_n),
\]
with $\vec{f}$ being a set of $l$ functions $f_1, f_2\, \ldots,f_l $
and $\vec{x}_n$ denoting the vector $\vec{x}$ at a discrete time $t=n$
(integer).

\end{enumerate}

Let us now define the term \textit{chaos}. In the literature there are many
definitions. A brief and concise presentation of them can be found for example
in \cite{LC_04}. We adopt here one of the most famous definitions of chaos due
to Devaney \cite[p.~50]{Devaney_1989}, which is based on the topological
approach of the problem.
\begin{definition}
  Let $V$ be a set and $\vec{f}:V\rightarrow V$ a map on this set. We
  say that $\vec{f}$ is \texttt{chaotic} on $V$ if
\begin{enumerate}
\item $\vec{f}$ has sensitive dependence on initial conditions.
\item $\vec{f}$ is topologically transitive.
\item periodic points are dense in $V$.
\end{enumerate}
\label{def:chaos}
\end{definition}
Let us explain in more detail the hypothesis of this definition.
\begin{definition}
  $\vec{f}:V\rightarrow V$ has \texttt{sensitive dependence on initial
    conditions} if there exists $\delta >0$ such that, for any $\vec{x} \in V$
    and any neighborhood $\Delta$ of $\vec{x}$, there exist $\vec{y} \in
    \Delta$ and $n \geq 0$, such that $|\vec{f}^n(\vec{x})-\vec{f}^n(\vec{y})|>
    \delta$, where $\vec{f}^n$ denotes $n$ successive applications of
    $\vec{f}$.
\label{def:deviation}
\end{definition}
Practically this definition implies that there exist points arbitrarily close
to $\vec{x}$ which eventually separate from $\vec{x}$ by at least $\delta$
under iterations of $\vec{f}$. We point out that not all points near $\vec{x}$
need eventually move away from $\vec{x}$ under iteration, but there must be at
least one such point in every neighborhood of $\vec{x}$.
\begin{definition}
  $\vec{f}:V\rightarrow V$ is said to be \texttt{topologically transitive} if
    for any pair of open sets $U, \, W \subset V$ there exists $n>0$ such that
    $\vec{f}^n(U)\cap W \neq \emptyset $.
\label{def:trasitivity}
\end{definition}
This definitions implies the existence of points which eventually move under
iteration from one arbitrarily small neighborhood to any other. Consequently,
the dynamical system cannot be decomposed into two disjoint invariant open
sets.

From Definition \ref{def:chaos} we see that a chaotic system possesses three
ingredients: a) unpredictability because of the sensitive dependence on
initial conditions, b) indecomposability, because it cannot be decomposed into
noninteracting subsystems due to topological transitivity, and c) an element
of regularity because it has periodic points which are dense.

Usually, in physics and applied sciences, people focus on the first hypothesis
of Definition \ref{def:chaos} and use the notion of chaos in relation to the
sensitive dependence on initial conditions. The most commonly employed method
for distinguishing between regular and chaotic motion, which quantifies the
sensitive dependence on initial conditions, is the evaluation of the mLCE
$\chi_1$. If $\chi_1 > 0$ the orbit is chaotic. This method was initially
developed at the late 70's based on theoretical results obtained at the end of
the 60's.

The concept of the LCEs has been widely presented in the literature from a
practical point of view, i.~e.~the description of particular numerical
algorithms for their computation
\cite{F_84b,ER_85,GPL_90,LichtenbergL_1992,DV_95}. Of course, there also exist
theoretical studies on the LCEs, which are mainly focused on the problem of
their existence, starting with the pioneer work of Oseledec \cite{O_68}. In
that paper the Multiplicative Ergodic Theorem (MET), which provided the
theoretical basis for the numerical computation of the LCEs, was stated and
proved. The MET was the subject of several theoretical studies afterwards
\cite{Ra_79,R_79b,JPS_87,W_93}. A combination of important theoretical and
numerical results on LCEs can be found in the seminal papers of Benettin et
al.~\cite{BGGS_80a,BGGS_80b}, written almost 30 years ago, where an explicit
method for the computation of all LCEs was developed.

In the present report we focus our attention both on the theoretical framework
of the LCEs, as well as on the numerical techniques developed for their
computation. Our goal is to provide a survey of the basic results on these
issues obtained over the last 40 years, after the work of Oseledec
\cite{O_68}. To this end, we present in detail the mathematical theory of the
LCEs and discuss its significance without going through tedious mathematical
proofs. In our approach, we prefer to present the definitions of various
quantities and to state the basic theorems that guarantee the existence of the
LCE, citing at the same time the papers where all the related mathematical
proofs can be found. We also describe in detail the various numerical
techniques developed for the evaluation of the maximal, of few or even of all
LCEs, and explain their practical implementation. We do not restrict our
presentation to the so--called \textit{standard method} developed by Benettin
et al.~\cite{BGGS_80b}, as it is usually done in the literature (see
e.~g.~\cite{F_84b,ER_85,LichtenbergL_1992}), but we include in our study
modern techniques for the computation of the LCEs like the discrete and
continuous methods based on the singular value decomposition (SVD) and the QR
decomposition procedures.

In our analysis we deal with finite--dimensional dynamical systems and in
particular with autonomous Hamiltonian systems and symplectic maps defined on
a compact manifold, meaning that we exclude cases with escapes in which the
motion can go to infinity. We do not consider the rather exceptional cases of
completely chaotic systems and of integrable ones, i.~e.~systems that can be
solved explicitly to give their variables as single--valued functions of time,
but we consider the most general case of `systems with divided phase space'
\cite[p.~19]{Contopoulos_2002} for which \textit{regular}\footnote[1]{Regular
orbits are often called \textit{ordered orbits} (see
e.~g.~\cite[p.~18]{Contopoulos_2002}).} (quasiperiodic) and \textit{chaotic
orbits} co--exist. In such systems one sees both regular and chaotic
domains. But the regular domains contain a dense set of unstable periodic
orbits, which are followed by small chaotic regions. On the other hand, the
chaotic domains contain stable periodic orbits that are followed by small
islands of stability. Thus, the regular and chaotic domains are intricately
mixed. However, there are regions where order is predominant, and other
regions where chaos is predominant.

Although in our report the theory of LCEs and the numerical techniques for
their evaluation are presented mainly for \textit{conservative systems},
i.e.~system that preserve the phase space volume, these techniques are not
valid only for such models.  For completeness sake, we also briefly discuss at
the end of the report the computation of LCEs for \textit{dissipative
systems}, for which the phase space volume decreases on average, and for time
series.

We tried to make the paper self--consistent by including definitions of the
used terminology and brief overviews of all the necessary mathematical
notions. In addition, whenever it was considered necessary, some illustrative
examples have been added to the text in order to clarify the practical
implementation of the presented material. Our aim has been to make this review
of use for both the novice and the more experienced practitioner interested in
LCEs. To this end, the reader who is interested in reading up on detailed
technicalities is provided with numerous signposts to the relevant literature.

Throughout the text bold lowercase letters denote vectors, while matrices are
represented, in general, by capital bold letters. We also note that the most
frequently used abbreviations in the text are: LCE(s), Lyapunov Characteristic
Exponent(s); $p$--LCE, Lyapunov Characteristic Exponent of order $p$; mLCE,
maximal Lyapunov Characteristic Exponent; $p$--mLCE, maximal Lyapunov
Characteristic Exponent of order $p$; MET, multiplicative ergodic theorem;
SVD, Singular Value Decomposition; PSS, Poincar\'{e} surface of section; FLI,
fast Lyapunov indicator; GALI, generalized alignment index.

The paper is organized as follows: 

In Section \ref{Hams_maps} we present the basic concepts of Hamiltonian
systems and symplectic maps, emphasizing on the evolution of orbits, as well
as of deviation vectors about them. In particular, we define the so--called
variational equations for Hamiltonian systems and the tangent map for
symplectic maps, which govern the time evolution of deviation vectors. We also
provide some simple examples of dynamical systems and derive the corresponding
set of variational equations and the corresponding tangent map.

Section \ref{early} contains some historical notes on the first attempts for
the application of the theoretical results of Oseledec \cite{O_68} for the
actual computation of the LCEs. We recall how the notion of exponential
divergence of nearby orbits was eventually quantified by the computation of
the mLCE, and we refer to the papers where the mLCE or the spectrum
of LCEs were computed for the first time.

The basic theoretical results on the LCEs are presented in Section
\ref{Theory} following mainly the milestone papers of Oseledec \cite{O_68} and
Benettin et al.~\cite{BGGS_80a,BGGS_80b}. In Section \ref{Def_LCE} the basic
definitions and theoretical results of LCEs of various orders are
presented. The practical consequences of these results on the computation of
the LCEs of order 1 and of order $p>1$ are discussed in Sections
\ref{Theory_comp_1} and \ref{Theory_comp_p} respectively. Then, in Section
\ref{Oseledec} the MET of Oseledec \cite{O_68} is stated in its various forms,
while its consequences on the spectrum of LCEs for conservative dynamical
systems are discussed in Section \ref{Spectrum_properties}.

Section \ref{maximal} is devoted to the computation of the mLCE $\chi_1$,
which is the oldest chaos indicator used in the literature. In Section
\ref{compute_maximal} the method for the computation of the mLCE is discussed
in great detail and the theoretical basis of its evaluation is explained. The
corresponding algorithm is presented in Section \ref{maximal_algorithm}, while
the behavior of $\chi_1$ for regular and chaotic orbits is analyzed in Section
\ref{maximal_algorithm_behavior}.

In Section \ref{allLCEs} the various methods for the computation of part or of
the whole spectrum of LCEs are presented. In particular, in Section
\ref{allLCEs_Benettin} the standard method developed in
\cite{SN_79,BGGS_80b}, is presented in great detail, while the corresponding
algorithm is given in Section \ref{Algorithm_all_LCEs}. In Section \ref{QR}
the connection of the standard method with the discrete QR decomposition
technique is discussed and the corresponding QR algorithm is given, while
Section \ref{Dis_Cont} is devoted to the presentation of other techniques for
computing few or all  LCEs, which are based on the SVD and QR decomposition
algorithms.

In Section \ref{other_chaos_methods} we briefly refer to various chaos
detection techniques based on the analysis of deviation vectors, as well as
to a second category of chaos indicators based on the analysis of the time
series constructed by the coordinates of the orbit under consideration. The
relation of two chaos indicators, namely the fast Lyapunov indicator (FLI) and
the generalized alignment index (GALI), to the computation of the LCEs is
also discussed.

Although the main topic of our presentation is the theory and the computation
of the LCEs for conservative dynamical systems, in the last section of our
report some complementary issues related to other types of dynamical systems
are concisely presented. In particular, Section \ref{LCEs_dis} is devoted to
the computation of the LCEs for dissipative systems, while in Section
\ref{LCEs_time_series} some basic features on the numerical computation of the
LCEs from a time series are presented.

Finally, in the appendix \ref{Wedge} we present
some basic elements of the exterior algebra theory in connection to the
evaluation of wedge products, which are needed for the computation of the
volume elements appearing in the definitions of the various LCEs.

\section{Autonomous Hamiltonian systems and symplectic maps}
\label{Hams_maps}

In our study we consider two main types of conservative dynamical systems:
\begin{enumerate}
\item Continuous systems corresponding to an \textit{autonomous Hamiltonian
    system} of $N$ degrees ($N$D) of freedom having a Hamiltonian function
\begin{equation}
  H(q_1,q_2, \ldots, q_N,p_1,p_2, \ldots, p_N)=h=\mbox{constant},
\label{eq:Ham_gen}
\end{equation}
where $q_i$ and $p_i$, $i=1,2,\ldots,N$ are the generalized coordinates and
conjugate momenta respectively. An orbit in the $l=2N$--dimensional phase
space $\cal{S}$ of this system is defined by a vector
\[
  \vec{x}(t)=(q_1(t),q_2(t), \ldots, q_N(t),p_1(t),p_2(t), \ldots, p_N(t)),
\]
with $x_i=q_i$, $x_{i+N}=p_i$, $i=1,2,\ldots,N$. The time evolution of this
orbit is governed by the Hamilton equations of motion, which in matrix form
are given by
\begin{equation}
  \dot{\vec{x}}= \vec{f}(\vec{x})=\left[ \begin{array}{cc} \frac{\partial
      H}{\partial \vec{p}}\,\, & - \frac{\partial H}{\partial \vec{q}}
\end{array} \right]^{\mathrm{T}}= \textbf{J}_{2N}\cdot\textbf{DH}, 
\label{eq:Hameq_gen}
\end{equation}
with $ \vec{q}=(q_1(t),q_2(t), \ldots, q_N(t))$, $ \vec{p}=(p_1(t),p_2(t),
\ldots, p_N(t))$, and
\[
  \textbf{DH}=\left[ \begin{array}{cccccccc} \frac{\partial H}{\partial q_1}
        &\frac{\partial H}{\partial q_2}& \cdots & \frac{\partial H}{\partial
        q_N} & \frac{\partial H}{\partial p_1} &\frac{\partial H}{\partial
        p_2}& \cdots & \frac{\partial H}{\partial p_N}
\end{array}
\right]^{\mathrm{T}}.
\]

Matrix $\textbf{J}_{2N}$ has the following block form
\[
\textbf{J}_{2N}= \left[ \begin{array}{cc} \textbf{0}_{N} & \textbf{I}_{N} \\
-\textbf{I}_{N} & \textbf{0}_{N}
\end{array}
 \right] ,
\]
with $\textbf{I}_{N}$ being the $N\times N$ identity matrix and
$\textbf{0}_{N}$ being the $N\times N$ matrix with all its elements equal to
zero. The solution of (\ref{eq:Hameq_gen}) is formally written with respect to
the induced flow \cal{$\Phi$}$^t: \cal{S}\rightarrow \cal{S}$ as
\begin{equation}
\vec{x}(t)= \mbox{\cal{$\Phi$}}^t\left( \vec{x}(0)\right) .
\label{eq:flow}
\end{equation}

\item Symplectic maps of $l=2N$ dimensions having the form
\begin{equation}
\vec{x}_{n+1}=\vec{f}(\vec{x}_n).
\label{eq:map_gen}
\end{equation}
A \textit{symplectic map} is an area--preserving map whose \textit{Jacobian
matrix}
\[
\textbf{M}= \textbf{Df}(\vec{x})=\frac{\partial\vec{f}}{\partial\vec{x}}
=\left[
  \begin{array}{cccc} \frac{\partial f_1}{\partial x_1} & \frac{\partial
      f_1}{\partial x_2} & \cdots & \frac{\partial f_1}{\partial x_{2N}} \\
    \frac{\partial f_2}{\partial x_1} & \frac{\partial f_2}{\partial x_2} &
    \cdots & \frac{\partial f_2}{\partial x_{2N}} \\ \vdots & \vdots & &
    \vdots \\ \frac{\partial f_{2N}}{\partial x_1} & \frac{\partial
    f_{2N}}{\partial x_2} & \cdots & \frac{\partial f_{2N}}{\partial x_{2N}}
\end{array}
 \right] ,
\]
satisfies
\begin{equation}
\textbf{M}^{\mathrm{T}}\cdot \textbf{J}_{2N} \cdot \textbf{M} = \textbf{J}_{2N}.
\label{eq:symplectic_condition}
\end{equation}
The state of the system at the discrete time $t=n$ is given by
\begin{equation}
\vec{x}_n= \mbox{\cal{$\Phi$}}^n\left( \vec{x}_0\right)= \left(
\vec{f}\right)^n \left( \vec{x}_0\right),
\label{eq:flow_map}
\end{equation}
where $\left( \vec{f}\right)^n \left( \vec{x}_0\right)= \vec{f}(\vec{f}(\cdots
\vec{f}(\vec{x}_0)\cdots ))$, $n$ times.
\end{enumerate}

\subsection{Variational equations and tangent map}
\label{Var_tang}

Let us now turn our attention to the (continuous or discrete) time evolution
of deviation vectors $\vec{w}$ from a given reference orbit of a dynamical
system. These vectors evolve on the \textit{tangent space} $\cal{T}_{\vec{x}}
\cal{S}$ of $\cal{S}$. We denote by $d_{\vec{x}}\mbox{\cal{$\Phi$}}^t$ the
linear mapping which maps the tangent space of $\cal{S}$ at point $\vec{x}$
onto the tangent space at point \cal{$\Phi$}$^t (\vec{x})$, and so we have
$d_{\vec{x}}\mbox{\cal{$\Phi$}}^t:\cal{T}_{\vec{x}} \cal{S} \rightarrow
\cal{T}\mbox{$_{\mbox{\cal{$\Phi$}}^t(\vec{x})}\cal{S}$}$ with
\begin{equation}
\vec{w}(t)=d_{\vec{x}}\mbox{\cal{$\Phi$}}^t\, \vec{w}(0),
\label{eq:t_flow}
\end{equation}
where $\vec{w}(0)$, $\vec{w}(t)$ are deviation vectors with respect to the
given orbit at times $t=0$ and $t>0$ respectively.

In the case of the Hamiltonian system (\ref{eq:Ham_gen}) an initial deviation
vector $\vec{w}(0)=(\delta x_1(0),\delta x_2(0),\ldots, \delta x_{2N}(0)) $
from the solution $\vec{x}(t)$ (\ref{eq:flow}) evolves on the tangent space
$\cal{T}_{\vec{x}} \cal{S}$ according to the so--called \textit{variational
equations}
\begin{equation}
  \dot{\vec{w}}=\textbf{Df}(\vec{x}(t)) \cdot\vec{w} =
  \frac{\partial\vec{f}}{\partial\vec{x}} (\vec{x}(t)) \cdot\vec{w} =\left[
  \textbf{J}_{2N}\cdot\textbf{D$^2$H}(\vec{x}(t)) \right] \cdot\vec{w}=:
  \textbf{A}(t) \cdot\vec{w} \, ,
\label{eq:var}
\end{equation}
with $\textbf{D$^2$H}(\vec{x}(t))$ being the Hessian matrix of Hamiltonian
(\ref{eq:Ham_gen}) calculated on the reference orbit $\vec{x}(t)$
(\ref{eq:flow}), i.~e.
\[
\textbf{D$^2$H}(\vec{x}(t))_{i,j} = \left. \frac{\partial^2 H}{\partial x_i
\partial x_j}\right|_{\mbox{\cal{$\Phi$}}^t\left(
\vec{x}(0)\right)}\,\,\,,\,\,\, i,j=1,2,\ldots,2N.
\]
We underline that equations (\ref{eq:var}) represent a set of \textit{linear
differential equations} with respect to $\vec{w}$, having time dependent
coefficients since, matrix $\textbf{A}(t)$ depends on the particular reference
orbit, which is a function of time $t$. The solution of (\ref{eq:var}) can be
written as
\begin{equation}
\vec{w}(t)=\textbf{Y}(t)\cdot\vec{w}(0) ,
\label{eq:w_ham}
\end{equation}
where $\textbf{Y}(t)$ is the so--called \textit{fundamental matrix} of
solutions of (\ref{eq:var}), satisfying the equation
\begin{equation}
\dot{\textbf{Y}}(t)= \textbf{Df}(\vec{x}(t)) \cdot \textbf{Y}(t)= \textbf{A}(t)
\cdot \textbf{Y}(t)\,\,\,,\,\,\, \mbox{with} \,\,\,
\textbf{Y}(0)=\textbf{I}_{2N}.
\label{eq:Y_ham}
\end{equation}

In the case of the symplectic map (\ref{eq:map_gen}) the evolution of a
deviation vector $\vec{w}_n$, with respect to a reference orbit $\vec{x}_n$,
is given by the corresponding \textit{tangent map}
\begin{equation}
  \vec{w}_{n+1}=\textbf{Df}(\vec{x}_n) \cdot\vec{w}_n =
  \frac{\partial\vec{f}}{\partial\vec{x}} (\vec{x}_n)\cdot\vec{w}_n =:
  \textbf{M}_n \cdot\vec{w}_n .
\label{eq:w_map}
\end{equation}
Thus, the evolution of the initial deviation vector $\vec{w}_0$ is
given by
\begin{equation}
\vec{w}_n= \textbf{M}_{n-1} \cdot \textbf{M}_{n-2} \cdot \ldots \cdot\textbf{M}_0
\cdot \vec{w}_0=:\textbf{Y}_n \cdot \vec{w}_0,
\label{eq:w0_map}
\end{equation}
with $\textbf{Y}_n$ satisfying the relation
\begin{equation}
  \textbf{Y}_{n+1}= \textbf{M}_n \cdot \textbf{Y}_n= \textbf{Df}(\vec{x}_n) \cdot
  \textbf{Y}_n,\,\,\, \mbox{with} \,\,\, \textbf{Y}_0=\textbf{I}_{2N}.
\label{eq:Y_map}
\end{equation}

\subsection{Simple examples of dynamical systems}
\label{variational_examples}

As representative examples of dynamical systems we consider a) the well--known
2D H\'{e}non--Heiles system \cite{HH_64}, having the
Hamiltonian function
\begin{equation}
H_2 = \frac{1}{2} (p_x^2+p_y^2) + \frac{1}{2} (x^2+y^2) + x^2 y -
\frac{1}{3} y^3, 
\label{eq:HH}
\end{equation}
with equations of motion
\begin{equation}
  \dot{\vec{x}}=\left[ \begin{array}{c} \dot{x} \\ \dot{y} \\ \dot{p}_x \\
      \dot{p}_y
    \end{array} \right] = \textbf{J}_{4} \cdot \textbf{DH}_2 = \textbf{J}_{4} \cdot \left[ \begin{array}{c}
      x+2xy\\ y+x^2-y^2 \\ p_x \\ p_y
\end{array} \right] \Rightarrow 
\left\lbrace 
\begin{array}{ccl}
\dot{x}&=& p_x \\
\dot{y}&=& p_y \\
\dot{p}_x&=& -x-2xy \\
\dot{p}_y&=& -y-x^2+y^2  
\end{array}
\right.  ,
\label{eq:HHeq}
\end{equation}
and b) the 4--dimensional (4d) symplectic map
\begin{equation}
\begin{array}{ccl}
x_{1,n+1}&=& x_{1,n}+x_{3,n}\\ x_{2,n+1}&=& x_{2,n}+x_{4,n}\\ x_{3,n+1}&=&
x_{3,n}-\nu \sin(x_{1,n+1})-\mu [1-\cos(x_{1,n+1}+x_{2,n+1})] \\ x_{4,n+1}&=&
x_{4,n}-\kappa \sin(x_{2,n+1})-\mu [1-\cos(x_{1,n+1}+x_{2,n+1})]
\end{array} \,\, (\mbox{mod} \,2\pi),
\label{eq:4dmap}
\end{equation}
with parameters $\nu$, $\kappa$ and $\mu$. All variables are given (mod
$2\pi$), so $x_{i,n} \in [−\pi, \pi)$, for $i = 1, 2, 3, 4$.  This map is a
variant of Froeschl\'{e}'s 4d symplectic map \cite{F_72} and its behavior has
been studied in \cite{CG_88,SCP_97}. It is easily seen that its Jacobian
matrix satisfies equation (\ref{eq:symplectic_condition}).

\subsection{Numerical integration of variational equations}
\label{integration_variational}

When dealing with Hamiltonian systems the variational equations (\ref{eq:var})
have to be integrated simultaneously with the Hamilton equations of motion
(\ref{eq:Hameq_gen}). Let us clarify the issue by looking to a specific
example. The variational equations of the 2D Hamiltonian (\ref{eq:HH}) are
\begin{equation}
\begin{array}{c}
\dot{\vec{w}}=\left[ \begin{array}{c}
\dot{\delta x} \\ \dot{\delta y} \\ \dot{\delta p}_x \\ \dot{\delta p}_y 
\end{array} \right]  =  \left[ \begin{array}{cccc}
0 & 0 & 1 &0 \\
0 & 0 & 0 &1 \\
-1-2y & -2x & 0 &0 \\
-2x & -1+2y & 0 &0 
\end{array} \right] \cdot \left[ \begin{array}{c}
\delta x \\ \delta y \\ \delta p_x \\ \delta p_y 
\end{array} \right]\Rightarrow     \\
\left\lbrace 
\begin{array}{ccl}
\dot{\delta x}&=& \delta p_x \\
\dot{\delta y}&=& \delta p_y \\
\dot{\delta p}_x&=& (-1-2y) \delta x + (-2x) \delta y \\
\dot{\delta p}_y&=&  (-2x) \delta x + (-1+2y) \delta y  
\end{array}
\right.  .
\end{array}
\label{eq:HH_var_eq}
\end{equation}
This system of differential equations is linear with respect to $\delta x$,
$\delta y$, $\delta p_x$, $\delta p_y$, but it cannot be integrated
independently of system (\ref{eq:HHeq}) since the $x$ and $y$ variables appear
explicitly in it. Thus, if we want to follow the time evolution of an initial
deviation vector $\vec{w}(0)$ with respect to a reference orbit with initial
condition $\vec{x}(0)$, we are obliged to integrate simultaneously the whole
set of differential equations (\ref{eq:HHeq}) and (\ref{eq:HH_var_eq}).

A numerical scheme for integrating the variational equations (\ref{eq:var}),
which exploits their linearity and is particularly useful when we need to
evolve more than one deviation vectors is the following. Solving the Hamilton
equations of motion (\ref{eq:Hameq_gen}) by any numerical integration scheme
we obtain the time evolution of the reference orbit (\ref{eq:flow}). In
practice this means that we know the values $\vec{x}(t_i)$ for $t_i=i\, \Delta
t$, $i=0,1,2,\ldots$, where $\Delta t$ is the integration time step. Inserting
this numerically known solution to the variational equations (\ref{eq:var}) we
end up with a linear system of differential equations with constant
coefficients for every time interval $[t_i, t_i+\Delta t)$, which can be solved
explicitly.

For example, in the particular case of Hamiltonian (\ref{eq:HH}), the system
of variational equations (\ref{eq:HH_var_eq}) becomes
\begin{equation}
\begin{array}{ccl}
\dot{\delta x}&=& \delta p_x \\ \dot{\delta y}&=& \delta p_y \\ \dot{\delta
p}_x&=& \left[ -1-2y(t_i)\right] \delta x + \left[ -2x(t_i)\right] \delta y \\
\dot{\delta p}_y&=& \left[ -2x(t_i)\right] \delta x + \left[ -1+2y(t_i)\right]
\delta y
\end{array}\,\,\, ,\,\,\, \mbox{for    }t\in [t_i,t_i+\Delta t),
\label{eq:HH_var_eq_explicit}
\end{equation}
which is a linear system of differential equations with constant coefficients
and thus, easily solved. In particular, equations (\ref{eq:HH_var_eq_explicit})
can by considered as the Hamilton equations of motion corresponding to the
Hamiltonian function
\begin{equation}
\begin{array}{c}
H_V(\delta x,\delta y,\delta p_x, \delta p_y)= \\ \\ \frac{1}{2} \left(\delta
p_x^2+\delta p_y^2\right) + \frac{1}{2} \left\lbrace
\left[1+2y(t_i)\right]\delta x^2 + \left[1-2y(t_i)\right] \delta y^2 +
2\left[2x(t_i) \right] \delta x \delta y\right\rbrace.
\end{array} 
\label{eq:HH_var}
\end{equation}

The Hamiltonian formalism (\ref{eq:HH_var}) of the variational equations
(\ref{eq:HH_var_eq_explicit}) is a specific example of a more general
result. In the case of the usual Hamiltonian function
\begin{equation}
H(\vec{q},\vec{p} )= \frac{1}{2} \sum_{i=1}^{N} p_i^2+ V(\vec{q}),
\label{eq:HH_usual}
\end{equation}
with $V(\vec{q})$ being the potential function, the variational equations
(\ref{eq:var}) for the time interval $[t_i, t_i+\Delta t)$ take the form (see
e.~g.~\cite{BFS_79})
\[
\dot{\vec{w}}=\left[ \begin{array}{c}
\dot{\vec{\delta q}} \\ 
\dot{\vec{\delta p}}
\end{array} \right] = \left[
\begin{array}{cc}
\textbf{0}_{N} & \textbf{I}_{N} \\
-\textbf{D$^2$V}(\vec{q}(t_i)) & \textbf{0}_{N}
\end{array}\right] \cdot \left[ \begin{array}{c}
 \vec{\delta q} \\ 
 \vec{\delta p}
\end{array} \right]
\]
with $ \vec{\delta q}=(\delta q_1(t),\delta q_2(t), \ldots, \delta q_N(t))$, $
\vec{\delta p}=(\delta p_1(t),\delta p_2(t) \ldots, \delta p_N(t))$, and
\[
\textbf{D$^2$V}(\vec{q}(t_i))_{jk}=\left. \frac{\partial^2
V(\vec{q})}{\partial q_j \partial q_k} \right|_{\vec{q}(t_i)}\,\,\, ,\,\,\,
j,k=1,2,\ldots,N.
\]
Thus, the tangent dynamics of (\ref{eq:HH_usual}) is represented by
the Hamiltonian function (see  e.~g.~\cite{PP_08})
\[
H_V( \vec{ \delta q}, \vec{\delta p})=\frac{1}{2} \sum_{j=1}^{N} \delta p_i^2+
\frac{1}{2}\sum_{j,k}^{N} \textbf{D$^2$V}(\vec{q}(t_i))_{jk} \delta q_j \delta
q_k .
\]

\subsection{Tangent dynamics of symplectic maps}
\label{Tangent_map_evol}

In the case of symplectic maps, the dynamics on the tangent space, which is
described by the tangent map (\ref{eq:w_map}), cannot be considered separately
from the phase space dynamics determined by the map (\ref{eq:map_gen})
itself. This is because the tangent map depends explicitly on the reference
orbit $\vec{x}_n$.

For example, the tangent map of the $4$d map  (\ref{eq:4dmap}) is
\begin{equation}
\begin{array}{ccl}
\delta x_{1,n+1}&=& \delta x_{1,n}+ \delta x_{3,n}\\ \delta x_{2,n+1}&=&
\delta x_{2,n}+\delta x_{4,n}\\ \delta x_{3,n+1}&=& a_n \delta x_{1,n} +b_n
\delta x_{2,n} +(1+a_n) \delta x_{3,n} +b_n \delta x_{4,n} \\ \delta
x_{4,n+1}&=& b_n \delta x_{1,n} +c_n \delta x_{2,n} +b_n \delta x_{3,n}
+(1+c_n) \delta x_{4,n}
\end{array} ,
\label{eq:4dmap_t}
\end{equation}
with
\[
\begin{array}{l}
a_n= -\nu \cos(x_{1,n+1})-\mu \sin(x_{1,n+1}+x_{2,n+1})\\
b_n= -\mu \sin(x_{1,n+1}+x_{2,n+1})\\
c_n= -\kappa \cos(x_{2,n+1})-\mu \sin(x_{1,n+1}+x_{2,n+1})
\end{array} ,
\]
which explicitly depend on $x_{1,n}$, $x_{2,n}$, $x_{3,n}$, $x_{4,n}$. Thus,
the evolution of a deviation vector requires the simultaneous iteration of
both the map (\ref{eq:4dmap}) and the tangent map (\ref{eq:4dmap_t}).

\section{Historical introduction: The early days of LCEs}
\label{early}

Prior to the discussion of the theory of the LCEs and the presentation of the
various algorithms for their computation, it would be interesting to go back
in time and see how the notion of LCEs, as well as the nowadays taken for
granted techniques for evaluating them, were formed.

The LCEs are asymptotic measures characterizing the average rate of growth (or
shrinking) of small perturbations to the orbits of a dynamical system, and
their concept was introduced by Lyapunov \cite{Lyapunov_1892}. Since then they
have been extensively used for studying dynamical systems. As it has already
been mentioned, one of the basic features of chaos is the sensitive dependence
on initial conditions and the LCEs provide quantitative measures of response
sensitivity of a dynamical system to small changes in initial conditions. For
a chaotic orbit at least one LCE is positive, implying exponential divergence
of nearby orbits, while in the case of regular orbits all LCEs are
zero. Therefore, the presence of positive LCEs is a signature of chaotic
behavior. Usually the computation of only the mLCE $\chi_1$ is sufficient for
determining the nature of an orbit, because $\chi_1>0$ guarantees that the
orbit is chaotic.

\begin{figure}
\centerline{ 
\begin{tabular}{cc}
\includegraphics[height=5.1cm,width=6.cm]{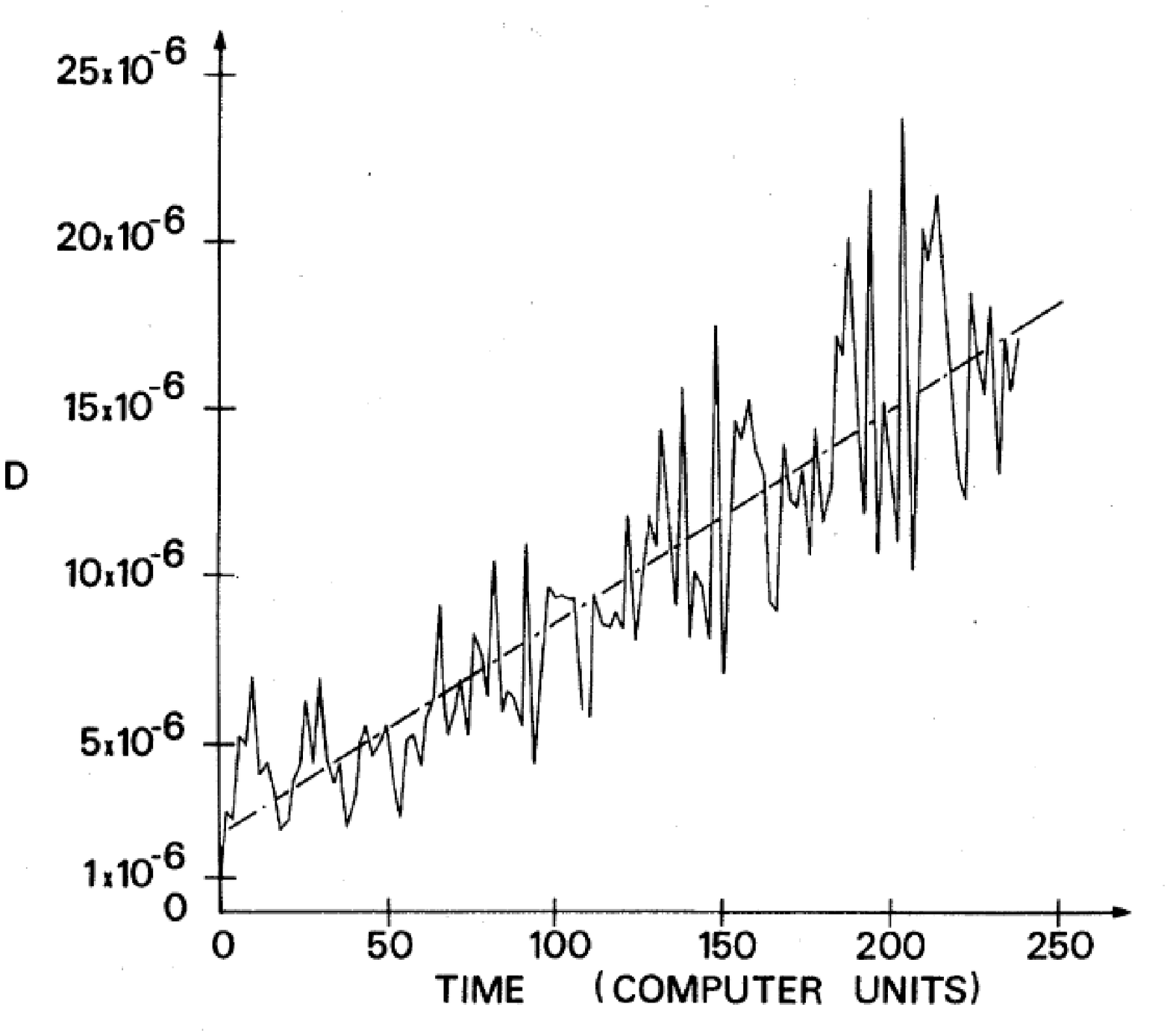} &
\includegraphics[height=5.2cm,width=6.cm]{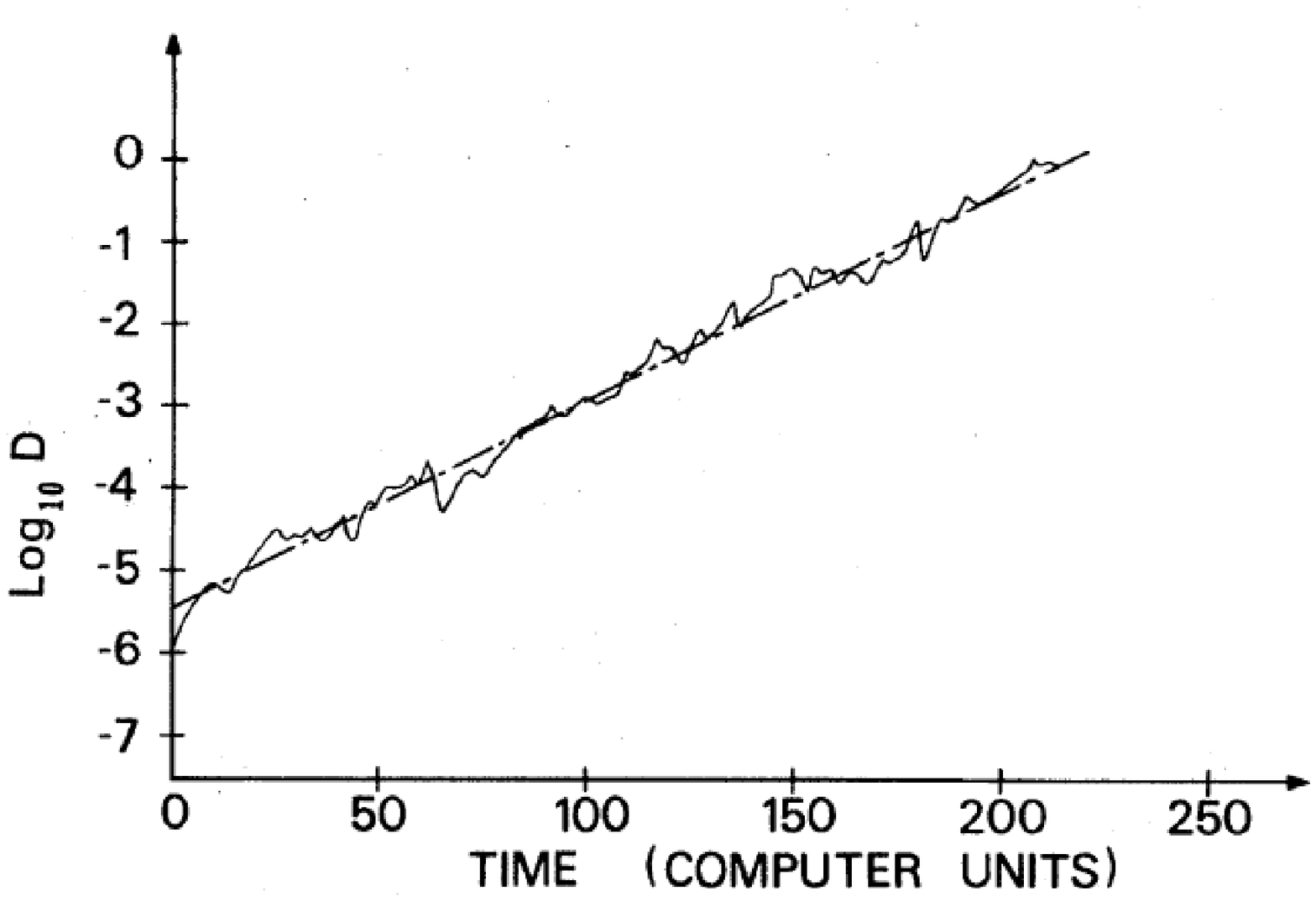}\end{tabular}}
\caption{Typical behavior of the time evolution of the distance $D$ between
  two initially close orbits in the case of regular and chaotic orbits. The
  particular results are obtained for a $2$D Hamiltonian system describing a
  Toda lattice of two particles with unequal masses (see \cite{CF_75} for more
  details). The initial Euclidian distance of the two orbits in the
  $4$--dimensional phase space is $D_0=10^{-6}$. $D$ exhibits a linear (on the
  average) growth when the two orbits are initially located in a region of
  regular motion (left panel), while it grows exponentially in the case of
  chaotic orbits (right panel).  The big difference in the values of $D$
  between the two cases is evident since the two panels have the same
  horizontal (time) axis but different vertical ones. In particular, the
  vertical axis is linear in the left panel and logarithmic in the right panel
  (after \cite{CF_75}).}
\label{f:CF_75}
\end{figure}

Characterization of the chaoticity of an orbit in terms of the divergence of
nearby orbits was introduced by H\'{e}non and Heiles \cite{HH_64} and further
used by several authors (e.~g. \cite{FL_70,F_70,
F_72,SF_73,CF_75,CDGS_76}). In these studies two initial points were chosen
very close to each other, having phase space distance of about
$10^{-7}-10^{-6}$, and were evolved in time. If the two initial points were
located in a region of regular motion their distance increased approximately
linearly with time, while if they were belonging to a chaotic region the
distance exhibited an exponential increase in time (Figure \ref{f:CF_75}).

Although the theory of LCEs was applied to characterize chaotic motion by
Oseledec \cite{O_68}, quite some time passed until the connection between LCEs
and exponential divergence was made clear \cite{BGS_76,P_77}. It is worth
mentioning that Casartelli et al.~\cite{CDGS_76} defined a quantity, which
they called `stochastic parameter', in order to quantify the exponential
divergence of nearby orbits, which was realized afterwards in \cite{BGS_76} to
be an estimator of the mLCE for $t \rightarrow \infty$.

\begin{figure}
\centerline{ 
\begin{tabular}{cc}
\includegraphics[scale=0.305]{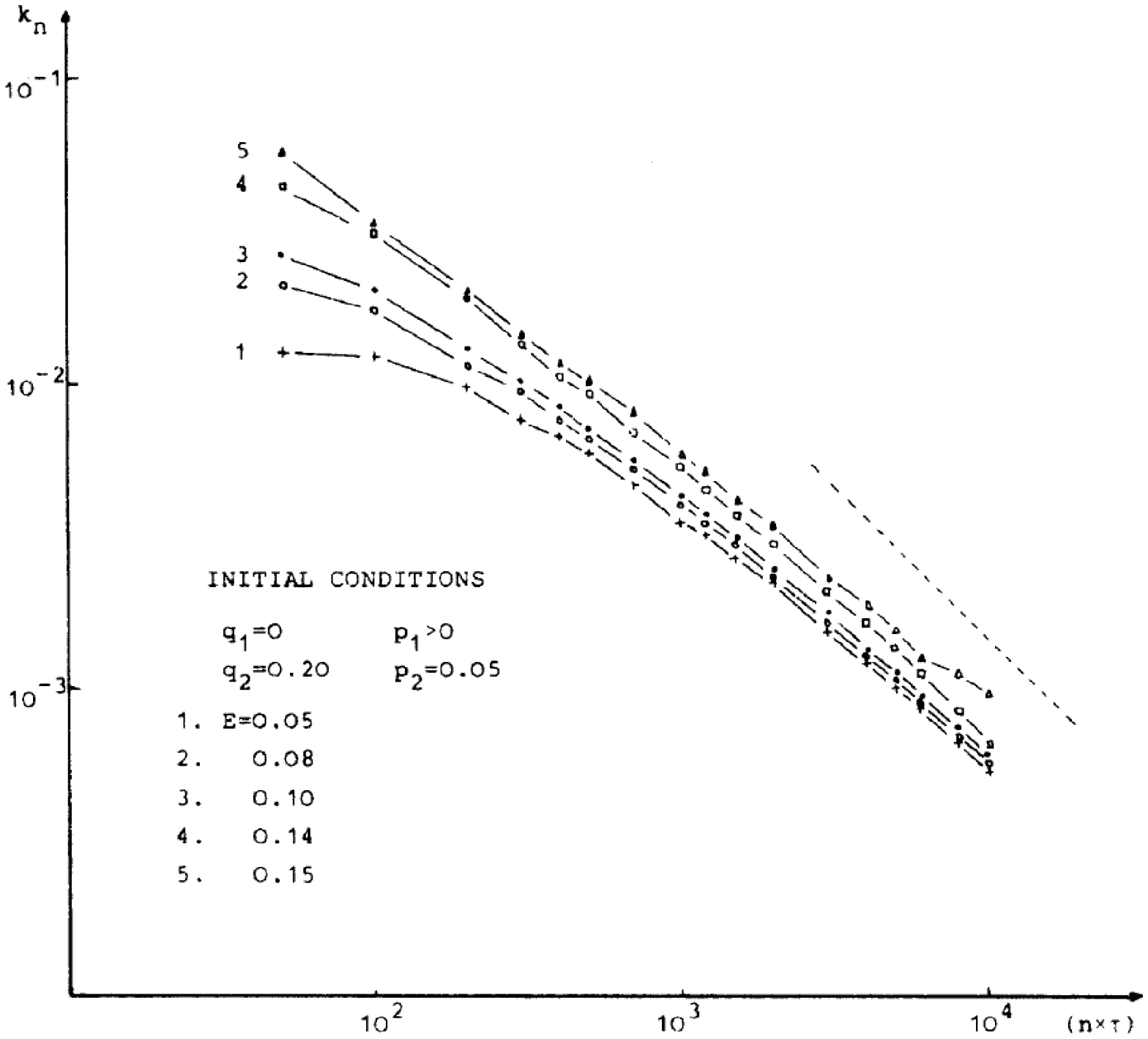}
\includegraphics[scale=0.305]{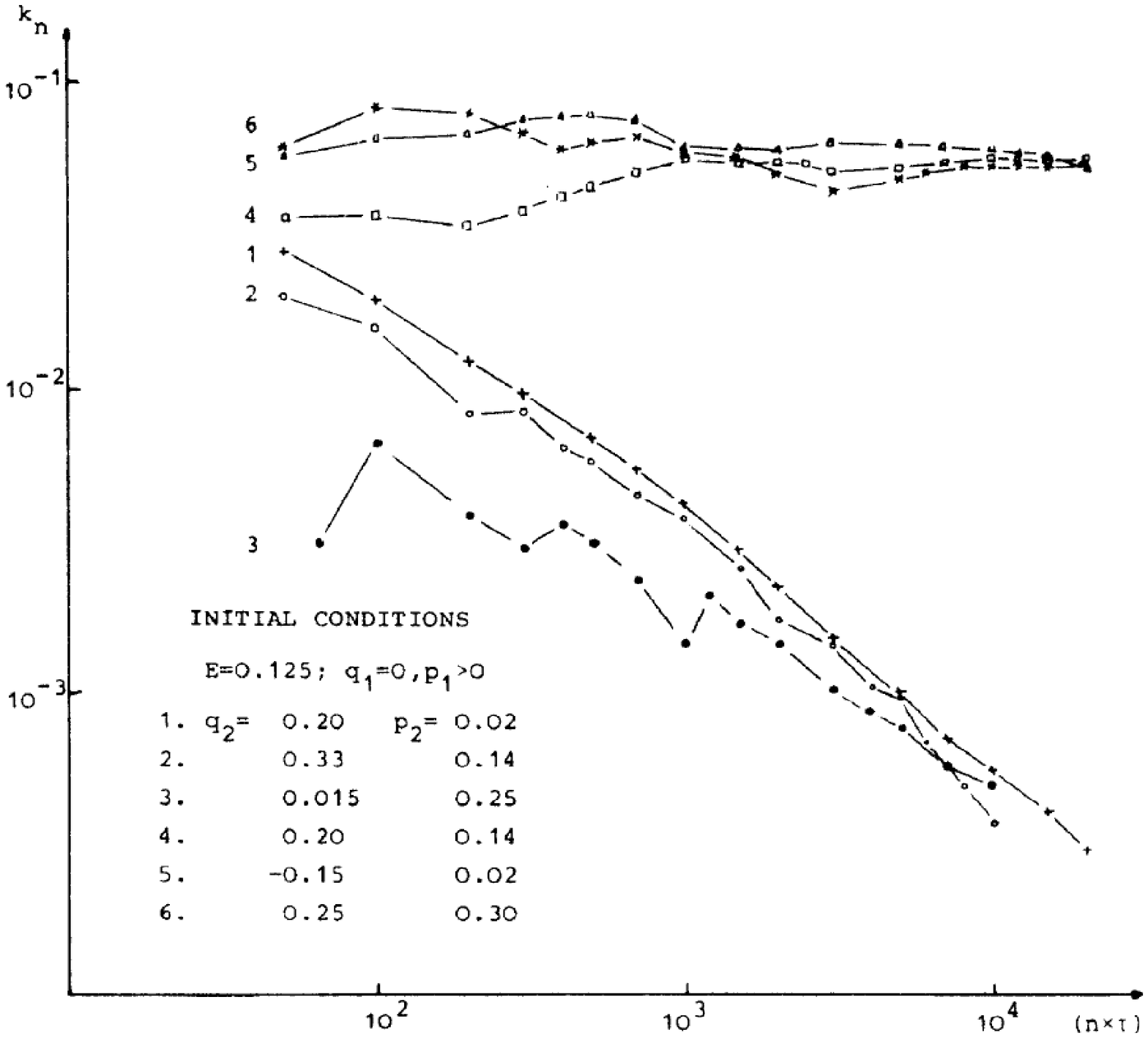}\end{tabular}}
\caption{Evolution of $X_1(t)$ (denoted as $k_n$) with respect to time $t$
  (denoted by $n\times \tau$) in log--log scale for several orbits of the
  H\'{e}non--Heiles system (\ref{eq:HH}). In the left panel $X_1(t)$ is
  computed for 5 different regular orbits at different energies $H_2$ (denoted
  as $E$) and it tends to zero following a power law $\propto t^{-1}$. A
  dashed straight line corresponding to a function proportional to $t^{-1}$ is
  also plotted. In the right panel the evolution of $X_1(t)$ is plotted for
  three regular orbits (curves 1--3) and three chaotic ones (curves 4--6) for
  $H_2=0.125$. Note that the values of the initial conditions given in the two
  panels correspond to $q_1=x$, $q_2=y$, $p_1=p_x$, $p_2=p_y$ in (\ref{eq:HH})
  (after \cite{BGS_76}).}
\label{f:BGS_76}
\end{figure}

So, the mLCE $\chi_1$ was estimated for the first time in \cite{BGS_76}, as
the limit for $t \rightarrow \infty$ of an appropriate quantity $X_1(t)$,
which was obtained from the evolution of the phase space distance of two
initially close orbits. In this paper some nowadays well--established
properties of $X_1(t)$ were discussed, like for example the fact that $X_1(t)$
tends to zero in the case of regular orbits following a power law $\propto
t^{-1}$, while it tends to nonzero values in the case of chaotic orbits
(Figure \ref{f:BGS_76}).
The same algorithm was immediately applied for the computation of the mLCE of
a dissipative system, namely the Lorenz system \cite{NS_77}.

The next improvement of the computational algorithm for the evaluation of the
mLCE was introduced in \cite{CGG_78}, where the variational equations were
used for the time evolution of deviation vectors instead of the previous
approach of the simultaneous integration of two initially close orbits. This
more direct approach constituted a significant improvement for the computation
of the mLCE since it allowed the use of larger integration steps, diminishing
the real computational time and also eliminated the problem of choosing a
suitable initial distance between the nearby orbits.

In \cite{BGGS_78} a theorem was formulated, which led directly to the
development of a numerical technique for the computation of some or even of all
LCEs, based on the time evolution of more than one deviation vectors, which are
kept linearly independent through a Gram--Schmidt orthonormalization
procedure (see also \cite{BG_79}). This method was explained in more detail in
\cite{SN_79}, where it was applied to the study of the Lorenz system and was
also presented in \cite{BFS_79}, where it was applied to the study of an $N$D
Hamiltonian system with $N$ varying from 2 to 10.

\begin{figure}
\centerline{ 
\begin{tabular}{cc}
\includegraphics[scale=0.3]{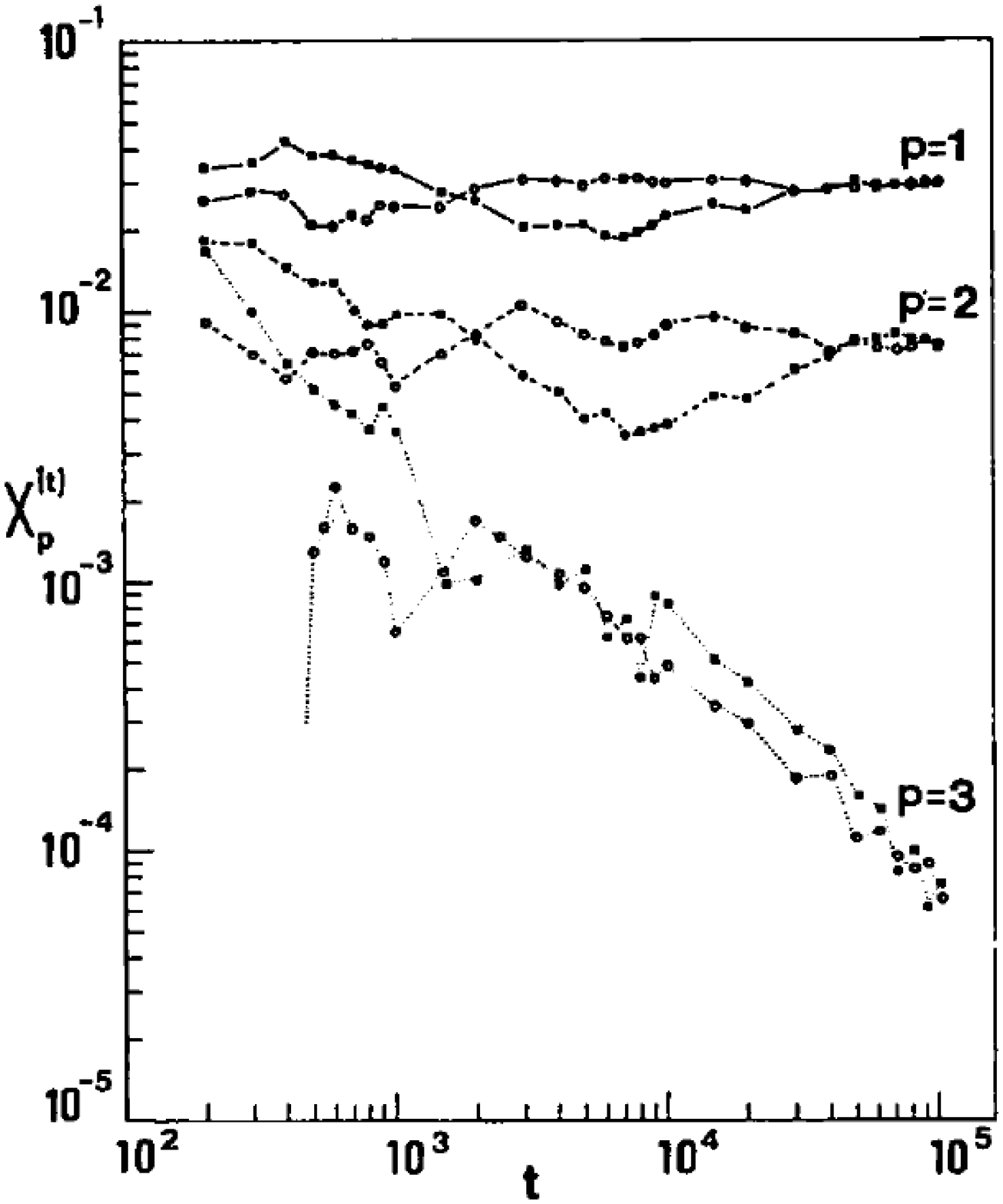} &
\includegraphics[scale=0.305]{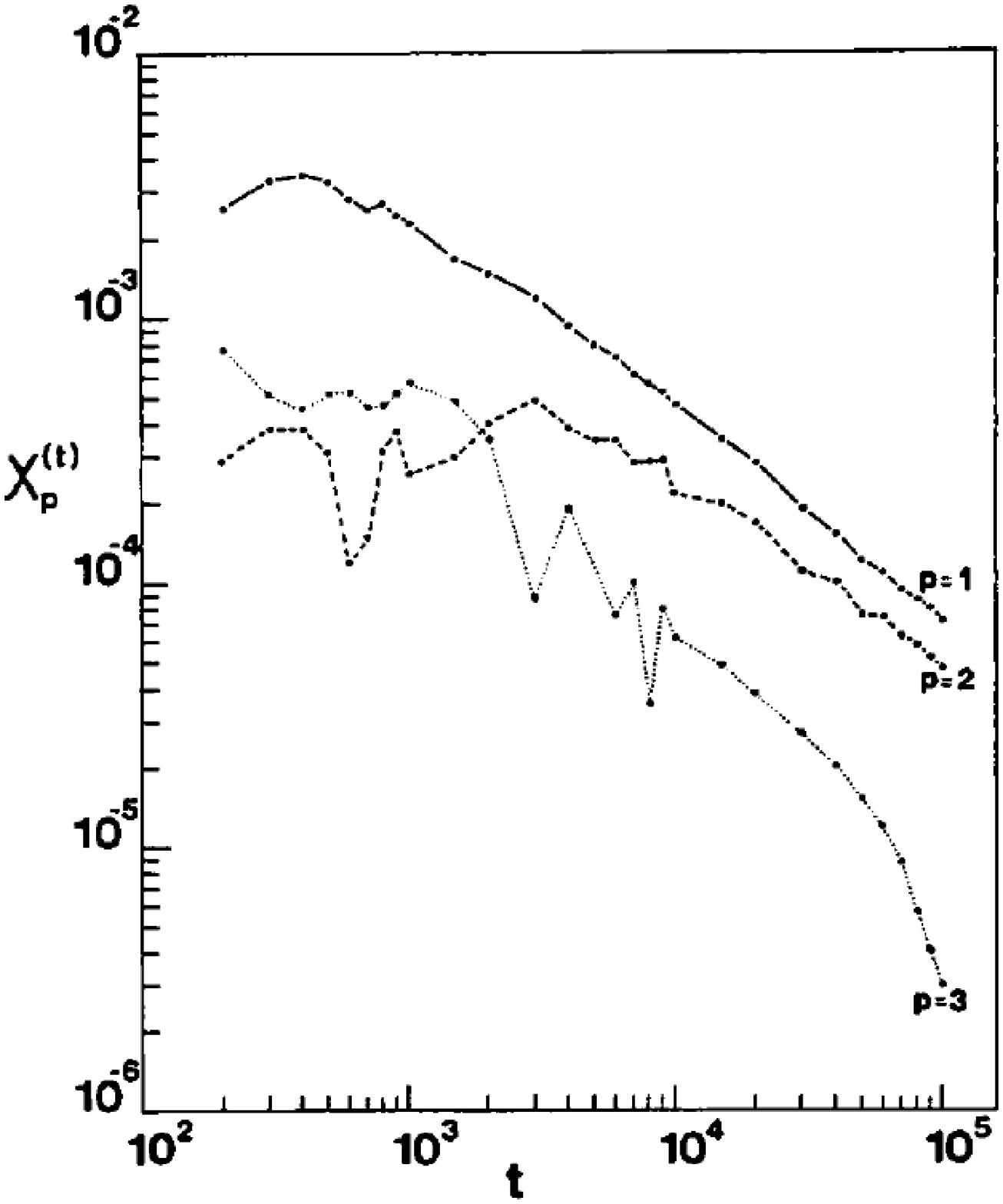}\end{tabular}}
\caption{Time evolution of appropriate quantities denoted by $X_p^{(t)}$,
  $p=1,2,3$, having respectively as limits for $t\rightarrow\infty$ the first
  three LCEs $\chi_1$, $\chi_2$, $\chi_3$, for two chaotic orbits (left panel)
  and one regular orbit (right panel) of the 3D Hamiltonian system initially
  studied in \cite{CGG_78} (see \cite{BGGS_80b} for more details). In both
  panels $X_3^{(t)}$ tends to zero implying that $\chi_3=0$. This is due to
  the fact that Hamiltonian systems have at least one vanishing LCE, namely
  the one corresponding to the direction along the flow (this property is
  explained in Section \ref{Spectrum_properties}). On the other hand, $\chi_1$
  and $\chi_2$ seem to get nonzero values (with $\chi_1 > \chi_2$) for chaotic
  orbits, while they appear to vanish for regular orbits (after
  \cite{BGGS_80b}). }
\label{f:BGGS_80b}
\end{figure}

The theoretical framework, as well as the numerical method for the
computation of the maximal, some or even all LCEs were given in the seminal
papers of Benettin et al.~\cite{BGGS_80a,BGGS_80b}. In \cite{BGGS_80b} the
complete set of LCEs was calculated for several different Hamiltonian
systems, including four and six dimensional maps. In Figure \ref{f:BGGS_80b}
we show the results of \cite{BGGS_80b} concerning the 3D Hamiltonian system of
\cite{CGG_78}. The importance of the papers of Benettin et
al. \cite{BGGS_80a,BGGS_80b} is reflected by the fact that almost all methods
for the computation of the LCEs are more or less based on them. Immediately
the ideas presented in \cite{BGGS_80a,BGGS_80b} were used for the computation
of the LCEs for a variety of dynamical systems like infinite--dimensional
systems described by delay differential equations \cite{F_82}, dissipative
systems \cite{ER_85}, conservative systems related to Celestial Mechanics
problems \cite{F_84,F_85}, as well as for the determination of the LCEs from a
time series \cite{WSSV_85,SS_85}.

\section{Lyapunov Characteristic Exponents: Theoretical treatment}
\label{Theory}

In this section we define the LCEs of various orders presenting also the basic
theorems which guarantee their existence and provide the theoretical
background for their numerical evaluation. In our presentation we basically
follow the fundamental papers of Oseledec \cite{O_68} and of Benettin et
al.~\cite{BGGS_80a} where all the theoretical results of the current section
are explicitly proved.

We consider a continuous or discrete dynamical system defined on a
differentiable manifold $\cal{S}$. Let \cal{$\Phi$}$^t(\vec{x})$ denote the
state at time $t$ of the system which at time $t=0$ was at $\vec{x}$ (see
equations (\ref{eq:flow}) and (\ref{eq:flow_map}) for the continuous and
discrete case respectively). For the action of \cal{$\Phi$}$^t$ over two
successive time intervals $t$ and $s$ we have the following composition law
\[
\mbox{\cal{$\Phi$}}^{t+s}=\mbox{\cal{$\Phi$}}^t\circ \mbox{\cal{$\Phi$}}^s.
\]

The tangent space at $\vec{x}$ is mapped onto the tangent space at
\cal{$\Phi$}$^t(\vec{x})$ by the differential
$d_{\vec{x}}\mbox{\cal{$\Phi$}}^t$ according to equation
(\ref{eq:t_flow}). The action of \cal{$\Phi$}$^t(\vec{x})$ is given by
equation (\ref{eq:w_ham}) for continuous systems and by equation
(\ref{eq:w0_map}) for discrete ones. Thus, the action of
$d_{\vec{x}}\mbox{\cal{$\Phi$}}^t$ on a particular initial deviation vector
$\vec{w}$ of the tangent space is given by the multiplication of matrix
$\textbf{Y}(t)$ for continuous systems or $\textbf{Y}_n$ for discrete systems
with vector $\vec{w}$.  From equations (\ref{eq:w_ham}) and (\ref{eq:w0_map})
we see that the action of $d_{\vec{x}}\mbox{\cal{$\Phi$}}^t$ over two
successive time intervals $t$ and $s$ satisfies the  composition law
\begin{equation}
d_{\vec{x}}\mbox{\cal{$\Phi$}}^{t+s}=d_{\mbox{\cal{$\Phi$}}^s (\vec{x})}
\mbox{\cal{$\Phi$}}^t \circ d_{\vec{x}}\mbox{\cal{$\Phi$}}^s.
\label{eq:d_Phi_ts} 
\end{equation} 
This equation can be written in the form 
\begin{equation}
\textbf{R} (t+s,\vec{x})=\textbf{R}(t, \mbox{\cal{$\Phi$}}^s(\vec{x})) \cdot
\textbf{R}(s,\vec{x}),
\label{eq:R} 
\end{equation} 
where $\textbf{R}(t,\vec{x})$ is the matrix corresponding to
$d_{\vec{x}}\mbox{\cal{$\Phi$}}^t$. We note that since
$\textbf{Y}(0)=\textbf{Y}_0=\textbf{I}_{2N}$ we get
$d_{\vec{x}}\mbox{\cal{$\Phi$}}^0\vec{w}=\vec{w}$ and
$\textbf{R}(0,\vec{x})=\textbf{I}_{2N}$. A function $\textbf{R}(t,\vec{x})$
satisfying relation (\ref{eq:R}) is called a \textit{multiplicative cocycle}
with respect to the dynamical system \cal{$\Phi$}$^t$.

Let $\cal{S}$ be a measure space with a normalized measure $\mu$ such
that
\begin{equation}
\mu(\mbox{$\cal{S}$})=1\,\,\, , \,\,\, \mu\left(\mbox{\cal{$\Phi$}}^t \cal{A}
\right) = \mu(\cal{A})
\label{eq:measure} 
\end{equation} 
for $\mbox{$\cal{A}$} \subset \cal{S}$. Suppose also that a smooth Riemannian
metric $\|\, \|$ is defined on $\cal{S}$. We consider the multiplicative
cocycle $\textbf{R}(t,\vec{x})$ corresponding to
$d_{\vec{x}}\mbox{\cal{$\Phi$}}^t$ and we are interested in its asymptotic
behavior for $t \rightarrow \pm \infty$. Since, as mentioned by Oseledec
\cite{O_68}, the case $t \rightarrow + \infty$ is analogous to the case $t
\rightarrow - \infty$, we restrict our treatment to the case $t \rightarrow +
\infty$, where time is increasing. In order to clarify what we are practically
interested in let us consider a nonzero vector $\vec{w}$ of the tangent space
$\cal{T}_{\vec{x}} \cal{S}$ at $\vec{x}$. Then the quantity
\[
\lambda_t (\vec{x})=\frac{\| d_{\vec{x}}\mbox{\cal{$\Phi$}}^t \vec{w} \|}{\|
\vec{w} \|}
\]
is called the \textit{coefficient of expansion in the direction of
$\vec{w}$}. If
\begin{displaymath}
\limsup_{t\rightarrow  \infty} \frac{1}{t} \ln \lambda_t(\vec{x}) > 0
\end{displaymath}
we say that exponential diverge occurs in the direction of $\vec{w}$. Of
course the basic question we have to answer is whether the
\textit{characteristic exponent} (also called \textit{characteristic exponent
of order 1})
\begin{displaymath}
\lim_{t\rightarrow  \infty} \frac{1}{t} \ln \lambda_t(\vec{x})
\end{displaymath} 
exists.

We will answer this question in a more general framework without restricting
ourselves to multiplicative cocycles. So, the results presented in the
following Section \ref{Def_LCE}, are valid for a general class of matrix
functions, a subclass of which contains the multiplicative cocycles which are
of more practical interest to us, since they describe the time evolution of
deviation vectors for the dynamical systems we study.

\subsection{Definitions and basic theorems}
\label{Def_LCE}

Let $\textbf{A}_t$ be an $n \times n$ matrix function defined either on the
whole real axis or on the set of integers, such that
$\textbf{A}_0=\textbf{I}_n$, for each time $t$ the value of function
$\textbf{A}_t$ is a nonsingular matrix and $\| \textbf{A}_t \|$ the usual
2--norm of $\textbf{A}_t$\footnote[2]{The 2--norm $\| \textbf{A} \|$ of an $n
\times n$ matrix \textbf{A} is induced by the 2--norm of vectors, i.~e.~the
usual Euclidean norm $\| \vec{x}\| = \left( \sum_{i=1}^{n} x_i^2
\right)^{1/2}$, by
\begin{displaymath}
\| \textbf{A} \| = \max_{\vec{x} \neq 0} \frac{\| \textbf{A} \vec{x}\|}{\|
\vec{x}\|}
\end{displaymath}
and is equal to the largest eigenvalue of matrix
$\sqrt{\textbf{A}^{\mathrm{T}} \textbf{A} }$.}. In particular, we consider only
matrices $\textbf{A}_t$ satisfying
\begin{equation}
\max \left\lbrace \| \textbf{A}_t \|,\| \textbf{A}_t^{-1} \| \right\rbrace
\leq e^{ct}
\label{eq:max_At} 
\end{equation} 
with $c>0$ a suitable constant.

\begin{definition}
  Considering a matrix function $\textbf{A}_t$ as above and a nonzero 
  vector $\vec{w}$ of the Euclidian space $\mathbb{R}^n$ the quantity
\begin{equation}
\chi(\textbf{A}_t,\vec{w}) = \limsup_{t\rightarrow \infty} \frac{1}{t} \ln \|
\textbf{A}_t \vec{w} \|
\label{eq:1_LCE} 
\end{equation} 
is called the \texttt{1-dimensional Lyapunov Characteristic Exponent} or the
\texttt{Lyapunov Characteristic Exponent of order 1 (1-LCE)} of
$\textbf{A}_t$ with respect to vector $\vec{w}$.
\label{def:1-LCE}
\end{definition}
For simplicity we will usually refer to 1--LCEs as LCEs. 

We note that the value of the norm $\| \vec{w}\|$ does not influence the value
of $\chi(\textbf{A}_t,\vec{w})$. For example, considering a vector $\beta
\vec{w}$, with $\beta \in \mathbb{R}$ a nonzero constant, instead of $
\vec{w}$ in Definition \ref{def:1-LCE}, we get the extra term $\ln |\beta| /t$
(with $|\,\,|$ denoting the  absolute value) in equation (\ref{eq:1_LCE})
whose limiting value for $t\rightarrow \infty$ is zero and thus does not
change the value of $\chi(\textbf{A}_t,\vec{w})$. More importantly, the value
of the LCE is independent of the norm appearing in equation
(\ref{eq:1_LCE}). This can be easily seen as follows: Let us consider a second
norm $\|\,\,\|'$ satisfying the inequality
\begin{displaymath}
\beta_1 \| \vec{w}\| \leq \| \vec{w}\|' \leq \beta_2 \| \vec{w}\|
\end{displaymath}
for some positive real numbers $\beta_1$, $\beta_2$, and for all vectors $
\vec{w}$. Such norms are called \textit{equivalent} (see e.g.~\cite[\S
5.4.7]{HornJ_1985}). Then, by the above--mentioned argument it is easily seen
that the use of norm $\|\,\,\|'$ in (\ref{eq:1_LCE}) leaves unchanged the
value of $\chi(\textbf{A}_t,\vec{w})$. Since all norms of finite dimensional
vector spaces are equivalent, we conclude that the LCEs do not depend on the
chosen norm.

Let $\vec{w}_i$, $i=1,2,\ldots,p$ be a set of linearly independent vectors in
$\mathbb{R}^n$, $E^p$ be the subspace generated by all $\vec{w}_i$ and
$\mbox{vol}_p( \textbf{A}_t, E^p)$ be the volume of the $p$--parallelogram
having as edges the $p$ vectors $\textbf{A}_t \vec{w}_i$. This volume is
computed as the norm of the wedge product of these vectors (see Appendix
\ref{Wedge} for the definition of the wedge product and the actual evaluation
of the volume)
\begin{displaymath}
\mbox{vol}_p( \textbf{A}_t, E^p)= \| \textbf{A}_t \vec{w}_1 \wedge
\textbf{A}_t \vec{w}_2 \wedge \cdots \wedge \textbf{A}_t \vec{w}_p \|.
\end{displaymath} 
Let also $\mbox{vol}_p( \textbf{A}_0, E^p)$ be the volume of the initial
$p$--parallelogram defined by all $\vec{w}_i$, since $\textbf{A}_0$ is the
identity matrix. Then the quantity
\[
\lambda_t (E^p)=\frac{\mathrm{vol}_p( \textbf{A}_t, E^p) }{ \mathrm{vol}_p(
\textbf{A}_0, E^p)}
\]
is called the \textit{coefficient of expansion in the direction of $E^p$} and
it depends only on $E^p$ and not on the choice of the linearly independent set
of vectors. Obviously for an 1--dimensional subspace $E^1$ the coefficient of
expansion is $\| \textbf{A}_t \vec{w}_1 \|/ \|\vec{w}_1\| $.  If the limit
\begin{displaymath}
\lim_{t\rightarrow  \infty} \frac{1}{t} \ln \lambda_t(E^p)
\end{displaymath} 
exits it is called the \textit{characteristic exponent of order p in the
direction of $E^p$}.

\begin{definition}
  Considering the linearly independent set $\vec{w}_i$, $i=1,2,\ldots,p$ and
  the corresponding subspace $E^p$ of $\mathbb{R}^n$ as above, the
  \texttt{p-dimensional Lyapunov Characteristic Exponent} or the
  \texttt{Lyapunov Characteristic Exponent of order p (p-LCE)} of
  $\textbf{A}_t$ with respect to subspace $E^p$ is defined as
\begin{equation}
\chi(\textbf{A}_t,E^p) = \limsup_{t\rightarrow \infty} \frac{1}{t} \ln
\mathrm{vol}_p( \textbf{A}_t, E^p).
\label{eq:p_LCE} 
\end{equation} 
\label{def:p-LCE}
\end{definition}
Similarly to the case of the 1--LCE, the value of the initial volume
$\mbox{vol}_p( \textbf{A}_0, E^p)$, as well as the used norm, do not influence
the value of $\chi(\textbf{A}_t,E^p)$.

From (\ref{eq:max_At}) and the Hadamard inequality (see e.~g.~\cite{O_68}),
according to which the Euclidean volume of a $p$--parallelogram does not
exceed the product of the lengths of its sides, we conclude that the LCEs of
equations (\ref{eq:1_LCE}) and (\ref{eq:p_LCE}) are finite.

From the definition of the LCE it follows that
\begin{displaymath}
\chi(\textbf{A}_t,c_1\vec{w}_1+c_2\vec{w}_2) \leq \max \left\lbrace
\chi(\textbf{A}_t,\vec{w}_1), \chi(\textbf{A}_t,\vec{w}_2) \right\rbrace
\end{displaymath} 
for any two vectors $\vec{w}_1,\vec{w}_2\in \mathbb{R}^n$ and $c_1, c_2 \in
\mathbb{R}$ with $c_1, c_2 \neq 0$, while the Hadamard inequality implies that
if $\vec{w}_i$, $i=1,2,\ldots,n$ is a basis of $\mathbb{R}^n$ then
\begin{equation}
\sum_{i=1}^n \chi(\textbf{A}_t,\vec{w}_i) \geq \limsup_{t\rightarrow \infty}
\frac{1}{t} \ln |\det \textbf{A}_t |
\label{eq:s_chi}
\end{equation} 
where $\det \textbf{A}_t$ is the determinant of matrix $ \textbf{A}_t$. 

It can
be shown that for any $r \in \mathbb{R}$ the set of vectors $\left\lbrace
\vec{w} \in \mathbb{R}^n : \chi(\textbf{A}_t,\vec{w}) \leq r\right\rbrace$ is
a vector subspace of $\mathbb{R}^n$ and that the function
$\chi(\textbf{A}_t,\vec{w})$ with $\vec{w} \in \mathbb{R}^n$, $\vec{w} \neq 0$
takes at most $n$ different values, say
\begin{equation}
\nu_1  > \nu_2  > \cdots >\nu_s  \,\,\, \mbox{with} \,\,\, 1\leq s\leq n. 
\label{eq:val_lyap}
\end{equation} 
For the subspaces
\begin{equation}
L_i=\left\lbrace \vec{w} \in \mathbb{R}^n : \chi(\textbf{A}_t,\vec{w}) \leq
\nu_i \right\rbrace
\label{eq:L_i_sub}
\end{equation} 
we have 
\begin{equation}
\mathbb{R}^n = L_1 \supset L_2 \supset \cdots \supset L_{s} \supset L_{s+1}
\stackrel{\mbox{def}}{=} \left\lbrace 0\right\rbrace ,
\label{eq:L_i_sequence}
\end{equation} 
with $L_{i+1} \neq L_{i}$ and $\chi (\textbf{A}_t,\vec{w})=\nu_i$ if and only
if $\vec{w} \in L_i \setminus L_{i+1}$ for $i=1,2, \ldots, s$. So in
descending order each LCE `lives' in a space of dimensionality less than that
of the preceding exponent. Such a structure of linear spaces with decreasing
dimension, each containing the following one, is called a \textit{filtration}.

\begin{definition}
  A basis $\vec{w}_i$, $i=1,2,\ldots,n$ of $\mathbb{R}^n$ is called normal if
  $\sum_{i=1}^n \chi(\textbf{A}_t,\vec{w}_i)$ attains a minimum at this
  basis. In other words, the basis $\vec{w}_i$, is a \texttt{normal basis} if
\begin{displaymath}
\sum_{i=1}^n \chi(\textbf{A}_t,\vec{w}_i) \leq \sum_{i=1}^n
\chi(\textbf{A}_t,\vec{g}_i)
\end{displaymath} 
where $\vec{g}_i$, $i=1,2,\ldots,n$ is any other basis of
$\mathbb{R}^n$.
\label{def:normal_basis}
\end{definition}
A normal basis $\vec{w}_i$, $i=1,2,\ldots,n$ is not unique but the numbers $
\chi(\textbf{A}_t,\vec{w}_i)$ depend only on $\textbf{A}_t$ and not on the
particular normal basis and are called the LCEs of function $\textbf{A}_t$. By
a possible permutation of the vectors of a given normal basis we can always
assume that $\chi(\textbf{A}_t,\vec{w}_1) \geq \chi(\textbf{A}_t,\vec{w}_2)
\geq \cdots \geq \chi(\textbf{A}_t,\vec{w}_n)$.

\begin{definition}
  Let $\vec{w}_i$, $i=1,2,\ldots,n$ be a normal basis of $\mathbb{R}^n$ and
  $\chi_1 \geq \chi_2 \geq \cdots \geq \chi_n$, with $\chi_i \equiv
  \chi(\textbf{A}_t,\vec{w}_i)$, $i=1,2,\ldots,n$, the LCEs of these
  vectors. Assume that value $\nu_i$, $i=1,2,\ldots,s$ appears exactly
  $k_i=k_i(\nu_i)>0$ times among these numbers. Then $k_i$ is called the
  \texttt{multiplicity} of value $\nu_i$ and the collection $(\nu_i,k_i)$
  $i=1,2,\ldots,s$ is called the \texttt{spectrum of LCEs}.
\label{def:spectrum}
\end{definition}

In order to clarify the used notation we stress that $\chi_i$, $i=1,2,\ldots,
n$ are the $n$ (possibly nondistinct) LCEs, satisfying $\chi_1 \geq \chi_2
\geq \cdots \geq \chi_n$, while $\nu_i$, $i=1,2,\ldots, s$ represent the $s$
($1\leq s \leq n$), different values the LCEs have, with $\nu_1 > \nu_2 >
\cdots >\nu_s$.

\begin{definition}
  The matrix function $\textbf{A}_t$ is called \texttt{regular} as
  $t\rightarrow \infty$ if for each normal basis $\vec{w}_i$, $i=1,2,\ldots,n$
  it holds that
\begin{displaymath}
\sum_{i=1}^n \chi(\textbf{A}_t,\vec{w}_i) = \liminf_{t\rightarrow \infty}
\frac{1}{t} \ln |\det \textbf{A}_t |,
\end{displaymath} 
which, due to (\ref{eq:s_chi}) leads to 
\begin{displaymath}
\liminf_{t\rightarrow \infty} \frac{1}{t} \ln |\det \textbf{A}_t | =
\limsup_{t\rightarrow \infty} \frac{1}{t} \ln |\det \textbf{A}_t |
\end{displaymath}
guaranteeing that the limit
\begin{displaymath}
\lim_{t\rightarrow \infty} \frac{1}{t} \ln |\det \textbf{A}_t | 
\end{displaymath}
exists, is finite and equal to
\begin{displaymath}
\lim_{t\rightarrow \infty} \frac{1}{t} \ln |\det \textbf{A}_t |= \sum_{i=1}^n
\chi(\textbf{A}_t,\vec{w}_i) = \sum_{i=1}^s k_i \nu_i.
\end{displaymath} 
\label{def:regular}
\end{definition}

We can now state a very important theorem for the LCEs:
\begin{theorem}
  If the matrix function $\textbf{A}_t$ is regular then the LCEs of all orders
  are given by equations (\ref{eq:1_LCE}) and (\ref{eq:p_LCE}) where
  the $\displaystyle \limsup_{t\rightarrow \infty}$ is substituted by
  $\displaystyle \lim_{t\rightarrow \infty}$
\begin{equation}
\chi(\textbf{A}_t,\vec{w}) = \lim_{t\rightarrow \infty} \frac{1}{t} \ln \|
\textbf{A}_t \vec{w} \|
\label{eq:1_LCE_f} 
\end{equation} 
\begin{equation}
\chi(\textbf{A}_t,E^p) = \lim_{t\rightarrow \infty} \frac{1}{t} \ln
\mathrm{vol}_p( \textbf{A}_t, E^p).
\label{eq:p_LCE_f} 
\end{equation} 
In particular, for any $p$--dimensional subspace $E^p \subseteq
\mathbb{R}^n$ we have
\begin{equation}
 \chi(\textbf{A}_t,E^p) = \sum_{j=1}^p \chi_{i_j}.
\label{eq:sum_chi_1}
\end{equation} 
with a suitable sequence $1 \leq i_1 \leq i_2 \leq \cdots \leq i_p \leq n$.
\label{Theorem:existense}
\end{theorem}
The part of the theorem concerning equations (\ref{eq:1_LCE_f}) and
(\ref{eq:p_LCE_f}) was proved by Oseledec in \cite{O_68}, while equation
(\ref{eq:sum_chi_1}), although was not explicitly proved in \cite{O_68}, can
be considered as a rather easily proven byproduct of the results presented
there. Actually, the validity of equation (\ref{eq:sum_chi_1}) was shown in
\cite{BGGS_80a}.

\subsection{Computing LCEs of order 1}
\label{Theory_comp_1}

Let us now discuss how we can use Theorem \ref{Theorem:existense} for the
numerical computation of LCEs, starting with the computation of LCEs of order
1.

As we have already mentioned in (\ref{eq:val_lyap}), the LCE  takes at
most $n$ different values $\nu_i$ , $i=1,2,\ldots, s$, $1 \leq s \leq n$. If
we could know a priori the sequence (\ref{eq:L_i_sequence}) of subspaces $L_i$
$i=1,2,\ldots, s$ of $\mathbb{R}^n$ we would, in principle, be able to compute
the values $\nu_i$ of all LCEs. This could be done by taking an initial
 vector $\vec{w}_i \in L_i\setminus L_{i+1}$ and compute
\begin{equation}
\nu_i= \lim_{t \rightarrow \infty} \frac{1}{t} \ln \|\textbf{A}_t \vec{w}_i
\|\;\;, \;\; i=1,2,\ldots,s.
\label{eq:num_x_i}
\end{equation} 
Now apart from $L_1=\mathbb{R}^n$ all the remaining subspaces $L_i$,
$i=2,3,\ldots, s$ have positive codimension codim$(L_i)$ ($=\dim \mathbb{R}^n
- \dim L_i >0$) and thus, vanishing Lebesgue measure. Then a random choice of
$\vec{w} \in \mathbb{R}^n$ would lead to the computation of $\chi_1$ from
(\ref{eq:num_x_i}), because, in principle $\vec{w}$ will belong to $L_1$ and
not to the subspaces $L_i$ $i=2,\ldots, s$. Let us consider a simple example
in order to clarify this statement.

\begin{figure}
\centerline{ 
\begin{tabular}{cc}
\includegraphics[scale=0.7]{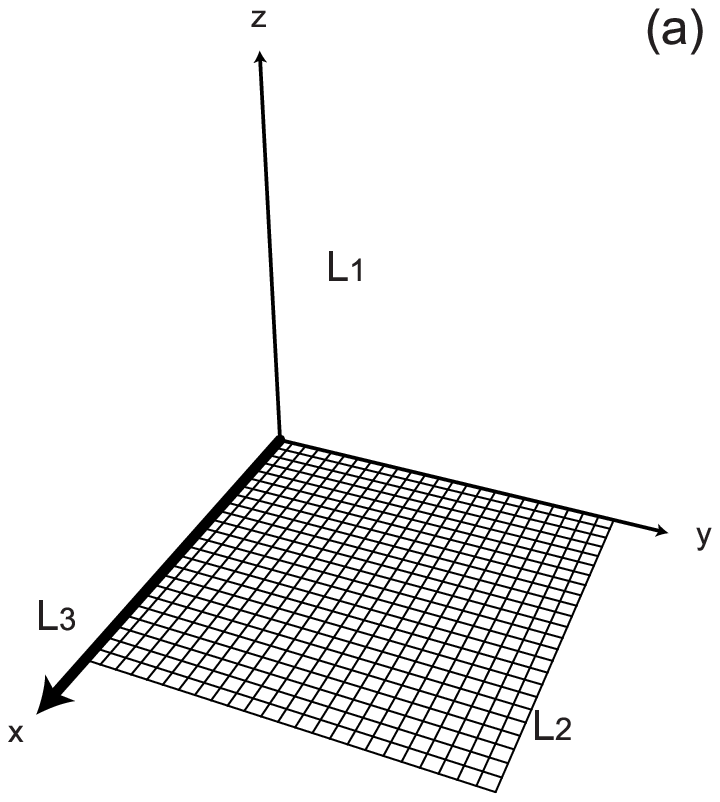} &
\includegraphics[scale=0.7]{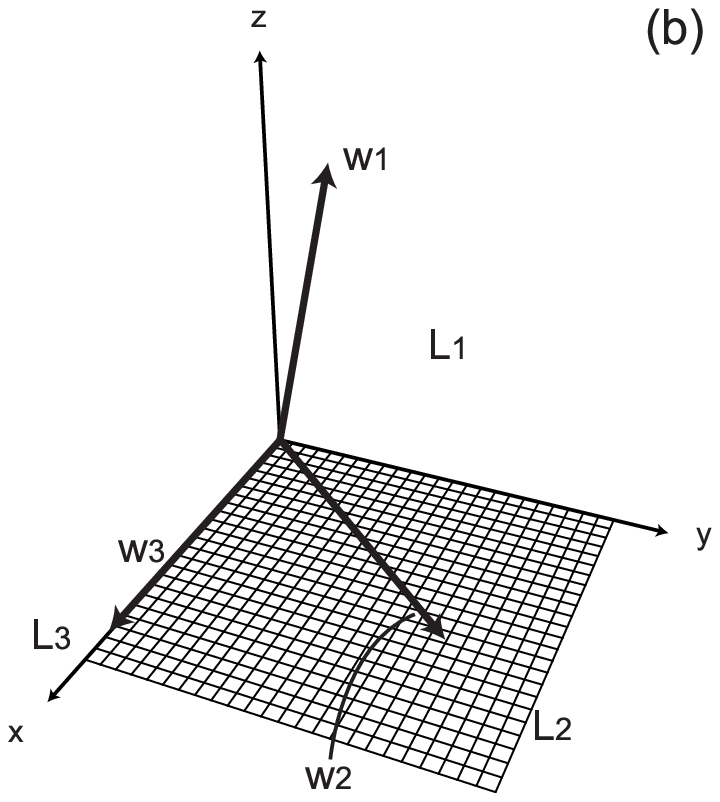}\end{tabular}}
\caption{(a) A schematic representation of the sequence of subspaces
  (\ref{eq:L_i_sequence}) where $L_1$ identifies with $\mathbb{R}^3$, $L_2
  \subset L_1$ is represented by the $xy$ plane and the $x$ axis is considered
  as the final subspace $L_3 \subset L_2$. (b) A random choice of a vector in
  $L_1\equiv \mathbb{R}^3$ will result with probability one to a vector
  belonging to $L_1$ and not to $L_2$, like vector $\vec{w}_1$.  Vectors
  $\vec{w}_2$, $\vec{w}_3$ belonging respectively to $L_2 \setminus L_3$ and
  to $L_3$ are not random since their coordinates should satisfy certain
  conditions. In particular, the $z$ coordinate of $\vec{w}_2$ should be zero,
  while both the $z$ and $y$ coordinate of $\vec{w}_3$ should vanish. The use
  of $\vec{w}_1$, $\vec{w}_2$, $\vec{w}_3$ in (\ref{eq:num_x_i}) leads to the
  computation of $\chi_1$, $\chi_2$ and $\chi_3$ respectively. }
\label{f:R3_a}
\end{figure}

Suppose that $L_1$ is the usual $3$--dimensional space $\mathbb{R}^3$, $L_2
\subset L_1$ is a particular 2--dimensional plane of $\mathbb{R}^3$, e.~g.~the
plane $z=0$, $L_3 \subset L_2$ is a particular 1--dimensional line e.~g.~the
$x$ axis (Figure \ref{f:R3_a}(a)) and the corresponding LCEs are $\chi_1 >
\chi_2 > \chi_3$ with multiplicities $k_1=k_2=k_3=1$.
For this case we have $\dim L_1=3$, $\dim L_2=2$, $\dim L_3=1$ and
codim$(L_1)=0$, codim$(L_2)=1$, codim$(L_3)=2$. Concerning the measures $\mu$
of these subspaces of $\mathbb{R}^3$, it is obvious that
$\mu(L_2)=\mu(L_3)=0$, since the measure of a surface or of a line in the
$3$--dimensional space $\mathbb{R}^3$ is zero.

If we randomly choose a vector $\vec{w} \in \mathbb{R}^3$ it will belong to
$L_1$ and not to $L_2$, i.~e.~having its $z$ coordinate different from zero
and thus, equation (\ref{eq:num_x_i}) would lead to the computation of the mLCE
$\chi_1$. Vector $\vec{w}_1$ in Figure \ref{f:R3_a}(b) represents such a
random choice.  In order to compute $\chi_2$ from (\ref{eq:num_x_i}) we should
choose vector $\vec{w}$ not randomly but in a specific way. In particular, it
should belong to $L_2$ but not to $L_3$, so its $z$ coordinate should be equal
to zero. Thus this vector should have the form $\vec{w}=(w_1,w_2,0)$ with
$w_1, w_2 \in \mathbb{R}$, $w_2 \neq 0$, like vector $\vec{w}_2$ in Figure
\ref{f:R3_a}(b). Our choice will become even more specific if we would like to
compute $\chi_3$ because in this case $\vec{w}$ should be of the form
$\vec{w}=(w_1,0,0)\neq \vec{0}$ with $w_1 \in \mathbb{R}$. Vector $\vec{w}_3$
of Figure \ref{f:R3_a}(b) is a choice of this kind.

From this example it becomes evident that a random choice of vector $\vec{w}$
in (\ref{eq:num_x_i}) will lead to the computation of the largest LCE $\chi_1$
with probability one. One more comment concerning the numerical implementation
of equation (\ref{eq:num_x_i}) should be added here. Even if in some special
examples one could happen to know a priori the subspaces $L_i$ $i=1,2,\ldots,
s$, so that one could choose $\vec{w} \in L_i \setminus L_{i+1}$ with $i\neq
1$ then the computational errors would eventually lead to the numerical
computation of $\chi_1$. Such an example was presented in \cite{BGGS_80b}.

\subsection{Computing LCEs of order $p>1$}
\label{Theory_comp_p}

Let us now turn our attention to the computation of $p$--LCEs with
$p>1$. Equation (\ref{eq:sum_chi_1}) of Theorem \ref{Theorem:existense}
actually tells us that the $p$--LCE $\chi(\textbf{A}_t,E^p)$ can take at most
$ \left( \begin{array}{c} n \\p
  \end{array} \right)$ distinct values, i.~e.~as many as all the possible sums of $p$ 1--LCEs out of $n$ are. Now, as the  choice of a random vector $\vec{w} \in \mathbb{R}^n$, or in other words, of a random 1--dimensional subspace of
$\mathbb{R}^n$ produced by $\vec{w}$, leads to the computation of the maximal
1--LCE, the random choice of a $p$--dimensional subspace $E^p$ of
$\mathbb{R}^n$, or equivalently the random choice of $p$ linearly independent
vectors $\vec{w}_i$ $i=1,2,\ldots, p$, leads to the computation of the
\textit{maximal $p$--LCE ($p$--mLCE)} which is equal to the sum of the $p$
largest 1--LCEs
\begin{equation}
 \chi(\textbf{A}_t,E^p) = \sum_{i=1}^p \chi_{i}.
\label{eq:pLCE_sum}
\end{equation} 
This relation was formulated explicitly in \cite{BGGS_78,BG_79} and proved in
\cite{BGGS_80a} but was implicitly contained in \cite{O_68}. The practical
importance of equation (\ref{eq:pLCE_sum}) was also clearly explained in
\cite{SN_79}. Benettin et al.~\cite{BGGS_80a} gave a more rigorous form to the
notion of the random choice of $E^p$, which is essential for the derivation of
(\ref{eq:pLCE_sum}), by introducing a condition that subspace $E^p$ should
satisfy. They named this condition \textit{Condition R} (at random). According
to Condition R a $p$--dimensional space $E^p \subset \mathbb{R}^n$ is chosen
at random if for all $j=2,3,\ldots,s$ we have
\begin{equation}
\dim (E^p \cap L_j) = \max \left\lbrace 0, p-\sum_{i=1}^{j-1} k_i\right\rbrace 
\label{eq:condR}
\end{equation} 
where $L_j$ belongs to the sequence of subspaces (\ref{eq:L_i_sequence}) and
$k_i$ is the multiplicity of the LCE $\nu_i$ (Definition \ref{def:spectrum}).

In order to clarify these issues let us consider again the example presented
in Figure \ref{f:R3_a}, where we have three distinct values for the 1--LCEs
$\chi_1> \chi_2> \chi_3$ with multiplicities $k_1=k_2=k_3=1$. In this case the
2--LCE can take one of the three possible values $\chi_1 + \chi_2$, $\chi_2 +
\chi_3$, $\chi_1 + \chi_3$, while the 3--LCE takes only one possible value,
namely $\chi_1 + \chi_2 + \chi_3$.

The computation of the 2--LCE requires the choice of two linearly independent
vectors $\vec{w}_1$, $\vec{w}_2$ and the application of equation
(\ref{eq:p_LCE_f}).  The two vectors $\vec{w}_1$, $\vec{w}_2$ define a
2--dimensional plane $E^2$ in $\mathbb{R}^3$ and $\chi(\textbf{A}_t,E^2)$
practically measures the time rate of the coefficient of expansion of the
surface of the parallelogram having as edges the vectors $\textbf{A}_t
\vec{w}_1$, $\textbf{A}_t \vec{w}_2$.

\begin{figure}
\centerline{ 
\begin{tabular}{cc}
\includegraphics[scale=0.65]{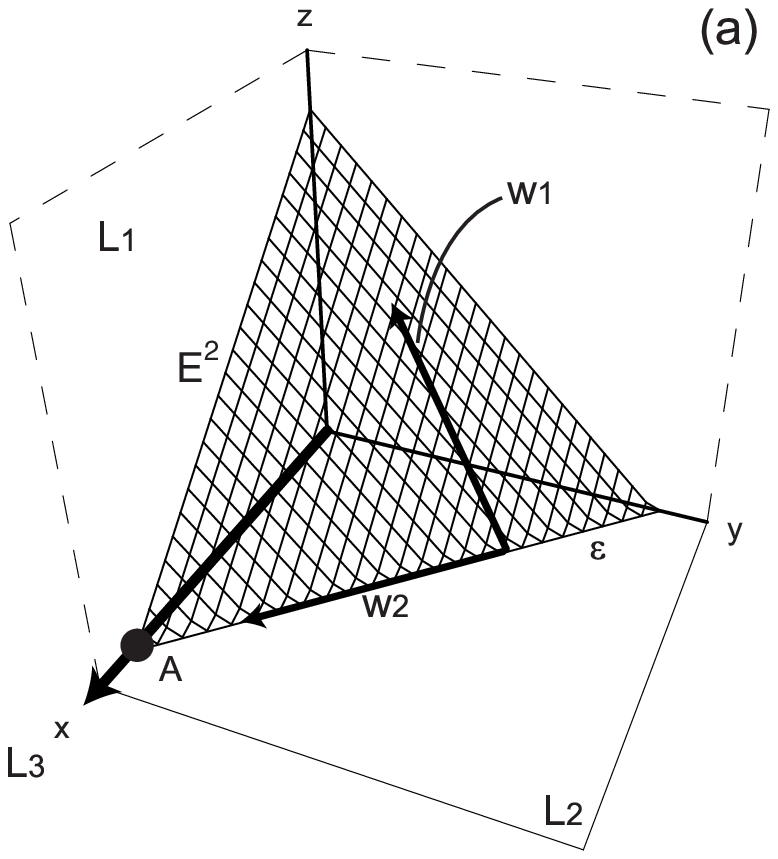} &
\includegraphics[scale=0.65]{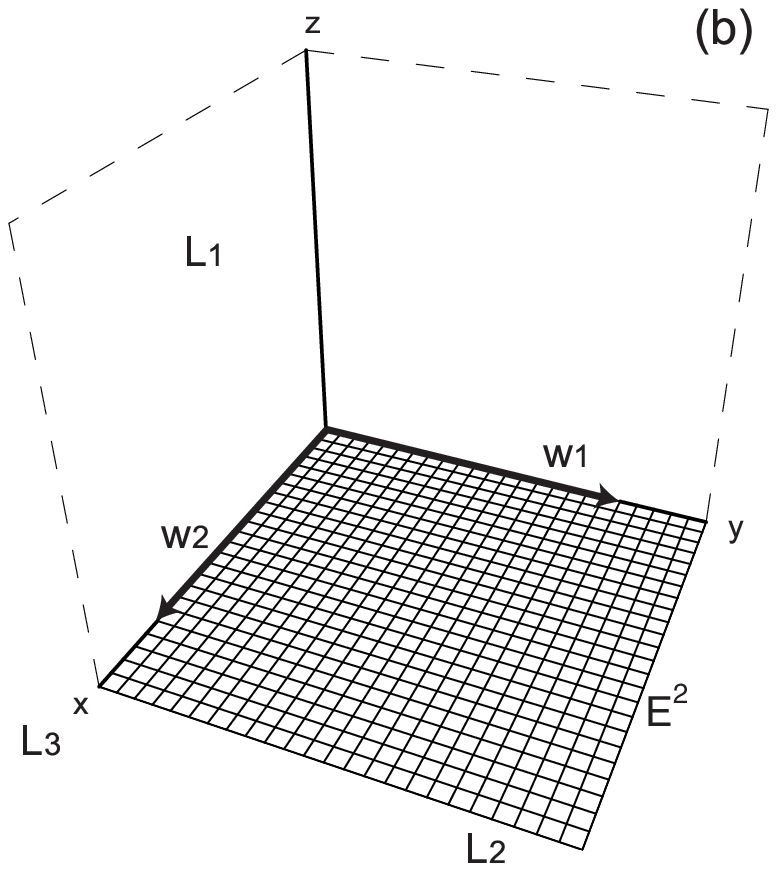}\\
\multicolumn{2}{c}{\includegraphics[scale=0.65]{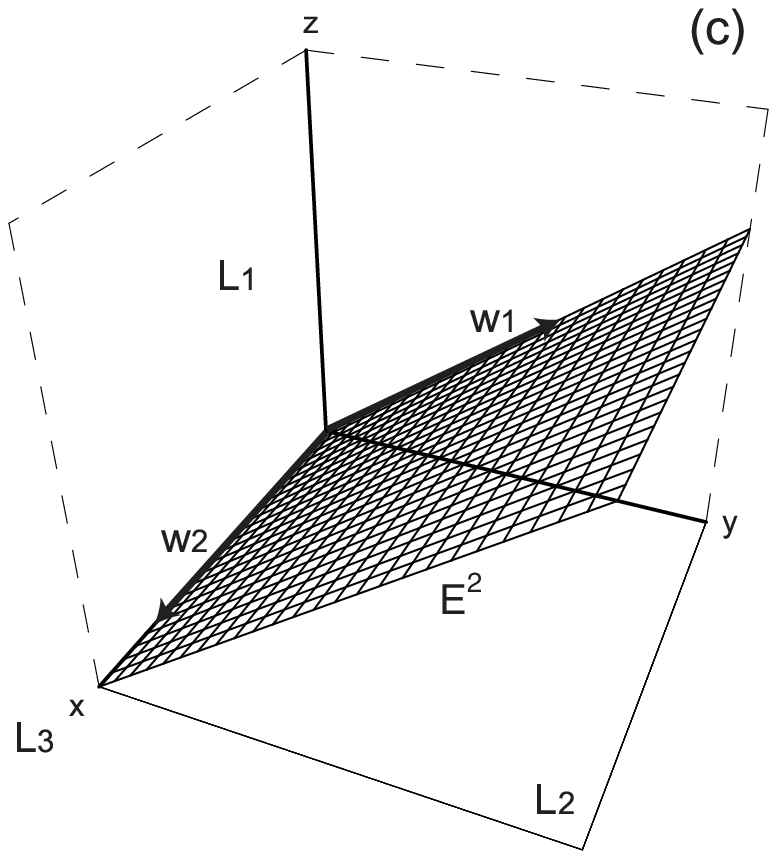}}
\end{tabular}}
\caption{ Possible choices of the 2--dimensional space $E^2$ for the
  computation of the 2--LCE in the example of Figure \ref{f:R3_a}, where
  $\mathbb{R}^3$ is considered as the tangent space of a hypothetical
  dynamical system. In each panel the chosen `plane' $E^2$ is drawn, as well
  as one of its possible basis constituted of vectors $\vec{w}_1$,
  $\vec{w}_2$. (a) a random choice of $E^2$ leads to a plane intersecting
  $L_2$ along line $\epsilon$ ($\dim (E^2 \cap L_2) = 1$) and $L_1$ at point A
  ($\dim (E^2 \cap L_3) = 0$). In this case equation (\ref{eq:p_LCE_f}) gives
  $\chi(\textbf{A}_t,E^2) = \chi_1+\chi_2$. More carefully made choices of
  $E^2$ (which are obviously not made at random) results to configurations
  leading to the computation of $\chi_2+\chi_3$ (b) and $\chi_1+\chi_3$ (c)
  from equation (\ref{eq:p_LCE_f}). In these cases $E^2$ does not satisfy
  Condition R (\ref{eq:condR}) since $\dim (E^2 \cap L_2) = 2$, $\dim (E^2
  \cap L_3) = 1$ in (b) and $\dim (E^2 \cap L_2) = 1$, $\dim (E^2 \cap L_3) =
  1$ in (c).  }
\label{f:R3_b}
\end{figure}

By choosing the two vectors $\vec{w}_1$, $\vec{w}_2$ randomly we define a
random plane $E^2$ in $\mathbb{R}^3$ which intersects the subspace $L_2$
(plane $xy$) along a line, i.~e.~$\dim (E^2 \cap L_2) = 1$ and the subspace
$L_3$ ($x$ axis) at a point, i.~e.~$\dim (E^2 \cap L_3) = 0$ (Figure
\ref{f:R3_b}(a)).
This random choice of plane $E^2$ satisfies Condition R (\ref{eq:condR}) and
thus, equation (\ref{eq:p_LCE_f}) leads to the computation of the 2--mLCE,
namely $\chi_1 + \chi_2$. This result can be also understood in the following
way. Plane $E^2$ in Figure \ref{f:R3_b}(a) can be considered to be spanned by
two vectors $\vec{w}_1$, $\vec{w}_2$ such that $\vec{w}_1 \in L_1$ but not in
its subspace $L_2$ and $\vec{w}_2 \in L_2 $ but not in its subspace
$L_3$. Then the expansion of $\vec{w}_1 \in L_1 \setminus L_2$ is determined
by the LCE $\chi_1$ and the expansion of $\vec{w}_2 \in L_2 \setminus L_3$ by
the LCE $\chi_2$. These 1--dimensional expansion rates result to an expansion
rate equal to $\chi_1 + \chi_2$ for the surface defined by the two vectors.

Other more carefully designed choices of the $E^2$ subspace lead to the
computation of the other possible values of the 2--LCE. If for example
$\vec{w}_1 \in L_2 \setminus L_3$ and $\vec{w}_2 \in L_3$ (Figure
\ref{f:R3_b}(b)) we have $E^2=L_2$ with $\dim (E^2 \cap L_2) = 2$ and $\dim
(E^2 \cap L_3) = 1$. In this case the expansion of $\vec{w}_1$ is determined
by the LCE $\chi_2$ and of $\vec{w}_2$ by $\chi_3$, and so the computed 2--LCE
is $\chi_2 + \chi_3$. Finally, a choice of $E^2$ of the form presented in
Figure \ref{f:R3_b}(c) leads to the computation of $\chi_1 + \chi_3$. In this
case the plane $E^2$ is defined by $\vec{w}_1 \in L_1 \setminus L_2$ and
$\vec{w}_2 \in L_3$ and intersects subspaces $L_2$ and $L_3$ along the line
corresponding to $L_3$, i.~e.~$\dim (E^2 \cap L_2) = 1$ and $\dim (E^2 \cap
L_3) = 1$. It can be easily checked that for the last two choices of $E^2$
(Figures \ref{f:R3_b}(b) and (c)), for which the computed 2--LCE does not take
its maximal possible value, Condition R (\ref{eq:condR}) is not satisfied, as
one should have expected from the fact that these choices correspond to
carefully designed configurations and not to a random process.

Similarly to the case of the computation of the 1--LCEs we note that, even if
in some exceptional case one could know a priori the subspaces $L_i$
$i=1,2,\ldots, s$, so that one could choose $\vec{w}_i$ $i=1,2,\ldots, p$ to
span a particular subspace $E^p$ in order to compute a specific value of the
$p$--LCE, smaller than $\sum_{i=1}^p \chi_i$ (like in Figures \ref{f:R3_b}(b)
and (c)), the inevitable computational errors would eventually lead to the
numerical computation of the maximal possible value of the $p$--LCE.

Summarizing we point out that the practical implementation of Theorem
\ref{Theorem:existense} guarantees that a random choice of $p$ initial vectors
$\vec{w}_i$ $i=1,2,\ldots, p$ with $1\leq p \leq n$ generates a space $E^p$
which satisfies Condition R (\ref{eq:condR}) and leads to the actual
computation of the corresponding $p$--mLCE, namely $\chi_1+\chi_2+\ldots
+\chi_p$. This statement, which was originally presented in
\cite{BGGS_78,BG_79}, led to the
standard algorithm for the computation of all LCEs presented in
\cite{BGGS_80b}. This algorithm is analyzed in Section \ref{allLCEs_Benettin}.

\subsection{The Multiplicative Ergodic Theorem}
\label{Oseledec}

After presenting results concerning the existence and the computation of the
LCEs of all orders for a general matrix function $\textbf{A}_t$, let us
restrict our study to the case of multiplicative cocycles
$\textbf{R}(t,\vec{x})$, which are matrix functions satisfying equation
(\ref{eq:R}). The multiplicative cocycles arise naturally in discrete and
continuous dynamical systems as was explained in the beginning of Section
\ref{Theory}.

In particular, we consider the multiplicative cocycle
$d_{\vec{x}}\mbox{\cal{$\Phi$}}^t$ which maps the tangent space at $\vec{x}
\in \cal{S}$ to the tangent space at \cal{$\Phi$}$^t(\vec{x})\in \cal{S}$ for
a dynamical system defined on the differentiable manifold $\cal{S}$. We recall
that $\cal{S}$ is a measure space with a normalized measure $\mu$ and that
\cal{$\Phi$}$^t$ is a diffeomorphism on $\cal{S}$, i.~e.~\cal{$\Phi$}$^t$ is
a measurable bijection of $\cal{S}$ which preserves the measure $\mu$
(\ref{eq:measure}) and whose inverse is also measurable. We remark that in
measure theory we disregard sets of measure 0. In this sense \cal{$\Phi$}$^t$
is called measurable if it becomes measurable upon disregarding from $\cal{S}$
a set of measure 0. Quite often we will us the expression \textit{`for almost
all $\vec{x}$ with respect to measure $\mu$'} for the validity of a statement,
implying that the statement is true for all points $\vec{x}$ with the possible
exception of a set of points with measure 0.

A basic property of the multiplicative cocycles is their regularity, since
Theorem \ref{Theorem:existense} guarantees the existence of characteristic
exponents and the finiteness of the LCEs of all orders for regular
multiplicative cocycles. Thus, it is important to determine specific
conditions that  multiplicative cocycles should fulfill in order to be
regular. Such conditions were first provided by Oseledec \cite{O_68} who also
formulated and proved the so--called \textit{Multiplicative Ergodic Theorem}
(MET), which is often referred as \textit{Oseledec's theorem}.

The MET gives information about the dynamical structure of a multiplicative
cocycle $\textbf{R}(t,\vec{x})$ and its asymptotic behavior for $t \rightarrow
\infty$. The application of the MET for the particular multiplicative cocycle
$d_{\vec{x}}\mbox{\cal{$\Phi$}}^t$ provides the theoretical framework for the
computation of the LCEs for dynamical systems. The MET is one of the
milestones in the study of ergodic properties of dynamical systems and it can
be considered as a sort of a spectral theorem for random matrix products
\cite{R_79a}.  As a testimony to the importance of this theorem one can find
several alternative proofs for it in the literature. The original proof of
Oseledec \cite{O_68} applies both to continuous and discrete systems. In view
to the application to algebraic groups, Raghunathan \cite{Ra_79} devised a
simple proof of the MET, which nevertheless could not guarantee the finiteness
of all LCEs. Although Raghunathan's results apply only to maps, an extension
to flows, following the ideas of Oseledec, was given by Ruelle
\cite{R_79b}. Benettin et al. \cite{BGGS_80a} proved a somewhat different
version of the theorem being mainly interested to its application on
Hamiltonian flows and symplectic maps. Alternative proofs can also be found in
\cite{JPS_87,W_93}.

In \cite{O_68} Oseledec proved that a multiplicative cocycle
$\textbf{R}(t,\vec{x})$ is regular and thus, the MET is applicable to it, if
it satisfies the condition
\begin{equation}
  \sup_{|t|\leq 1} \ln^+ \|\textbf{R}^{\pm}(t,\vec{x}) \| \in L^1(\cal{S},
  \mu), \,\,\footnote[3]{We recall that a measurable function
  $f:\cal{S}\rightarrow\mathbb{R}$ (or $\mathbb{C}$) of the measure space
  $(\cal{S}, \mu)$ belongs to the space $L^1(\cal{S}, \mu)$ if its absolute
  value has a finite Lebesgue integral, i.~e.~
\begin{displaymath}
  \int |f| d\mu < \infty\,\,\,.
\end{displaymath}}
\label{eq:O_condition}
\end{equation} 
where $\ln^+ a =\max \left\lbrace 0, \ln a \right\rbrace $. From
(\ref{eq:O_condition}) we obtain the estimate
\begin{equation}
\|  \textbf{R}(t,\vec{x}) \| \leq e^{J(\vec{x}) |t|}
\label{eq:O_condition_2}
\end{equation} 
for $t\rightarrow \pm \infty$ for almost all $\vec{x}$ with respect to $\mu$,
where $J(\vec{x})$ is a measurable function. From (\ref{eq:O_condition_2}) it
follows that $\textbf{R}(t,\vec{x})$, considered as a function of $t$ for
fixed $\vec{x}$, satisfies equation (\ref{eq:max_At}). Benettin et
al. \cite{BGGS_80a} considered a slightly different version of the MET with
respect to the one presented in \cite{O_68}. Their version was adapted to the
framework of a continuous or discrete dynamical system with \cal{$\Phi$}$^t$
being a diffeomorphism of class $C^1$, i.~e.~both \cal{$\Phi$}$^t$ and its
inverse are continuously differentiable. They formulated the MET for the
particular multiplicative cocycle $d_{\vec{x}}\mbox{\cal{$\Phi$}}^t$, which
they proved to be regular. Since our presentation is mainly focused on
autonomous Hamiltonian systems and symplectic maps we will also state the MET
for the specific cocycle $d_{\vec{x}}\mbox{\cal{$\Phi$}}^t$. The version of
the MET we present is mainly based on \cite{O_68, R_79b,BGGS_80a} and combines
different formulations of the theorem given by various authors over the years.

\begin{theorem}[Multiplicative Ergodic Theorem -- MET] Consider a
  dynamical system as follows:  Let its phase space $\cal{S}$ be an
  $n$--dimensional compact manifold with a normalized measure $\mu$,
  $\mu(\mbox{$\cal{S}$})=1$ and a smooth Riemannian metric $\|\, \|$. Consider
  also a measure--preserving diffeomorphism \cal{$\Phi$}$^t$ of class $C^1$
  satisfying
\begin{displaymath}
\mbox{\cal{$\Phi$}}^{t+s}=\mbox{\cal{$\Phi$}}^t\circ \mbox{\cal{$\Phi$}}^s\,\,
,
\end{displaymath} 
with $t$ denoting time and having real (continuous system) or integer
(discrete system) values. Then for almost all $\vec{x} \in \cal{S}$, with
respect to measure $\mu$ we have:
\begin{enumerate}
\item The family of multiplicative cocycles
  $d_{\vec{x}}\mbox{\cal{$\Phi$}}^t:\cal{T}_{\vec{x}} \cal{S}
  \rightarrow
  \cal{T}\mbox{$_{\mbox{\cal{$\Phi$}}^t(\vec{x})}\cal{S}$}$, where
  $\cal{T}_{\vec{x}} \cal{S}$ denotes the tangent space of $\cal{S}$
  at point $\vec{x}$, is regular.

\item The LCEs of all orders exist and are independent of the choice of the
  Riemannian metric of $\cal{S}$.

In particular, for any $\vec{w} \in \cal{T}_{\vec{x}}
\cal{S}$ the finite limit 
\begin{equation}
\chi(\vec{x},\vec{w}) = \lim_{t\rightarrow \infty} \frac{1}{t} \ln \|
d_{\vec{x}}\mbox{\cal{$\Phi$}}^t \vec{w} \|
\label{eq:MET_02} 
\end{equation} 
exists and defines the LCE of order 1 (1--LCE). There exists at least one
normal basis $\vec{v}_i$, $i=1,2,\ldots,n$ of $\cal{T}_{\vec{x}} \cal{S} $ for
which the corresponding (possibly nondistinct) 1--LCEs
$\chi_i(\vec{x})=\chi(\vec{x},\vec{v}_i)$ are ordered as
\begin{equation}
\chi_1(\vec{x})  \geq \chi_2(\vec{x})  \geq \cdots \geq \chi_n(\vec{x}). 
\label{eq:MET_03}
\end{equation} 
Assume that the value $\nu_i(\vec{x})$, $i=1,2,\ldots,s$ with $s=s(\vec{x})$,
$1\leq s\leq n$ appears exactly $k_i(\vec{x})=k_i(\vec{x},\nu_i)>0$ times
among these numbers. Then the spectrum of LCEs
$(\nu_i(\vec{x}),k_i(\vec{x}))$, $i=1,2,\ldots,s$ is a measurable function of
$\vec{x}$, and as $\vec{w} \neq 0$ varies in $\cal{T}_{\vec{x}} \cal{S}$,
$\chi(\vec{x},\vec{w})$ takes one of these $s$ different values
\begin{equation}
\nu_1(\vec{x})  > \nu_2(\vec{x})  > \cdots >\nu_s(\vec{x}).
\label{eq:MET_04}
\end{equation} 
It also holds
\begin{equation}
  \sum_{i=1}^s k_i(\vec{x}) \nu_i(\vec{x}) = \lim_{t\rightarrow \infty}
  \frac{1}{t} \ln |\det d_{\vec{x}}\mbox{\cal{$\Phi$}}^t |.
\label{eq:MET_05}
\end{equation} 

For any $p$-dimensional ($1 \leq p \leq n$) subspace $E^p \subseteq
\cal{T}_{\vec{x}} \cal{S} $, generated by a linearly independent set
$\vec{w}_i$, $i=1,2,\ldots,p$ the finite limit
\begin{equation}
\chi(\vec{x},E^p) = \lim_{t\rightarrow \infty} \frac{1}{t} \ln
\mathrm{vol}_p( d_{\vec{x}}\mbox{\cal{$\Phi$}}^t, E^p),
\label{eq:MET_06} 
\end{equation} 
where $\mathrm{vol}_p( d_{\vec{x}}\mbox{\cal{$\Phi$}}^t, E^p)$ is the volume
of the $p$--parallelogram having as edges the vectors
$d_{\vec{x}}\mbox{\cal{$\Phi$}}^t \vec{w}_i$, exists and defines the LCE of
order $p$ ($p$--LCE). The value of $\chi(\vec{x},E^p)$ is equal to the sum of
$p$ 1--LCEs $\chi_i(\vec{x})$, $i=1,2,\ldots,n$.

\item The set of vectors 
\begin{displaymath}
  L_i(\vec{x})=\left\lbrace \vec{w} \in \cal{T}_{\vec{x}} \cal{S} :
    \chi(\vec{x},\vec{w}) \leq \mbox{$\nu_{i}(\vec{x})$} \right\rbrace \,\,\,
    , \,\,\, 1\leq i\leq s
\end{displaymath} 
is a linear subspace of $ \cal{T}_{\vec{x}} \cal{S}$ satisfying 
\begin{equation}
  \cal{T}_{\vec{x}} \cal{S} \mbox{$= L_1(\vec{x}) \supset L_2(\vec{x}) \supset
    \cdots \supset L_{s}(\vec{x}) \supset L_{s+1}(\vec{x})
    \stackrel{\mbox{def}}{=} \left\lbrace 0\right\rbrace$}.
\label{eq:MET_08}
\end{equation} 
If $\vec{w}\in L_i(\vec{x}) \setminus L_{i+1}(\vec{x})$ then
$\chi(\vec{x},\vec{w}) = \nu_i(\vec{x})$ for $i=1,2,\ldots, s$. The
multiplicity $k_i(\vec{x})$ of values $\nu_i(\vec{x})$ is given by
$k_i(\vec{x})= \dim L_i(\vec{x}) - \dim L_{i+1}(\vec{x})$.

\item The symmetric positive--defined matrix
\begin{displaymath}
\Lambda_{\vec{x}}= \lim_{t\rightarrow \infty} \left(
\textbf{Y}^{\mathrm{T}}(t) \cdot \textbf{Y}(t) \right)^{1/2t}
\end{displaymath} 
exists. $\textbf{Y}(t)$ is the matrix corresponding to
$d_{\vec{x}}\mbox{\cal{$\Phi$}}^t$ and is defined by equations
(\ref{eq:Y_ham}) and (\ref{eq:Y_map}) for continuous and discrete dynamical
systems respectively. The logarithms of the eigenvalues of $\Lambda_{\vec{x}}$
are the $s$ distinct 1--LCEs (\ref{eq:MET_04}) of the dynamical system. The
corresponding eigenvectors are orthogonal (since $\Lambda_{\vec{x}}$ is
symmetric), and for the corresponding eigenspaces
$V_1(\vec{x}),V_2(\vec{x}),\ldots, V_s(\vec{x})$ we have
\begin{displaymath}
k_i(\vec{x})=\dim V_i(\vec{x})\,\,\, ,\,\,\, L_i(\vec{x})= \bigoplus_{r=i}^s
V_r(\vec{x}) \,\,\, \mbox{for} \,\,\, i=1,2,\ldots,s.
\end{displaymath}  
Thus, $\cal{T}_{\vec{x}} \cal{S}$ is decomposed as
\begin{displaymath}
\mbox{$\cal{T}$}_{\vec{x}} \mbox{$\cal{S}$} = V_1(\vec{x}) \oplus V_2(\vec{x})
\oplus \cdots \oplus V_s(\vec{x}),
\end{displaymath}  
and for every nonzero vector $\vec{w} \in V_i(\vec{x})$, $i=1,2,\ldots,s$, we
get
\begin{displaymath}
\chi(\vec{x},\vec{w})=\nu_i(\vec{x}).
\end{displaymath}  
\end{enumerate}
\label{Theorem:MET}
\end{theorem}

A short remark is necessary here. The regularity of
$d_{\vec{x}}\mbox{\cal{$\Phi$}}^t$, which guarantees the validity of equations
(\ref{eq:MET_02}) and (\ref{eq:MET_06}) and the finiteness of the LCEs of all
orders, should not be confused with the regular nature of orbits of the
dynamical system. Regular orbits have all their LCEs equal to zero (see also
the discussion in Section \ref{maximal_algorithm_behavior}).

Benettin et al. \cite{BGGS_78, BGGS_80a} have formulated also the following
theorem which provides the theoretical background for the numerical algorithm
they presented in \cite{BGGS_80b} for the computation of all LCEs.
\begin{theorem}
  Under the assumptions of the MET, the $p$--LCE of any $p$-dimensional
  subspace $E^p \subseteq \cal{T}_{\vec{x}} \cal{S}$ satisfying Condition R
  (\ref{eq:condR}), is equal to the sum of the $p$ largest 1--LCEs
  (\ref{eq:MET_03}):
\begin{equation}
\chi(\vec{x},E^p) = \lim_{t\rightarrow \infty} \frac{1}{t} \ln \mathrm{vol}_p(
d_{\vec{x}}\mbox{\cal{$\Phi$}}^t, E^p)= \sum_{i=1}^p \chi_i(\vec{x}).
\label{eq:pLCER} 
\end{equation} 
\label{theorem:condR}
\end{theorem}

\subsection{Properties of the spectrum of LCEs}
\label{Spectrum_properties}

Let us now turn our attention to the structure of the spectrum of LCEs for
$N$D autonomous Hamiltonian systems and for $2N$d symplectic maps, which are
the main dynamical systems we are interested in. Such systems preserve the
phase-space volume, and thus, the r.~h.~s.~of (\ref{eq:MET_05}) vanishes. So
for the sum of all the 1--LCEs we have
\begin{equation}
\sum_{i=1}^{2N} \chi_i(\vec{x})=0.
\label{eq:sumLCEs} 
\end{equation} 

The symplectic nature of these systems gives indeed more. It has been proved
in \cite{BGGS_80a} that the spectrum of LCEs consists of pairs of values
having opposite signs
\begin{equation}
\chi_i(\vec{x})= - \chi_{2N-i+1}(\vec{x}) \,\,\, , \,\,\, i=1,2,\ldots, N.
\label{eq:op_signs}
\end{equation} 
Thus, the spectrum of LCEs becomes
\begin{displaymath}
\chi_1(\vec{x}) \geq \chi_2(\vec{x}) \geq \cdots \geq \chi_N(\vec{x}) \geq
-\chi_N(\vec{x}) \geq \cdots \geq -\chi_{2}(\vec{x}) \geq -\chi_1(\vec{x}).
\end{displaymath} 

For autonomous Hamiltonian flows we can say something more. Let us first
recall that for a general differentiable flow on a compact manifold without
stationary points at least one LCE must vanish \cite{BGGS_80a,H_83}. This
follows from the fact that, in the direction along the flow a deviation vector
grows only linearly in time. So, in the case of a Hamiltonian flow, due to the
symmetry of the spectrum of LCEs (\ref{eq:op_signs}), at least two LCEs
vanish, i.~e.
\begin{displaymath}
\chi_N(\vec{x})= \chi_{N+1}(\vec{x})=0,
\end{displaymath} 
while the presence of any additional independent integral of motion leads to
the vanishing of another pair of LCEs.

Let us now study the particular case of a periodic orbit of period $T$, such
that $\Phi^T(\vec{x})=\vec{x}$, following \cite{BG_79,BFS_79}.  In this case
$d_{\vec{x}}\mbox{\cal{$\Phi$}}^T$ is a linear operator on the tangent space
$\cal{T}_{\vec{x}} \cal{S}$ so that for any deviation vector $\vec{w}(0) \in
\cal{T}_{\vec{x}} \cal{S}$ we have
\begin{equation}
\vec{w}(T)=\textbf{Y}\cdot \vec{w}(0) ,
\label{eq:w_ham_T}
\end{equation}
where $\textbf{Y}$ is the constant matrix corresponding to
$d_{\vec{x}}\mbox{\cal{$\Phi$}}^T$. Suppose that $\textbf{Y}$ has $2N$
(possibly complex) eigenvalues $\lambda_i$, $i=1,2,\ldots,2N$, whose magnitudes
can be ordered as
\begin{displaymath}
 | \lambda_1 | \geq | \lambda_2 | \geq \ldots \geq | \lambda_{2N} | .
\end{displaymath}
Let $\hat{\vec{w}}_i$, $i=1,2,\ldots,2N$, denote the corresponding unitary
eigenvectors. Then for $\vec{w}(0)=\hat{\vec{w}}_i$ equation
(\ref{eq:w_ham_T}) implies
\begin{equation}
\vec{w}(kT)=\lambda_i^k\hat{\vec{w}}_i\,\,\, , \,\,\, k=1,2,\ldots 
\label{eq:T_w_lambda}
\end{equation}
and so we conclude from (\ref{eq:MET_02}) that
\begin{displaymath}
\chi(\vec{x},\hat{\vec{w}}_i) = \frac{1}{T} \ln |\lambda_i| =\chi_i(\vec{x}),
\,\,\, i=1,2,\dots, 2N.
\end{displaymath} 

Furthermore for a deviation vector
\begin{displaymath}
\vec{w}(0)= c_1 \hat{\vec{w}}_1 + c_2 \hat{\vec{w}}_2 + \ldots + c_{2N}
\hat{\vec{w}}_{2N}
\end{displaymath} 
with $c_i \in \mathbb{R}$, $i=1,2,\ldots, 2N$, it follows from
(\ref{eq:T_w_lambda}) that the first nonvanishing coefficient $c_i$ eventually
dominates the evolution of $\vec{w}(t)$ and we get
$\chi(\vec{x},\vec{w})=\chi_i$. In this case we can define a filtration
similar to the one presented in (\ref{eq:MET_08}) by defining $L_1=\left[
\hat{\vec{w}}_1, \hat{\vec{w}}_2, \ldots, \hat{\vec{w}}_{2N} \right]=
\cal{T}_{\vec{x}} \cal{S} $, $L_2=\left[ \hat{\vec{w}}_2, \ldots,
\hat{\vec{w}}_{2N} \right] $, $\dots$, $L_{2N}=\left[\hat{\vec{w}}_{2N}
\right] $, $L_{2N+1}=\left[0 \right] $, where $\left[ \,\,\right]$ denotes the
linear space spanned by vectors $ \hat{\vec{w}}_1, \hat{\vec{w}}_2, \ldots,
\hat{\vec{w}}_{2N}$ and so on. It becomes evident that a random choice of an
initial deviation vector $\vec{w}(0) \in \cal{T}_{\vec{x}} \cal{S}$ will lead
to the computation of the mLCE $\chi_1(\vec{x})$ since, in general,
$\vec{w}(0) \in L_1 \setminus L_2$.

So, in the case of an unstable periodic orbit where $| \lambda_1 |>1$ we get
$\chi_1(\vec{x})>0$, which implies that nearby orbits diverge exponentially
from the periodic one. These orbits are not called chaotic, although their
mLCE is larger than zero, but simply `unstable'. In fact, unstable periodic
orbits exist also in integrable systems. Since the measure of periodic orbits
in a general dynamical system has zero measure, periodic orbits (stable and
unstable) are rather exceptional.

In the general case of a nonperiodic orbit we are no more allowed to use
concepts as eigenvectors and eigenvalues because the linear operator
$d_{\vec{x}}\mbox{\cal{$\Phi$}}^t$ maps $\cal{T}_{\vec{x}} \cal{S}$ into
$\cal{T}\mbox{$_{\mbox{\cal{$\Phi$}}^t(\vec{x})}\cal{S}$} \neq
\cal{T}_{\vec{x}} \cal{S}$, while eigenvectors are intrinsically defined only
for linear operators of a linear space into itself. Nevertheless, in the case
of nonperiodic orbits the MET proves the existence of the LCEs and of
filtration (\ref{eq:MET_08}). In a way, the MET provides an extention of the
linear stability analysis of periodic orbits to the case of nonperiodic
ones, although one should always keep in mind that the LCEs are related to the
real and positive eigenvalues of the symmetric, positive--defined matrix
$\textbf{Y}^{\mathrm{T}}(t) \cdot \textbf{Y}(t)$ \cite{GSO_87,MC_08}. On the
other hand, linear stability analysis involves the computation of the
eigenvalues of the nonsymmetric matrix $\textbf{Y}(t)$, which solves the
linearized equations of motion (\ref{eq:Y_ham}) for Hamiltonian flows or
(\ref{eq:Y_map}) for maps.  These eigenvalues are real or come in pairs of
complex conjugate pairs and, in general, they are not directly related to the
LCEs which are real numbers.

An important property of the LCEs is that they are constant in a connected
chaotic domain. This is due to the fact that every nonperiodic orbit in the
same connected chaotic domain covers densely this domain, thus, two different
orbits of the same domain are in a sense dynamically equivalent. The unstable
periodic orbits in this chaotic domain have in general LCEs that are different
from the constant LCEs of the nonperiodic orbits. This is due to the fact
that the periodic orbits do not visit the whole domain, thus, they cannot
characterize its dynamical behavior. In fact, different periodic orbits have
different LCEs.

\section{The maximal LCE}
\label{maximal}

From this point on, in order to simplify our notation, we will not explicitly
write the dependence of the LCEs on the specific point $\vec{x} \in
\cal{S}$. So, in practice, considering that we are referring to a specific
point $\vec{x} \in \cal{S}$, we denote by $\chi_i$ the LCEs of order 1 and by
$\chi_i^{(p)}$ the LCEs of order $p$.

For the practical determination of the chaotic nature of orbits a numerical
computation of the mLCE $\chi_1$ can be employed. If the studied orbit is
regular $\chi_1=0$, while if it is chaotic $\chi_1>0$, implying exponential
divergence of nearby orbits. The computation of the mLCE has been used
extensively as a chaos indicator after the introduction of numerical
algorithms for the determination of its value at  late 70's
\cite{BGS_76,NS_77,BS_78,CGG_78,BGGS_80b}.

Apart from using the mLCE as a criterion for the chaoticity or the regularity
of an orbit its value also attains a `physical' meaning and defines a specific
time scale for the considered dynamical system. In particular, the inverse of
the mLCE, which is called \textit{Lyapunov time}
\begin{equation}
t_L=\frac{1}{\chi_1}
\label{eq:lyap_time}
\end{equation} 
gives an estimate of the time needed for a dynamical system to become chaotic
and in practice measures the time needed for nearby orbits of the system to
diverge by $e$ (see e.~g~\cite[p.~508]{Contopoulos_2002}).

\subsection{Computation of the mLCE}
\label{compute_maximal}

The mLCE can be computed by the numerical implementation of equation
(\ref{eq:MET_02}). In Section \ref{Theory_comp_1} we showed that a random
choice of the initial deviation vector $\vec{w}(0) \in \cal{T}_{\vec{x}}
\cal{S}$ leads to the numerical computation of the mLCE. We recall that the
deviation vector $\vec{w}(t)$ at time $t>0$ is determined by the action of the
operator $d_{\vec{x}}\mbox{\cal{$\Phi$}}^t$ on the initial deviation vector
$\vec{w}(0)$ according to equation (\ref{eq:t_flow})
\begin{equation}
\vec{w}(t)=d_{\vec{x}}\mbox{\cal{$\Phi$}}^t\, \vec{w}(0).
\label{eq:t_flow_again}
\end{equation}
This equation represents the solution of the variational equations
(\ref{eq:var}) or the evolution of a deviation vector under the action of the
tangent map (\ref{eq:w_map}), and takes  the form (\ref{eq:w_ham}) and
(\ref{eq:w0_map}) respectively. We emphasize  that, both the variational
equations and the equations of the tangent map are linear with respect to the
tangent vector $\vec{w}$, i.~e.
\begin{equation}
  d_{\vec{x}}\mbox{\cal{$\Phi$}}^t \left(a \, \vec{w} \right) = a
  \,d_{\vec{x}}\mbox{\cal{$\Phi$}}^t \vec{w}, \,\,\, \mbox{for any} \,\,\, a
  \in \ \mathbb{R}.
\label{eq:linearity}
\end{equation} 

In order to evaluate the mLCE of an orbit with initial condition $\vec{x}(0)$,
one has to follow simultaneously the time evolution of the orbit itself and of
a deviation vector $\vec{w}$ from this orbit with initial condition
$\vec{w}(0)$. In the case of a Hamiltonian flow (continuous time) we solve
simultaneously the Hamilton equations of motion (\ref{eq:Hameq_gen}) for the
time evolution of the orbit and the variational equations (\ref{eq:var}) for
the time evolution of the deviation vector. In the case of a symplectic map
(discrete time) we iterate the map (\ref{eq:map_gen}) for the evolution of the
orbit simultaneously with the tangent map (\ref{eq:w_map}), which determines
the evolution of the tangent vector. The mLCE is then computed as the limit
for $t\rightarrow \infty$ of the quantity
\begin{equation}
X_1(t)= \frac{1}{t} \ln \frac{\| d_{\vec{x}(0)}\mbox{\cal{$\Phi$}}^t\,
\vec{w}(0) \|}{\| \vec{w}(0) \|} =\frac{1}{t} \ln \frac{\| \vec{w}(t) \|}{\|
\vec{w}(0) \|},
\label{eq:X_1_t}
\end{equation} 
often called \textit{finite time mLCE}. So, we have
\begin{equation}
\chi_1= \lim_{t\rightarrow \infty} X_1(t).
\label{eq:chi_1_t_limit}
\end{equation} 

The direct numerical implementation of equations (\ref{eq:X_1_t}) and
(\ref{eq:chi_1_t_limit}) for the evaluation of $\chi_1$ meets a severe
difficulty. If, for example, the orbit under study is chaotic, the norm $\|
\vec{w}(t) \|$ increases exponentially with increasing time $t$, leading to
numerical overflow, i.~e.~$\| \vec{w}(t) \|$ attains very fast extremely large
values that cannot be represented in the computer. This difficulty can be
overcome by a procedure which takes advantage of the linearity of
$d_{\vec{x}}\mbox{\cal{$\Phi$}}^t$ (\ref{eq:linearity}) and of the composition
law (\ref{eq:d_Phi_ts}). Fixing a small time interval $\tau$ we express time
$t$ with respect to $\tau$ as $t=k\tau$, $k=1,2,\ldots$. Then for the quantity
$X_1(t)$ we have
\begin{eqnarray}
X_1(k\tau)&=& \frac{1}{k\tau} \ln \frac{\| \vec{w}(k\tau) \|}{\| \vec{w}(0)
\|} \nonumber\\ &=& \frac{1}{k\tau} \ln \left( \frac{\| \vec{w}(k\tau) \|}{\|
\vec{w}((k-1)\tau) \|} \frac{\| \vec{w}((k-1)\tau) \|}{\| \vec{w}((k-2)\tau)
\|} \cdots \frac{\| \vec{w}(2\tau) \|}{\| \vec{w}(\tau) \|} \frac{\|
\vec{w}(\tau) \|}{\| \vec{w}(0) \|} \right) \nonumber\\ &=& \frac{1}{k\tau}
\sum_{i=1}^{k} \ln \frac{\| \vec{w}(i\tau) \|}{\| \vec{w}((i-1)\tau) \|}
\Rightarrow \nonumber\\ X_1(k\tau)&=&\frac{1}{k\tau} \sum_{i=1}^{k} \ln
\frac{\| d_{\vec{x}(0)}\mbox{\cal{$\Phi$}}^{i\tau}\, \vec{w}(0) \|}{\|
d_{\vec{x}(0)}\mbox{\cal{$\Phi$}}^{(i-1)\tau}\, \vec{w}(0) \| }.
\label{eq:X_1_t_sum1} 
\end{eqnarray}
Denoting by $D_0$ the norm of the initial deviation vector $\vec{w}(0)$
\begin{displaymath}
D_0=\| \vec{w}(0) \|,
\end{displaymath} 
we get for the evolved deviation vector at time $t=k\tau$
\begin{eqnarray}
d_{\vec{x}(0)}\mbox{\cal{$\Phi$}}^{i\tau}\, \vec{w}(0)&=&
d_{\vec{x}(0)}\mbox{\cal{$\Phi$}}^{\tau+(i-1)\tau}\,
\vec{w}(0)\stackrel{\mbox{\begin{scriptsize}(\ref{eq:d_Phi_ts})\end{scriptsize}}}{=}
d_{\mbox{\cal{$\Phi$}}^{(i-1)\tau} (\vec{x}(0))} \mbox{\cal{$\Phi$}}^{\tau}
\!\!\left( d_{\vec{x}(0)}\mbox{\cal{$\Phi$}}^{(i-1)\tau} \vec{w}(0) \right)
\nonumber\\
&\stackrel{\mbox{\begin{scriptsize}(\ref{eq:linearity})\end{scriptsize}}}{=}&\frac{\|
d_{\vec{x}(0)}\mbox{\cal{$\Phi$}}^{(i-1)\tau}\, \vec{w}(0) \|}{D_0}\,\,\,
d_{\mbox{\cal{$\Phi$}}^{(i-1)\tau} (\vec{x}(0))} \mbox{\cal{$\Phi$}}^{\tau}
\!\! \left( \frac{d_{\vec{x}(0)}\mbox{\cal{$\Phi$}}^{(i-1)\tau} \vec{w}(0)}{\|
d_{\vec{x}(0)}\mbox{\cal{$\Phi$}}^{(i-1)\tau}\, \vec{w}(0) \|} D_0 \right)
\Rightarrow \nonumber\\ \frac{d_{\vec{x}(0)}\mbox{\cal{$\Phi$}}^{i\tau}\,
\vec{w}(0)}{\| d_{\vec{x}(0)}\mbox{\cal{$\Phi$}}^{(i-1)\tau}\, \vec{w}(0)
\|}&=&\frac{d_{\mbox{\cal{$\Phi$}}^{(i-1)\tau} (\vec{x}(0))}
\mbox{\cal{$\Phi$}}^{\tau} \!\! \left(
\frac{d_{\vec{x}(0)}\mbox{\cal{$\Phi$}}^{(i-1)\tau} \vec{w}(0)}{\|
d_{\vec{x}(0)}\mbox{\cal{$\Phi$}}^{(i-1)\tau}\, \vec{w}(0) \|} D_0 \right)
}{D_0}.
\label{eq:evol_dx} 
\end{eqnarray}
Let us now denote by 
\begin{displaymath}
\hat{\vec{w}}((i-1)\tau)=\frac{d_{\vec{x}(0)}\mbox{\cal{$\Phi$}}^{(i-1)\tau}
\vec{w}(0)}{\| d_{\vec{x}(0)}\mbox{\cal{$\Phi$}}^{(i-1)\tau}\, \vec{w}(0) \|}
D_0,
\end{displaymath} 
the deviation vector at point \cal{$\Phi$}$^{(i-1)\tau}(\vec{x}(0))$ having
the same direction with $\vec{w}((i-1)\tau)$ and norm $D_0$, and by $D_i$ its
norm after its evolution for $\tau$ time units
\begin{displaymath}
D_i=\|d_{\mbox{\cal{$\Phi$}}^{(i-1)\tau} (\vec{x}(0))}
\mbox{\cal{$\Phi$}}^{\tau} \hat{\vec{w}}((i-1)\tau) \|.
\end{displaymath} 
Using this notation we derive from equation (\ref{eq:evol_dx})
\begin{equation}
\ln \frac{\| d_{\vec{x}(0)}\mbox{\cal{$\Phi$}}^{i\tau}\, \vec{w}(0)\|}{\|
d_{\vec{x}(0)}\mbox{\cal{$\Phi$}}^{(i-1)\tau}\, \vec{w}(0) \|}= \ln
\frac{D_i}{D_{0}} = \ln \alpha_i,
\label{eq:ratio_d}
\end{equation} 
with $\alpha_i$ being the local coefficient of expansion of the deviation
vector for a time interval of length $\tau$ when the corresponding orbit
evolves from position \cal{$\Phi$}$^{(i-1)\tau}(\vec{x}(0))$ to position
\cal{$\Phi$}$^{i\tau}(\vec{x}(0))$ ($\ln \alpha_i/\tau$ is also called
\textit{stretching number} \cite{VC_94}\cite[p.~257]{Contopoulos_2002}).

From equations (\ref{eq:chi_1_t_limit}), (\ref{eq:X_1_t_sum1}) and
(\ref{eq:ratio_d}) we conclude that the mLCE $\chi_1$ can be computed as
\begin{equation}
\chi_1=\lim_{k\rightarrow \infty}X_1(k \tau)=\lim_{k\rightarrow \infty}
\frac{1}{k \tau} \sum_{i=1}^k \ln \frac{D_i}{D_{0}}=\lim_{k\rightarrow
\infty} \frac{1}{k \tau} \sum_{i=1}^k \ln \alpha_i.
\label{eq:compute_x1}
\end{equation} 
Since the initial norm $D_0$ can have any arbitrary value, one usually set it
to $D_0=1$. Equation (\ref{eq:compute_x1}) implies that practically $\chi_1$
is the limit value, for $t\rightarrow \infty$, of the mean of the
stretching numbers along the studied orbit \cite{BGGS_80b,FFL_93,VC_94}.

\subsection{The numerical algorithm}
\label{maximal_algorithm}

In practice, for the evaluation of the mLCE we follow the evolution of a
unitary initial deviation vector $\hat{\vec{w}}(0)=\vec{w}(0)$, $\| \vec{w}(0)
\|=D_0=1$ and every $t=\tau$ time units we replace the evolved vector
$\vec{w}(k\tau)$, $k=1,2,\ldots$, by vector $\hat{\vec{w}}(k\tau)$ having the
same direction but norm equal to 1 ($\| \hat{\vec{w}}(k\tau) \|=1$). Before
each new renormalization the corresponding $\alpha_k$ is computed and $\chi_1$
is estimated from equation (\ref{eq:compute_x1}).

\begin{figure}
\centerline{ 
\begin{tabular}{c}
\includegraphics[scale=0.58]{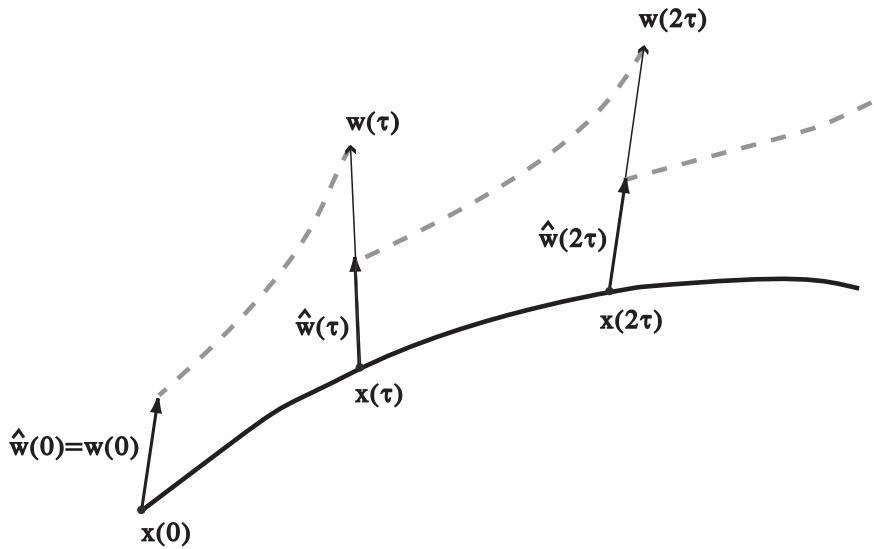} 
\end{tabular}}
\caption{ Numerical scheme for the computation of the mLCE $\chi_1$. The
  unitary deviation vector $\hat{\vec{w}}( (i-1) \tau )$, $i=1,2,\ldots$, is
  evolved according to the variational equations (\ref{eq:var}) (continuous
  time) or the equations of the tangent map (\ref{eq:w_map}) (discrete time)
  for $t=\tau$ time units. The evolved vector $\vec{w}(i\tau)$ is replaced by
  a unitary vector $\hat{\vec{w}}(i\tau)$ having the same direction with
  $\vec{w}(i\tau)$. For each successive time interval $[(i-1)\tau, i\tau]$ the
  quantity $\alpha_i=\|\vec{w}(i\tau)\|$ is computed and $\chi_1$ is estimated
  from equation (\ref{eq:compute_x1}).}
\label{f:1_LCE}
\end{figure}

More precisely at $t=\tau$ we have $\alpha_1=\|\vec{w}(\tau)\|$. Then we
define a unitary vector $ \hat{\vec{w}}(\tau)$ by renormalizing
$\vec{w}(\tau)$ and using it as an initial deviation vector we evolve it along
the orbit from $\vec{x}(\tau)$ to $\vec{x}(2\tau)$ according to equation
(\ref{eq:t_flow_again}), having
$\vec{w}(2\tau)=d_{\vec{x}(\tau)}\mbox{\cal{$\Phi$}}^{\tau}\,
\hat{\vec{w}}(\tau)$. Then we define $\alpha_2=\|\vec{w}(2\tau)\|$ and we
estimate $\chi_1$ (see Figure \ref{f:1_LCE}).
We iteratively apply the above described procedure until a good approximation
of $\chi_1$ is achieved. The algorithm for the evaluation of the mLCE $\chi_1$
is described in pseudo--code in Table \ref{tab:program_mLCE}.
\begin{table}
  \caption{The algorithm for the computation of the mLCE $\chi_1$ as the limit
  for $t \rightarrow \infty$ of $X_1(t)$ according to equation
  (\ref{eq:compute_x1}). The program computes the evolution of $X_1(t)$ as a
  function of time $t$ up to a given upper value of time $t=T_M$ or until
  $X_1(t)$ attains a very small value, smaller than a low threshold value
  $X_{1m}$.  }
\begin{tabular}{ll}
\hline
Input:  & 1. Hamilton equations of motion (\ref{eq:Hameq_gen}) and variational equations (\ref{eq:var}), or\\
        & \quad \quad equations of the map (\ref{eq:map_gen}) and of the tangent map (\ref{eq:w_map}).\\
        & 2. Initial condition for the orbit $\vec{x}(0)$.\\
        & 3. Initial \textit{unitary} deviation vector $\vec{w}(0)$.\\
        & 4. Renormalization time $\tau$.\\
        & 5. Maximal time: $T_M$ and minimum allowed value of $X_1(t)$: $X_{1m}$.\\
\hline
Step 1  & \textbf{Set} the stopping flag, $\textsl{SF} \gets 0$, and the counter, $k \gets 1$. \\
Step 2  & \textbf{While} $(\textsl{SF}=0)$\/ \textbf{Do}\\
        & \quad \quad \textbf{Evolve} the orbit and the deviation vector from time $t=(k-1)\tau$\\
        & \quad \quad  to $t=k \tau$, i.~e.~\textbf{Compute} $\vec{x}(k \tau)$ and $\vec{w}(k \tau)$. \\
Step 3  & \quad \quad \textbf{Compute} current value of $\alpha_k=\|\vec{w}(k \tau)\|$.\\
        & \quad \quad \textbf{Compute} and \textbf{Store} current value of $X_1(k\tau)=\frac{1}{k \tau} \sum_{i=1}^k \ln \alpha_i$.\\
Step 4  & \quad \quad Renormalize deviation vector by \textbf{Setting} $\vec{w}(k \tau) \gets \vec{w}(k \tau)/\alpha_k$. \\
Step 5  & \quad \quad \textbf{Set}  the counter $k \gets k+1$. \\ 
Step 6  & \quad \quad \textbf{If} $[ (k \tau > T_M)$ or $(X_1((k-1)\tau) < X_{1m})]$ \textbf{Then} \\
        & \quad \quad \quad \quad \textbf{Set}  $\textsl{SF} \gets 1$. \\
        & \quad \quad \textbf{End If}  \\
        & \textbf{End While}\\
Step 7  & \textbf{Report} the time evolution of $X_1(t)$.\\
\hline
\end{tabular} 
\label{tab:program_mLCE}
\end{table}

Instead of utilizing the variational equations or the tangent map for the
evolution of a deviation vector in the above described algorithm, one could
integrate equations (\ref{eq:Hameq_gen}) or iterate equations
(\ref{eq:map_gen}) for two orbits starting nearby and estimate $\vec{w}(t)$ by
difference. Indeed, this approach, influenced by the rough idea of divergence
of nearby orbits introduced in \cite{HH_64}, was initially adopted for the
computation of the mLCE \cite{BGS_76,NS_77,BS_78}. This technique was
abandoned after a while as it was realized that the use of explicit equations
for the evolution of deviation vectors was more reliable and efficient
\cite{CGG_78,SN_79,BGGS_80b}, although in some cases it is used also nowadays
(see e.~g.~\cite{WHZ_06}).

\subsection{Behavior of $X_1(t)$ for regular and chaotic orbits}
\label{maximal_algorithm_behavior}

Let us now discuss in more detail the behavior of the computational scheme for
the evaluation of the mLCE for the cases of regular and chaotic orbits.

The LCE of regular orbits vanish \cite{BGS_76,CCF_80} due to the linear
increase with time of the norm of deviation vectors. We illustrate this
behavior in the case of an $N$D Hamiltonian system, but a similar analysis can
be easily carried out for symplectic maps. In such systems regular orbits lie
on $N$--dimensional tori. If such tori are found around a stable periodic
orbit, they can be accurately described by $N$ formal integrals of motion in
involution, so that the system would appear locally integrable. This means
that we could perform a local transformation to action--angle variables,
considering as actions $J_1, J_2, \ldots, J_N$ the values of the $N$ formal
integrals, so that Hamilton's equations of motion, locally attain the form
\begin{equation}
\dot{J}_i  =  0, \,\,\, \dot{\theta}_i  =  \omega_i(J_1, J_2, \ldots, J_N),
\,\,\, i=1,2,\ldots,N. 
\label{eq:a-a}
\end{equation}
These equations can be easily integrated to give
\begin{displaymath}
J_i(t)  =  J_{i0},\,\,\, 
\theta_i(t)  = \theta_{i0} +\omega_i(J_{10}, J_{20}, \ldots,
J_{N0})\, t,
\,\,\, i=1,2,\ldots,N, 
\end{displaymath}
where $J_{i0}$, $\theta_{i0}$, $i=1,2,\ldots,N$ are the initial conditions of
the studied orbit.

By denoting as $\xi_i$, $\eta_i$, $i=1,2,\ldots,N$ small deviations of $J_i$
and $\theta_i$ respectively, the variational equations (\ref{eq:var}) of
system (\ref{eq:a-a}), describing the evolution of a deviation vector are
\begin{displaymath}
\dot{\xi}_i  =  0, \,\,\,
\dot{\eta}_i  =  \sum_{j=1}^N \omega_{ij}\cdot \xi_j, \,\,\,
i=1,2,\ldots,N, 
\end{displaymath}
where
\begin{displaymath}
\omega_{ij}=\left. \frac{\partial \omega_i}{\partial J_j} \right|_{\vec{J}_0}
 ,\,\,\, i,j=1,2,\ldots,N,
\end{displaymath}
and $\vec{J}_0=(J_{10}, J_{20}, \ldots, J_{N0})=\mbox{constant}$, represents
the $N$--dimensional vector of the initial actions. The solution of these
equations is:
\begin{equation}
\begin{array}{ccl}
\xi_i(t) & = & \xi_i(0) \\ \eta_i (t) & = & \eta_i (0) + \left[ \sum_{j=1}^N
\omega_{ij} \xi_j(0)\right]  t,
\end{array} \,\,\, i=1,2,\ldots,N.
\label{eq:twist_map_var_sol}
\end{equation}

From equations (\ref{eq:twist_map_var_sol}) we see that an initial deviation
vector $\vec{w}(0)$ with coordinates $\xi_i(0)$, $i=1,2,\ldots,N$ in the
action variables and $\eta_i(0)$, $i=1,2,\ldots,N$ in the angles,
i.~e.~$\vec{w}(0)=(\xi_1(0),\xi_2(0),\ldots,
\xi_N(0),\eta_1(0),\eta_2(0),\ldots, \eta_N(0))$, evolves in time in such a
way that its action coordinates remain constant, while its angle coordinates
increase linearly in time. This behavior implies an almost linear increase of
the norm of the deviation vector. To see this, let us assume that 
vector $\vec{w}(0)$ has initially unit magnitude, i.~e.
\begin{displaymath}
\sum_{i=1}^N\xi_i^2 (0) +  \sum_{i=1}^N\eta_i^2 (0)=1
\end{displaymath}
whence the time evolution of its norm is given by
\begin{displaymath}
\|\vec{w}(t)\|= \left\{ 1+ \left[\sum_{i=1}^N \left( \sum_{j=1}^N \omega_{ij}
\xi_j(0) \right)^2 \right] t^2 + \left[2\sum_{i=1}^N \left(\eta_i(0)
\sum_{j=1}^N \omega_{ij} \xi_j(0) \right) \right] t\right\} ^{1/2}.
\end{displaymath}
This implies that the norm for long times grows linearly with $t$
\begin{equation}
\| \vec{w}(t)\| \propto  t.
\label{eq:approx_norm}
\end{equation} 
So, from equation (\ref{eq:X_1_t}) we see that for long times $X_1(t)$ is of
the order $\mathcal{O}(\ln t/ t)$, which means that $X_1(t)$ tends
asymptotically to zero, as $t \rightarrow \infty$ like $t^{-1}$.
This asymptotic behavior is evident in numerical computations of the mLCE of
regular orbits, as we can see for example in the left panel of
Figure~\ref{f:BGS_76}.

The asymptotic behavior of $X_1(t)$ for regular orbits, described above,
represents a particular case of a more general estimation presented in
\cite{GSO_87}. In particular, Goldhirsch et al.~\cite{GSO_87} showed that, in
general, after some initial transient time the value of the mLCE $\chi_1$ is
related to its finite time estimation by
\begin{equation}
X_1(t)= \chi_1+\frac{b+z(t)}{t},
\label{eq:GSO_est}
\end{equation} 
where $b$ is a constant and $z(t)$ is a `noise' term of zero mean. According
to their analysis, this approximate formula is valid both for regular and
chaotic orbits. It is easily seen that from (\ref{eq:GSO_est}) we retrieve
again the asymptotic behavior $X_1(t) \propto t^{-1}$ for the case of regular
orbits ($\chi_1=0$).

\begin{figure}
\centerline{ 
\begin{tabular}{cc}
\includegraphics[scale=0.30]{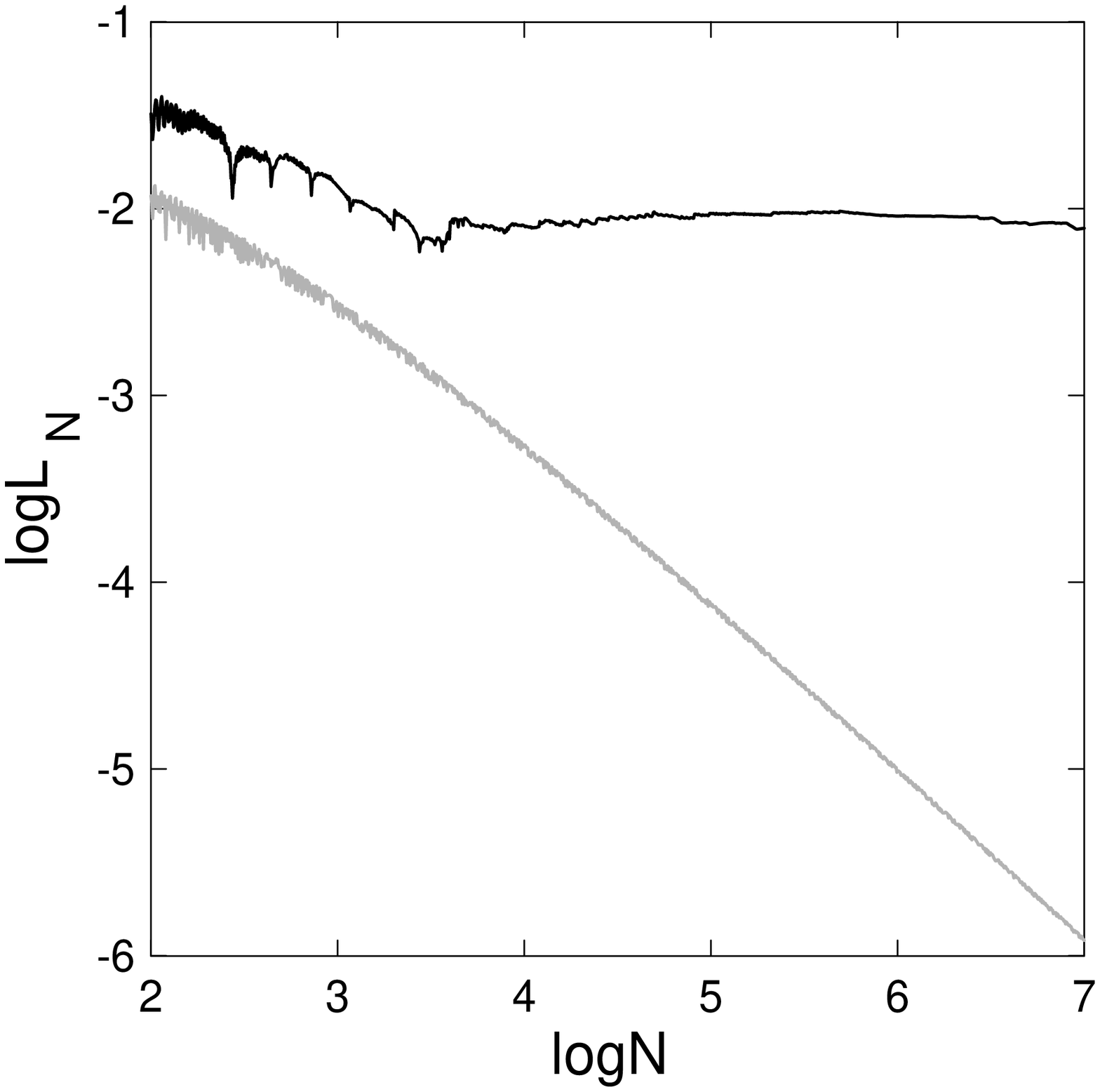} &
\includegraphics[scale=0.30]{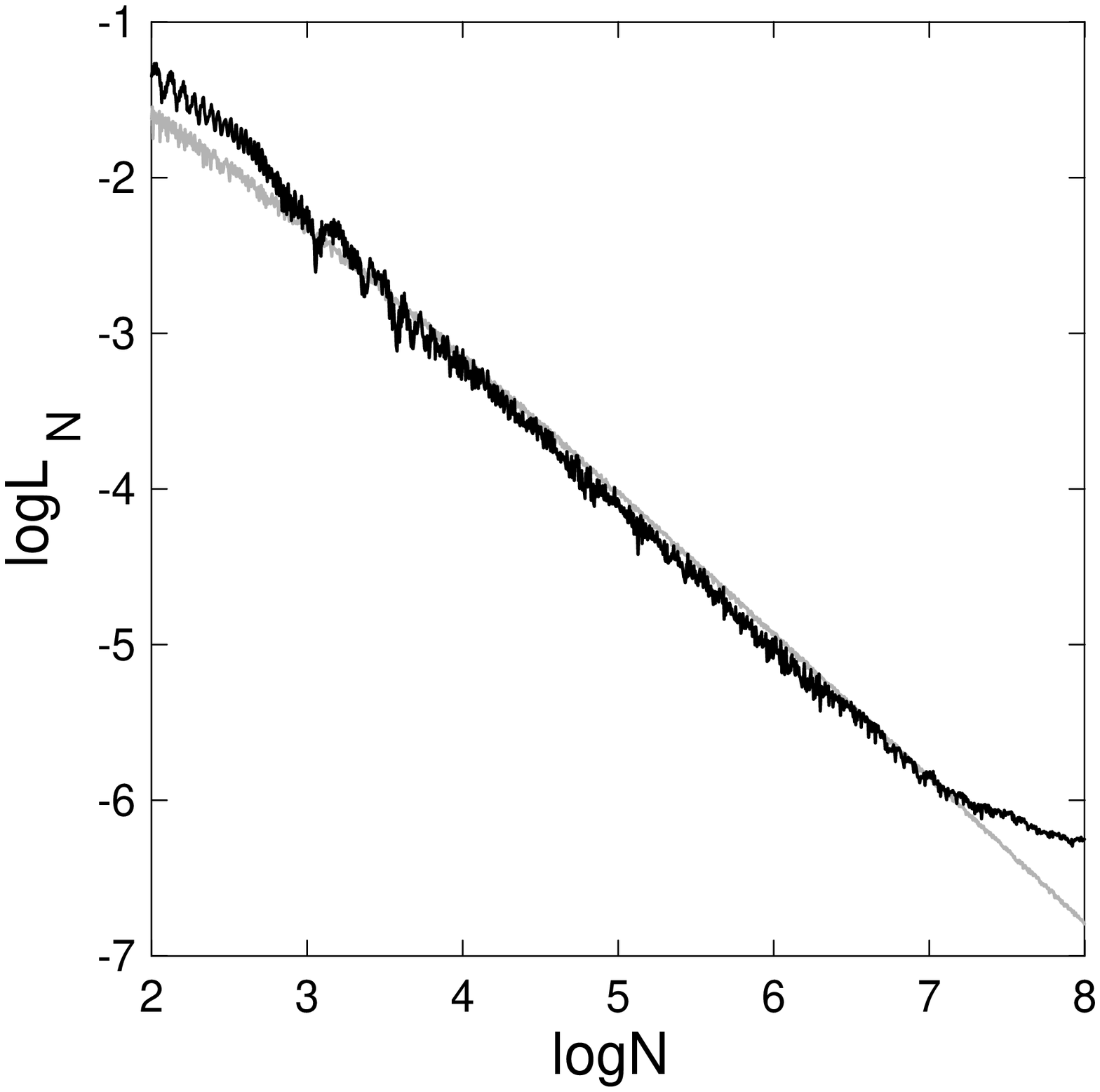}
\end{tabular}}
\caption{Evolution of $X_1(t)$ (denoted as $L_N$) with respect to the discrete
  time $t$ (denoted as $N$) in log--log scale for regular (grey curves) and
  chaotic (black curves) orbits of the 4d map (\ref{eq:4dmap}) (left panel)
  and of a 4d map composed of two coupled 2d standard maps (right panel) (see
  \cite{S_01} for more details). For regular orbits $X_1(t)$ tends to zero
  following a power law decay, $X_1(t)\propto t^{-1}$. For chaotic orbits
  $X_1(t)$ exhibits for some initial time interval the same power law decay
  before stabilizing to the positive value of the mLCE $\chi_1$. The length of
  this time interval is larger for smaller values of $\chi_1$. The chaotic
  orbits have $\chi_1 \approx 8\cdot 10^{-3}$ (left panel) and $\chi_1 \approx
  1.6\cdot 10^{-7}$ (right panel) (after \cite{S_01}).  }
\label{f:Skokos_01}
\end{figure}

In the case of chaotic orbits the variation of $X_1(t)$ is usually irregular
for relatively small $t$ and only for large $t$ the value of $X_1(t)$
stabilizes  and tends to a constant positive value which is the mLCE
$\chi_1$. If for example the value of $\chi_1$ is very small then initially,
for small and intermediate values of $t$, the term proportional to $t^{-1}$
dominates the r.~h.~s.~of equation (\ref{eq:GSO_est}) and $X_1(t) \propto
t^{-1}$. As $t$ grows the significance of term $(b+z(t))/t$ diminishes and
eventually the value of $\chi_1$ becomes dominant and $X_1(t)$ stabilizes. It
becomes evident that for smaller values of $\chi_1$ the larger is the time
required for $X_1(t)$ to reach its limiting value, and consequently $X_1(t)$
behaves as in the case of regular orbits, i.~e.~$X_1(t) \propto t^{-1}$ for
larger time intervals. This behavior is clearly seen in Figure
\ref{f:Skokos_01}
where the evolution of $X_1(t)$ of chaotic orbits with small mLCE is shown. In
particular, the values of the mLCE are $\chi_1 \approx 8\cdot 10^{-3}$ (left
panel) and $\chi_1 \approx 1.6\cdot 10^{-7}$ (right panel). In both panels the
evolution of $X_1(t)$ of regular orbits (following the power law $\propto
t^{-1}$) is also plotted in order to facilitate the comparison between the two
cases.

\section{Computation of the spectrum of LCEs}
\label{allLCEs}

While the knowledge of the mLCE $\chi_1$ can be used for determining the
regular ($\chi_1=0$) or chaotic ($\chi_1>0$) nature of orbits, the knowledge
of part, or of the whole spectrum of LCEs, provides additional information on
the underlying dynamics and on the statistical properties of the system, and
can be used for measuring the fractal dimension of strange attractors in
dissipative systems.

In Section \ref{Spectrum_properties} it was stated that, for Hamiltonian
systems the existence of an integral of motion results to a pair of zero
values in the spectrum of LCEs. As an example of such case we refer to the
Hamiltonian system studied in \cite{BFS_79}. This system has one more integral
of motion apart from the Hamiltonian function and so 4 LCEs were always found
to be equal to zero. Thus, the determination of the number of LCEs that vanish
can be used as an indicator of the number of the independent integrals of
motion that a dynamical system has.

It has been also stated in Section \ref{Spectrum_properties} that the spectrum
of the LCEs of orbits in a connected chaotic region is independent of their
initial conditions. So, we have a strong indication that two chaotic orbits
belong to connected chaotic regions if they exhibit the same spectrum. As an
example of this situation we refer to the case studied in \cite{ABS_06} of two
chaotic orbits of a $16$D Hamiltonian system having similar spectra of LCEs
but very different initial conditions.

Vice versa, the existence of different LCEs spectra of chaotic orbits provides
strong evidence that these orbits belong to different chaotic regions of the
phase space that do not communicate. In \cite{BGGS_80b} two chaotic orbits,
previously studied in \cite{CGG_78}, were found to have significantly
different spectra of LCEs and they were considered to belong to different
chaotic regions which were called the `big' (corresponding to the largest
$\chi_1$) and the `small' chaotic sea. It is worth mentioning that the
numerical results of \cite{BGGS_80b} suggested the possible existence of an
additional integral of motion for the `small' chaotic sea, since $\chi_2$
seemed to vanish. This assumption was in accordance to the results of
\cite{CGG_78} where such an integral was formally constructed.

The spectrum of LCEs is also related to two important quantities namely, the
metric entropy, also called \textit{Kolmogorov--Sinai (KS) entropy} $h$, and
the \textit{information dimension} $D_1$, which are trying to quantify the
statistical properties of dynamical systems. For the explicit definition of
these quantities, as well as detailed discussion of their relation to the
LCEs the reader is referred for example to \cite{BG_79,F_82,F_84b,ER_85}
\cite[p.~304--305]{LichtenbergL_1992} for the KS entropy and to
\cite{KY_79,F_82,FOY_83,GP_83,ER_85} for the information dimension.

In particular, Pesin \cite{P_77} showed that under suitable smoothness
conditions the relation between the KS entropy $h$ and the LCEs is given by
\begin{displaymath}
h=\int_{\cal{M}}\left[\sum_{\chi_i(\vec{x})>0} \chi_i(\vec{x})\right] d\mu ,
\end{displaymath} 
where the sum is extended over all positive LCEs and the integral is
defined over a specified region $\cal{M}$ of the phase space
$\cal{S}$.

Kaplan and Yorke \cite{KY_79} introduced a quantity, which they called the
\textit{Lyapunov dimension}
\begin{equation}
 D_L=j+\frac{ \sum_{i=1}^{j}\chi_i}{| \chi_{j+1} |}
\label{eq:L_dimension}
\end{equation} 
where $j$ is the largest integer for which
$\chi_1+\chi_2+\ldots+\chi_j\geq 0$.  The \textit{Kaplan--Yorke conjecture}
states that the information dimension $D_1$ is equal to the Lyapunov
dimension $D_L$, i.~e.
\begin{equation}
D_1=D_L,
\label{eq:KY_conj}
\end{equation} 
for a typical system, and thus, it can be used for the determination of the
fractal dimension of strange attractors. The meaning of the word `typical' is
that it is not hard to construct examples where equation (\ref{eq:KY_conj}) is
violated (see e.~g~\cite{FOY_83}). But the claim is that these examples are
pathological in that the slightest arbitrary change of the system restores the
applicability of (\ref{eq:KY_conj}) and that such violation has `zero
probability' of occurring in practice. The validity of the Kaplan--Yorke
conjecture has been proved in some cases \cite{Y_82,LY_88} although a general
proof has not been achieved yet. We note that in the case of a $2N$D
conservative system $D_L$ is equal to the dimension of the whole space,
i.~e.~$D_L=2N$, because $j=2N$ in (\ref{eq:L_dimension}) since
$\sum_{i=1}^{2N}\chi_i=0$ according to equation (\ref{eq:sumLCEs}).

So, it becomes evident that developing an efficient algorithm for the
numerical evaluation of few or of all LCEs is of great importance for the
study of dynamical systems. In this section we present the different methods
developed over the years for the computation of the spectrum of LCEs, focusing
on the method suggested by Benettin et al.~\cite{BGGS_80b}, the so--called
\textit{standard method}.

\subsection{The standard method for computing LCEs}
\label{allLCEs_Benettin}

The basis for the computation of few or even of all  LCEs is Theorem
\ref{theorem:condR}, which states that the computation of a $p$--LCE from
equation (\ref{eq:MET_06}), considering a random choice of $p$ ($1<p\leq 2N$)
linearly independent initial deviation vectors, leads to the evaluation of the
$p$--mLCE $\chi_1^{(p)}$, which is equal to the sum of the $p$ largest 1--LCEs
(\ref{eq:pLCER}).

In order to evaluate the $p$--mLCE of an orbit with initial condition
$\vec{x}(0)$, one has to follow simultaneously the time evolution of the orbit
itself and of $p$ linearly independent deviation vectors with initial
conditions $\vec{w}_1(0),\vec{w}_2(0),\ldots, \vec{w}_p(0)$ (using the
variational equations (\ref{eq:var}) or the equations of the tangent map
(\ref{eq:w_map})). Then, the $p$--mLCE is computed as the limit for $t
\rightarrow \infty$ of the quantity
\begin{eqnarray}
  X^{(p)}(t)&=&\frac{1}{t} \ln \frac{\mbox{vol}_p\left(
      d_{\vec{x}(0)}\mbox{\cal{$\Phi$}}^{t}\, \vec{w}_1(0),
      d_{\vec{x}(0)}\mbox{\cal{$\Phi$}}^{t}\, \vec{w}_2(0), \cdots,
      d_{\vec{x}(0)}\mbox{\cal{$\Phi$}}^{t}\, \vec{w}_p(0) \right)
      }{\mbox{vol}_p\left( \vec{w}_1(0), \vec{w}_2(0), \ldots, \vec{w}_p(0)
      \right)} \nonumber \\ &=&\frac{1}{t} \ln \frac{\| \vec{w}_1(t) \wedge
      \vec{w}_2(t) \wedge \cdots \wedge \vec{w}_p(t) \|}{\| \vec{w}_1(0)
      \wedge \vec{w}_2(0) \wedge \cdots \wedge \vec{w}_p(0) \|}= \frac{1}{t}
      \ln \frac{\left\| \bigwedge_{i=1}^{p} \vec{w}_i(t) \right\|}{\left\|
      \bigwedge_{i=1}^{p} \vec{w}_i(0) \right\|},
\label{eq:x_p_def_1}
\end{eqnarray}
which is also called the \textit{finite time $p$--mLCE}. So we have
\begin{equation}
\chi_1^{(p)}=\chi_1+\chi_2+\cdots+\chi_p=\lim_{t \rightarrow
\infty}X^{(p)}(t).
\label{eq:limit_pmLCE}
\end{equation}  
We recall that the quantity $\mbox{vol}_p\left( \vec{w}_1, \vec{w}_2, \ldots,
\vec{w}_p \right)$ appearing in the above definition is the volume of the
$p$--parallelogram having as edges the vectors $\vec{w}_1, \vec{w}_2, \cdots,
\vec{w}_p$ (see equations  (\ref{eq:volume_def}) and (\ref{eq:A1_norm_wedge})
in Appendix \ref{Wedge}).

The direct numerical implementation of equations (\ref{eq:x_p_def_1}) and
(\ref{eq:limit_pmLCE}) faces one additional difficulty apart from the fast
growth of the norm of deviation vectors discussed in Section
\ref{compute_maximal}. This difficulty is due to the fact that when at least
two vectors are involved (e.~g.~for the computation of $\chi_1^{(2)}$), the
angles between their directions become too small for numerical computations.

This difficulty can be overcome on the basis of the following simple remark:
an invertible linear map, as $d_{\vec{x}(0)}\mbox{\cal{$\Phi$}}^{t}$, maps a
linear $p$--dimensional subspace onto a linear subspace of the same dimension,
and the coefficient of expansion of any $p$--dimensional volume under the
action of any such linear map (like for example $\left\| \bigwedge_{i=1}^{p}
\vec{w}_i(t) \right\| / \left\| \bigwedge_{i=1}^{p} \vec{w}_i(0) \right\|$ in
our case) does not depend on the initial volume \cite{BGGS_80b}. Since the
numerical value of $\left\| \bigwedge_{i=1}^{p} \vec{w}_i(0) \right\|$ does
not depend on the choice of the orthonormal basis of the space (see Appendix
\ref{Wedge} for more details), in order to show the validity of this remark we
will consider an appropriate basis which will facilitate our calculations.

In particular, let us consider an orthonormal basis $\left\lbrace
\vec{\hat{e}}_1, \vec{\hat{e}}_2 , \ldots , \vec{\hat{e}}_p\right\rbrace $ of
the $p$--dimensional space $E^p \subseteq \mathcal{T}_{\vec{x}(0)}
\mathcal{S}$  spanned by $\left\lbrace \vec{w}_1(0), \vec{w}_2(0) , \ldots ,
\vec{w}_p(0)\right\rbrace $. This basis can be extended to an orthonormal
basis of the whole $2N$--dimensional space $\left\lbrace \vec{\hat{e}}_1,
\vec{\hat{e}}_2 , \ldots , \vec{\hat{e}}_p, \vec{\hat{e}}_{p+1}, \ldots ,
\vec{\hat{e}}_{2N} \right\rbrace $ and $E^p \subseteq \mathcal{T}_{\vec{x}(0)}
\mathcal{S}$ 
can be written as the direct sum of $E^p$ and of the $(2N-p)$--dimensional
subspace $E'$ spanned by $\left\lbrace \vec{\hat{e}}_{p+1}, \ldots ,
\vec{\hat{e}}_{2N} \right\rbrace $
\begin{displaymath}
\mathcal{T}_{\vec{x}(0)} \mathcal{S} = \mbox{$E^p \bigoplus E'$}.
\end{displaymath} 
Consider also the $2N \times p$ matrix $\textbf{W}(0)$ having as columns the
coordinates of vectors $\vec{w}_i(0)$, $i=1,2,\ldots,p$ with respect to the
complete orthonormal basis $\vec{\hat{e}}_j$, $j=1,2,\ldots,2N$, in analogy to
equation (\ref{eq:w_matrix}). Since $\vec{w}_i(0) \in E^p$ this matrix has the
form
\begin{displaymath}
\textbf{W}(0)=\left[ 
\begin{array}{c}
 \widetilde{\textbf{W}}(0) \\
\textbf{0}_{(2N-p)\times p}
\end{array}
\right] 
\end{displaymath} 
where $\widetilde{\textbf{W}}(0)$ is a square $p \times p$ matrix and
$\textbf{0}_{(2N-p)\times p}$ is the $(2N-p)\times p$ matrix with all its
elements equal to zero. Then, according to equations (\ref{eq:A1_norm_wedge})
and (\ref{eq:volume_def}) the volume of the initial $p$--parallelogram is
\begin{equation}
\left\| \bigwedge_{i=1}^{p} \vec{w}_i(0) \right\|= \left| \det
\widetilde{\textbf{W}}(0) \right|,
\label{eq:vol_par_0}
\end{equation} 
since $\det \widetilde{\textbf{W}}^{\mathrm{T}}(0) = \det
\widetilde{\textbf{W}}(0)$ for the square matrix $\widetilde{\textbf{W}}(0)$.

Each deviation vector is evolved according to equation (\ref{eq:t_flow}) and
it can be computed through equation (\ref{eq:w_ham}) or (\ref{eq:w0_map}),
with $\textbf{Y}(t)$ being the $2N\times 2N$ matrix representing the action of
$d_{\vec{x}(0)}\mbox{\cal{$\Phi$}}^t$. By doing a similar choice for the basis
of the $\cal{T}\mbox{$_{\mbox{\cal{$\Phi$}}^t(\vec{x}(0))}\cal{S}$}$ space,
equation (\ref{eq:w_matrix}) gives for the evolved vectors 
\begin{displaymath}
\left[
\begin{array}{cccc}
\vec{w}_1(t) & \vec{w}_2(t) & \cdots & \vec{w}_p(t) \end{array} \right] =
\left[
\begin{array}{cccc}
 \vec{\hat{e}}_1 & \vec{\hat{e}}_2 & \cdots &
\vec{\hat{e}}_p
\end{array} \right] \cdot \textbf{Y}(t) \cdot \textbf{W}(0) =\left[
\begin{array}{cccc}
 \vec{\hat{e}}_1 & \vec{\hat{e}}_2 & \cdots &
\vec{\hat{e}}_p
\end{array} \right] \cdot  \textbf{W}(t). 
\end{displaymath}
Writing $\textbf{Y}(t)$  as
\begin{displaymath}
\textbf{Y}(t)=\left[ 
\begin{array}{cc}
 \textbf{Y}_1(t) & \textbf{Y}_2(t)
\end{array} \right] 
\end{displaymath} 
where $\textbf{Y}_1(t)$ is the $2N \times p$ matrix formed from the first $p$
columns of $\textbf{Y}(t)$ and $\textbf{Y}_2(t)$ is the $2N \times (2N-p)$
matrix formed from the last $2N-p$ columns of $\textbf{Y}(t)$, $
\textbf{W}(t)$ assumes the form
\begin{displaymath}
\textbf{W}(t)=\textbf{Y}_1(t) \cdot \widetilde{\textbf{W}}(0).
\end{displaymath} 
Then from equation (\ref{eq:A1_norm_wedge}) we get
\begin{eqnarray}
\left\| \bigwedge_{i=1}^{p} \vec{w}_i(t) \right\| & = &\sqrt{ \det \left(
 \widetilde{\textbf{W}}^{\mathrm{T}}(0) \cdot \textbf{Y}_1^{\mathrm{T}}(t)
 \cdot \textbf{Y}_1(t) \cdot \widetilde{\textbf{W}}(0) \right) } \nonumber \\
 & = & \sqrt{ \det \widetilde{\textbf{W}}^{\mathrm{T}}(0) \det \left(
 \textbf{Y}_1^{\mathrm{T}}(t) \cdot \textbf{Y}_1(t) \right) \det
 \widetilde{\textbf{W}}(0) }\nonumber \\ & = & | \det
 \widetilde{\textbf{W}}(0) | \sqrt{\det \left( \textbf{Y}_1^{\mathrm{T}}(t)
 \cdot \textbf{Y}_1(t)\right)}.
\label{eq:vol_par_t}
\end{eqnarray} 

Thus, from equations (\ref{eq:vol_par_0}) and (\ref{eq:vol_par_t}) we conclude
that the coefficient of expansion 
\begin{displaymath} 
\frac{\left\| \bigwedge_{i=1}^{p} \vec{w}_i(t) \right\|}{\left\|
\bigwedge_{i=1}^{p} \vec{w}_i(0) \right\|} =\sqrt{\det \left(
\textbf{Y}_1^{\mathrm{T}}(t) \cdot \textbf{Y}_1(t) \right)}
\end{displaymath} 
does not depend on the initial volume but it is an intrinsic quantity of the
subspaces defined by the properties of
$d_{\vec{x}(0)}\mbox{\cal{$\Phi$}}^t$. Note that in the particular case of
$p=2N$ the coefficient of expansion is equal to $| \det \textbf{Y}(t) |$ in
accordance to equation (\ref{eq:MET_05}). An alternative way of expressing
this property is that, for two sets of linearly independent vectors
$\left\lbrace \vec{w}_1(0), \vec{w}_2(0) , \ldots , \vec{w}_p(0)\right\rbrace
$ and $\left\lbrace \vec{f}_1(0), \vec{f}_2(0) , \ldots ,
\vec{f}_p(0)\right\rbrace $ spanning the same $p$--dimensional subspace of
$\mathcal{T}_{\vec{x}(0)} \mathcal{S}$, the relation
\begin{equation} 
\frac{\left\| \bigwedge_{i=1}^{p} \vec{w}_i(t) \right\|}{\left\|
\bigwedge_{i=1}^{p} \vec{w}_i(0) \right\|} =\frac{\left\| \bigwedge_{i=1}^{p}
\vec{f}_i(t) \right\|}{\left\| \bigwedge_{i=1}^{p} \vec{f}_i(0) \right\|}
\label{eq:w_f_rate}
\end{equation} 
holds \cite{SN_79}. 

Let us now describe the method for the actual computation of the
$p$--mLCE. Similarly to the computation of the mLCE we fix a small time
interval $\tau$ and define quantity $X^{(p)}(t)$ (\ref{eq:x_p_def_1}) as
\begin{equation}
X^{(p)}(k\tau)=\frac{1}{k\tau} \sum_{i=1}^k \ln \frac{\|
\bigwedge_{j=1}^{p}d_{\vec{x}(0)}\mbox{\cal{$\Phi$}}^{i\tau}\,
\vec{w}_j(0)\|}{\| \bigwedge_{j=1}^{p}
d_{\vec{x}(0)}\mbox{\cal{$\Phi$}}^{(i-1)\tau}\, \vec{w}_j(0) \|} =
\frac{1}{k\tau} \sum_{i=1}^k \ln \gamma_i^{(p)}
\label{eq:x_p_def_2}
\end{equation} 
where $\gamma_i^{(p)}$, $i=1,2,\ldots$, is the coefficient of expansion of a
$p$--dimensional volume from $t=(i-1)\tau$ to $t=i \tau$. According to
equation (\ref{eq:w_f_rate}) $\gamma_i^{(p)}$ can be computed as the
coefficient of expansion of the $p$--parallelogram defined by any $p$ vectors
spanning the same $p$--dimensional space. A suitable choice for this set is to
consider an orthonormal set of vectors $\left\lbrace
\vec{\hat{w}}_1((i-1)\tau), \vec{\hat{w}}_2((i-1)\tau) , \ldots ,
\vec{\hat{w}}_p((i-1)\tau)\right\rbrace $ giving to equation
(\ref{eq:x_p_def_2}) the simplified form
\begin{equation}
X^{(p)}(k\tau)= \frac{1}{k\tau} \sum_{i=1}^k \ln \gamma_i^{(p)}=
\frac{1}{k\tau} \sum_{i=1}^k \ln
\left\|\bigwedge_{j=1}^{p}d_{\vec{x}((i-1)\tau)}\mbox{\cal{$\Phi$}}^{\tau}\,
\vec{\hat{w}}_j((i-1)\tau) \right\|.
\label{eq:x_p_def_3}
\end{equation} 

Thus, from equations (\ref{eq:limit_pmLCE}) and (\ref{eq:x_p_def_3}) we get
\begin{equation}
\chi_1^{(p)}=\chi_1+\chi_2+\cdots+\chi_p=\lim_{k \rightarrow
\infty}\frac{1}{k\tau} \sum_{i=1}^k \ln \gamma_i^{(p)}
\label{eq:limit_pmLCE_f}
\end{equation}  
for the computation of the $p$--mLCE. This equation is valid for $1 \leq p
\leq 2N$ since in the extreme case of $p=1$ it is simply reduced to equation
(\ref{eq:compute_x1}) with $\alpha_i \equiv \gamma_{i}^{(1)}$. In order to
estimate the values of $\chi_i$, $i=1,2,\ldots,p$, which is our actual goal,
we compute from (\ref{eq:limit_pmLCE_f}) all the $\chi_1^{(p)}$ quantities and
evaluate the LCEs from
\begin{equation}
\chi_i= \chi_1^{(i)} -  \chi_1^{(i-1)}, \,\,\, i=2,3,\ldots,p
\label{eq:comp_lce_1}
\end{equation} 
with $\chi_1^{(1)}\equiv\chi_1$ \cite{SN_79}.

Benettin et al.~\cite{BGGS_80b} noted that the $p$ largest 1--LCEs can be
evaluated at once by computing the evolution of just $p$ deviation vectors for
a particular choice of the orthonormalization procedure, namely performing the
Gram--Schmidt orthonormalization method.

Let us discuss the Gram--Schmidt orthonormalization method in some detail. Let
$\vec{w}_j(i\tau)$, $j=1,2,\ldots,p$ be the evolved deviation vectors
$\vec{\hat{w}}_j((i-1)\tau)$ from time $t=(i-1) \tau$ to $t=i \tau$. From this
set of linearly independent vectors we construct a new set of orthonormal
vectors $\vec{\hat{w}}_j(i\tau)$ from equations
\begin{eqnarray} \label{eq:GS_1}
\vec{u}_1(i\tau)&=&\vec{w}_1(i\tau), \,\,\, \gamma_{1i}= \|\vec{u}_1(i\tau)
\|, \,\,\, \vec{\hat{w}}_1(i\tau)= \frac{\vec{u}_1(i\tau)}{\gamma_{1i}}, \\
\vec{u}_2(i\tau)&=&\vec{w}_2(i\tau)-\langle\vec{w}_2(i\tau),
\vec{\hat{w}}_1(i\tau)\rangle \vec{\hat{w}}_1(i\tau), \,\,\, \gamma_{2i}=
\|\vec{u}_2(i\tau) \|, \,\,\, \vec{\hat{w}}_2(i\tau)=
\frac{\vec{u}_2(i\tau)}{\gamma_{2i}}, \nonumber \\
\vec{u}_3(i\tau)&=&\vec{w}_3(i\tau)-\langle\vec{w}_3(i\tau),
\vec{\hat{w}}_1(i\tau)\rangle \vec{\hat{w}}_1(i\tau) -\langle\vec{w}_3(i\tau),
\vec{\hat{w}}_2(i\tau)\rangle \vec{\hat{w}}_2(i\tau), \,\,\, \nonumber \\ & &
\gamma_{3i}= \|\vec{u}_3(i\tau) \|, \,\,\, \vec{\hat{w}}_3(i\tau)=
\frac{\vec{u}_3(i\tau)}{\gamma_{3i}}, \nonumber \\ \vdots \nonumber
\end{eqnarray} 
which are repeated up to the computation of $\vec{\hat{w}}_p(i\tau)$. We
remark that $\langle \vec{w}, \vec{u}\rangle$ denotes the usual inner product
of vectors $\vec{w}$, $\vec{u}$. The general form of the above equations,
which is the core of the Gram--Schmidt orthonormalization method, is
\begin{eqnarray}
\vec{u}_k(i\tau)&=&\vec{w}_k(i\tau)-\sum_{j=1}^{k-1} \langle\vec{w}_k(i\tau),
\vec{\hat{w}}_j(i\tau)\rangle \vec{\hat{w}}_j(i\tau), \,\,\, \nonumber \\ & &
\gamma_{ki}= \|\vec{u}_k(i\tau) \|, \,\,\, \vec{\hat{w}}_k(i\tau)=
\frac{\vec{u}_k(i\tau)}{\gamma_{ki}},
\label{eq:GS_general}
\end{eqnarray} 
for $ 1\leq k \leq p$.

As we will show in Section \ref{QR} the volume of the $p$--parallelogram
having as edges the vectors
$d_{\vec{x}((i-1)\tau)}\mbox{\cal{$\Phi$}}^{\tau}\,
\vec{\hat{w}}_j((i-1)\tau)= \vec{w}_j(i \tau)$, $j=1,2,\ldots,p$ is equal to
the volume of the $p$--parallelogram having as edges the vectors
$\vec{u}_j(i\tau)$, i.~e.
\begin{equation}
\left\|\bigwedge_{j=1}^{p} d_{\vec{x}((i-1)\tau)}\mbox{\cal{$\Phi$}}^{\tau}\,
\vec{\hat{w}}_j((i-1)\tau) \right\| = \left\| \bigwedge_{j=1}^{p}
\vec{u}_j(i\tau) \right\|.
\label{eq:vols_eq}
\end{equation} 
Since vectors $\vec{u}_j(i\tau)$ are normal to each other, the volume of their
$p$--parallelogram is equal to the product of their norms. This leads to
\begin{equation}
\gamma_i^{(p)}=\left\| \bigwedge_{j=1}^{p} \vec{u}_j(i\tau)
\right\|=\prod_{j=1}^p \gamma_{ji}.
\label{eq:comp_g}
\end{equation} 
Then, equation (\ref{eq:limit_pmLCE_f}) takes the form
\begin{displaymath}
\chi_1^{(p)}=\chi_1+\chi_2+\cdots+\chi_p=\lim_{k \rightarrow
\infty}\frac{1}{k\tau} \sum_{i=1}^k \ln \left( \prod_{j=1}^p \gamma_{ji}
\right).
\end{displaymath}  
Using now equation (\ref{eq:comp_lce_1}) we are able to evaluate the
1--LCE $\chi_p$ as
\begin{displaymath}
\chi_p= \chi_1^{(p)} - \chi_1^{(p-1)}= \lim_{k \rightarrow
\infty}\frac{1}{k\tau} \sum_{i=1}^k \ln \frac{\prod_{j=1}^p
\gamma_{ji}}{\prod_{j=1}^{p-1} \gamma_{ji}} = \lim_{k \rightarrow
\infty}\frac{1}{k\tau} \sum_{i=1}^k \ln \gamma_{pi}.
\end{displaymath} 

In conclusion we see that the value of the 1--LCE $\chi_p$ with
$1<p\leq 2N$ can be computed as the limiting value, for $t\rightarrow
\infty$, of the quantity
\begin{displaymath}
X_p(k \tau)=\frac{1}{k \tau} \sum_{i=1}^k \ln \gamma_{pi},
\end{displaymath} 
i.~e.
\begin{equation}
\chi_p=\lim_{k \rightarrow \infty} X_p(k \tau)=\lim_{k \rightarrow \infty}
\frac{1}{k \tau} \sum_{i=1}^k \ln \gamma_{pi},
\label{eq:chi_p}
\end{equation} 
where $\gamma_{ji}$, $j=1,2,\ldots,p$, $i=1,2,\ldots$ are quantities evaluated
during the successive orthonormalization procedures (equations (\ref{eq:GS_1})
and (\ref{eq:GS_general})). Note that for $p=1$ equation (\ref{eq:chi_p}) is
actually equation (\ref{eq:compute_x1}) with $\alpha_i \equiv \gamma_{1i}$.

\subsection{The numerical algorithm for the standard method}
\label{Algorithm_all_LCEs}

In practice, in order to compute the $p$ largest 1--LCEs with $1<p\leq 2N$ we
follow the evolution of $p$ initially orthonormal deviation vectors
$\vec{\hat{w}}_j(0)=\vec{w}_j(0)$ and every $t=\tau$ time units we replace the
evolved vectors $\vec{w}_j(k \tau)$ $j=1,2,\ldots,p$, $k=1,2,\dots$ by a new
set of orthonormal vectors produced by the Gram-Schmidt orthonormalization
method (\ref{eq:GS_general}). During the orthonormalization procedure the
quantities $\gamma_{jk}$ are computed and $\chi_1, \chi_2,\ldots, \chi_p$ are
estimated from equation (\ref{eq:chi_p}). This algorithm is described in
pseudo--code in Table \ref{tab:program_all_LCE} and can be used for the
computation of few or even all 1--LCEs.
\begin{table}
  \caption{\textbf{The standard method.} The algorithm for the computation of
  the $p$ largest LCEs $\chi_1, \chi_2,\ldots, \chi_p$ as limits for $t
  \rightarrow \infty$ of quantities $X_1(t), X_2(t), \ldots, X_p(t)$
  (\ref{eq:x_p_def_2}), according to equation (\ref{eq:chi_p}). The program
  computes the evolution of $X_1(t), X_2(t), \ldots, X_p(t)$ with respect to
  time $t$ up to a given upper value of time $t=T_M$ or until any of the
  quantities $X_1(t), X_2(t), \ldots, X_p(t)$ attain a very small value,
  smaller than a low threshold value $X_{m}$.  }
\begin{tabular}{ll}
\hline
Input:  & 1. Hamilton equations of motion (\ref{eq:Hameq_gen}) and variational equations (\ref{eq:var}), or\\
        & \quad \quad equations of the map (\ref{eq:map_gen}) and of the tangent map (\ref{eq:w_map}).\\
        & 2. Number of desired LCEs $p$.\\
        & 3. Initial condition for the orbit $\vec{x}(0)$.\\
        & 4. Initial \textit{orthonormal} deviation vectors $\vec{w}_1(0)$, $\vec{w}_2(0)$, $\ldots$, $\vec{w}_p(0)$.\\
        & 5. Renormalization time $\tau$.\\
        & 6. Maximal time: $T_M$ and minimum allowed value of $X_1(t)$, \\
        & \quad \quad $X_2(t)$, $\ldots$, $X_p(t)$: $X_{m}$.\\
\hline
Step 1  & \textbf{Set} the stopping flag, $\textsl{SF} \gets 0$, and the counter, $k \gets 1$. \\
Step 2  & \textbf{While} $(\textsl{SF}=0)$\/ \textbf{Do}\\
        & \quad \quad \textbf{Evolve} the orbit and the deviation vectors from time $t=(k-1)\tau$\\
        & \quad \quad  to $t=k \tau$, i.~e.~\textbf{Compute} $\vec{x}(k \tau)$ and $\vec{w}_1(k \tau)$, $\vec{w}_2(k \tau)$, $\ldots$, $\vec{w}_p(k \tau)$. \\
Step 3  & \quad \quad Perform the \textbf{Gram-Schmidt orthonormalization} procedure \\
        & \quad \quad according to equation (\ref{eq:GS_general}):\\
        & \quad \quad \textbf{Do} for $j=1$ to $p$ \\
        & \quad \quad \quad \quad  \textbf{Compute} current  vectors $\vec{u}_j(k \tau)$ and  values of $\gamma_{jk}$.\\
        & \quad \quad \quad \quad \textbf{Compute} and \textbf{Store} current values of $X_j(k\tau)=\frac{1}{k \tau} \sum_{i=1}^k \ln \gamma_{ji}$.\\
        & \quad \quad \quad \quad  \textbf{Set} $\vec{w}_j(k \tau) \gets \vec{u}_j(k \tau)/\gamma_{jk}$.\\
        & \quad \quad \textbf{End Do} \\
Step 4  & \quad \quad \textbf{Set}  the counter $k \gets k+1$. \\ 
Step 5  & \quad \quad \textbf{If} $[ (k \tau > T_M)$ or (Any of $X_j((k-1)\tau) < X_{m}$, $j=1,2,\ldots, p)]$ \textbf{Then} \\
        & \quad \quad \quad \quad \textbf{Set}  $\textsl{SF} \gets 1$. \\
        & \quad \quad \textbf{End If}  \\
        & \textbf{End While}\\
Step 6  & \textbf{Report} the time evolution of $X_1(t), X_2(t), \ldots, X_p(t)$.\\
\hline
\end{tabular} 
\label{tab:program_all_LCE}
\end{table}
A Fortran code of this algorithm can be found in \cite{WSSV_85}, while
\cite{S_96} contains a similar code developed for the computer algebra
platform ``Mathematica" (Wolfram Research Inc.).

\begin{figure}
\centerline{ 
\begin{tabular}{c}
\includegraphics[scale=0.47]{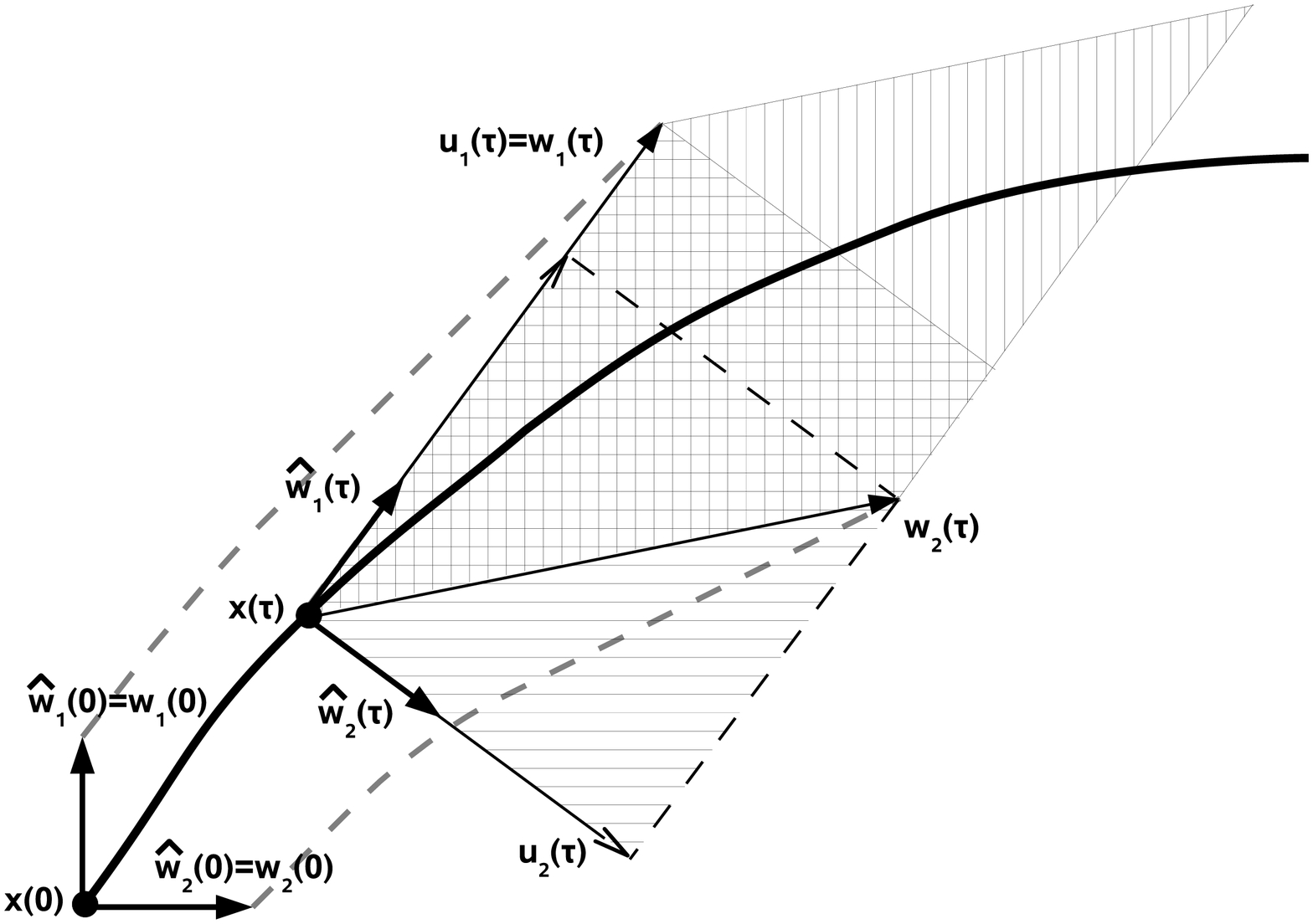} 
\end{tabular}}
\caption{ Numerical scheme for the computation of the 2 largest LCEs $\chi_1$,
  $\chi_2$ according to the standard method. The orthonormal deviation vectors
  $\vec{w}_1(0)$, $\vec{w}_2(0)$ are evolved according to the variational
  equations (\ref{eq:var}) (continuous time) or the equations of the tangent
  map (\ref{eq:w_map}) (discrete time) for $t= \tau$ time units. The evolved
  vectors $\vec{w}_1(\tau)$, $\vec{w}_2(\tau)$, are replaced by a set of
  orthonormal vectors $\vec{\hat{w}}_1(\tau)$, $\vec{\hat{w}}_2(\tau)$, which
  span the same 2--dimensional vector space, according to the Gram-Schmidt
  orthonormalization method (\ref{eq:GS_general}). Then these vectors are
  again evolved and the same procedure is iteratively applied. For each
  successive time interval $[(i-1)\tau, i\tau]$, $i=1,2,\ldots$, the quantities
  $\gamma_{1i}=\| \vec{u}_1(i\tau) \|$, $\gamma_{2i}=\| \vec{u}_2(i\tau) \|$
  are computed and $\chi_1$, $\chi_2$ are estimated from equation
  (\ref{eq:chi_p}).}
\label{f:2_LCE}
\end{figure}

Let us illustrate the implementation of this algorithm in the particular case
of the computation of the 2 largest LCEs $\chi_1$ and $\chi_2$. As shown in
Figure \ref{f:2_LCE}
we start our computation with two orthonormal deviation vectors $\vec{w}_1(0)$
and $\vec{w}_2(0)$ which are evolved to $\vec{w}_1(\tau)$, $\vec{w}_2(\tau)$
at $t=\tau$. Then according to the the Gram-Schmidt orthonormalization method
(\ref{eq:GS_1}) we define vectors $\vec{u}_1(\tau)$ and $\vec{u}_2(\tau)$. In
particular, $\vec{u}_1(\tau)$ coincides with $\vec{w}_1(\tau)$ while,
$\vec{u}_2(\tau)$ is the component of vector $\vec{w}_2(\tau)$ in the
direction perpendicular to vector $\vec{u}_1(\tau)$. The norms of these two
vectors define the quantities $\gamma_{11}=\| \vec{u}_1(\tau) \|$,
$\gamma_{21}=\| \vec{u}_2(\tau) \|$ needed for the estimation of $\chi_1$,
$\chi_2$ from equation (\ref{eq:chi_p}). Then vectors $\vec{\hat{w}}_1(\tau)$
and $\vec{\hat{w}}_2(\tau)$ are defined as unitary vectors in the directions
of $\vec{u}_1(\tau)$ and $\vec{u}_2(\tau)$ respectively. Since the unitary
vectors $\vec{\hat{w}}_1(\tau)$, $\vec{\hat{w}}_2(\tau)$ are normal by
construction they constitute the initial set of orthonormal vectors for the
next iteration of the algorithm. From Figure \ref{f:2_LCE} we easily see that
the parallelograms defined by vectors $\vec{w}_1(\tau)$, $\vec{w}_2(\tau)$ and
by vectors $\vec{u}_1(\tau)$ and $\vec{u}_2(\tau)$ have the same area. This
equality corresponds to the particular case $p=2$, $i=1$ of equation
(\ref{eq:vols_eq}). Evidently, since vectors $\vec{u}_1(\tau)$,
$\vec{u}_2(\tau)$ are perpendicular to each other, we have $\mbox{vol}_2\left(
\vec{u}_1(\tau), \vec{u}_2(\tau)\right)= \gamma_{11}\gamma_{21}$ in accordance
to equation (\ref{eq:comp_g}).

\subsection{Connection between the standard method and the QR decomposition}
\label{QR}

Let us rewrite  equations (\ref{eq:GS_1}) of the Gram-Schmidt
orthonormalization procedure, by solving them with respect to $\vec{w}_j(i
\tau)$, $j=1,2,\ldots,p$, with $1<p\leq 2N$
\begin{eqnarray} \label{eq:GS_QR_55}
\vec{w}_1(i\tau)&=&\gamma_{1i}\vec{\hat{w}}_1(i\tau) \\ \vec{w}_2(i\tau) &=
&\langle \vec{\hat{w}}_1(i\tau), \vec{w}_2(i\tau) \rangle
\vec{\hat{w}}_1(i\tau)+ \gamma_{2i} \vec{\hat{w}}_2(i\tau) \nonumber \\
\vec{w}_3(i\tau) &= &\langle \vec{\hat{w}}_1(i\tau), \vec{w}_3(i\tau) \rangle
\vec{\hat{w}}_1(i\tau)+ \langle \vec{\hat{w}}_2(i\tau), \vec{w}_3(i\tau)
\rangle \vec{\hat{w}}_2(i\tau)+\gamma_{3i} \vec{\hat{w}}_3(i\tau) \nonumber \\
\vdots \nonumber
\end{eqnarray} 
and get the general form
\begin{displaymath}
\vec{w}_k(i\tau)= \sum_{j=1}^{k-1} \langle \vec{\hat{w}}_j(i\tau),
\vec{w}_k(i\tau) \rangle \vec{\hat{w}}_j(i\tau)+\gamma_{ki}
\vec{\hat{w}}_k(i\tau),\,\,\, k=1,2,\ldots, p.
\end{displaymath} 
This set of equations can be rewritten in matrix form as follows:
\begin{eqnarray*}
\left[
\begin{array}{cccc}
 \vec{w}_1(i \tau) & \vec{w}_2(i \tau) & \cdots & \vec{w}_p(i \tau)
 \end{array} \right] = \left[
\begin{array}{cccc}
 \vec{\hat{w}}_1(i \tau) & \vec{\hat{w}}_2(i \tau) & \cdots &
\vec{\hat{w}}_p(i \tau)
\end{array} \right] \cdot \nonumber \\
\cdot \left[ \begin{array}{ccccc} \gamma_{1i} & \langle
\vec{\hat{w}}_1(i\tau), \vec{w}_2(i\tau) \rangle & \langle
\vec{\hat{w}}_1(i\tau), \vec{w}_3(i\tau) \rangle & \cdots & \langle
\vec{\hat{w}}_1(i\tau), \vec{w}_p(i\tau) \rangle \\ 0 & \gamma_{2i} & \langle
\vec{\hat{w}}_2(i\tau), \vec{w}_3(i\tau) \rangle & \cdots & \langle
\vec{\hat{w}}_2(i\tau), \vec{w}_p(i\tau) \rangle \\ 0 & 0 & \gamma_{3i} &
\cdots & \langle \vec{\hat{w}}_3(i\tau), \vec{w}_p(i\tau) \rangle \\ \vdots &
\vdots & \vdots & & \vdots \\ 0 & 0 & 0 & & \gamma_{pi} \\
\end{array}
 \right].
\end{eqnarray*}
So the $2N \times p$ matrix $\textbf{W}(i \tau) = \left[
\begin{array}{cccc}
  \vec{w}_1(i \tau) & \vec{w}_2(i \tau) & \cdots & \vec{w}_p(i \tau)
  \end{array}\right]$, having as columns the linearly independent deviation
  vectors $\vec{w}_j(i \tau)$, $j=1,2,\ldots,p$ is written as a product of the
  $2N \times p$ matrix $\textbf{Q}= \left[
\begin{array}{cccc}
  \vec{\hat{w}}_1(i \tau) & \vec{\hat{w}}_2(i \tau) & \cdots &
  \vec{\hat{w}}_p(i \tau)
  \end{array}\right]$, having as columns the coordinates of the orthonormal
  vectors $\vec{\hat{w}}_j(i \tau)$, $j=1,2,\ldots,p$ and satisfying
  $\textbf{Q}^{\mathrm{T}} \textbf{Q} = \textbf{I}_p$, and of an upper
  triangular $p\times p$ matrix $\textbf{R}(i \tau)$ with positive diagonal
  elements
\begin{displaymath}
\textbf{R}_{jj}(i \tau)=\gamma_{ji}, \,\,\, j=1,2,\ldots,p, \,\,\,
i=1,2,\ldots .
\end{displaymath} 
From equations (\ref{eq:GS_QR_55}) we easily see that $\langle
\vec{\hat{w}}_j(i\tau), \vec{w}_j(i\tau) \rangle=\gamma_{ji}$ and so matrix
$\textbf{R}(i \tau)$ can be also expressed as
\begin{displaymath}
\textbf{R}(i \tau)= \left[ \begin{array}{cccc} \langle \vec{\hat{w}}_1(i\tau),
\vec{w}_1(i\tau) \rangle & \langle \vec{\hat{w}}_1(i\tau), \vec{w}_2(i\tau)
\rangle & \cdots & \langle \vec{\hat{w}}_1(i\tau), \vec{w}_p(i\tau) \rangle \\
0 & \langle \vec{\hat{w}}_2(i\tau), \vec{w}_2(i\tau) \rangle & \cdots &
\langle \vec{\hat{w}}_2(i\tau), \vec{w}_p(i\tau) \rangle \\ \vdots & \vdots &
& \vdots \\ 0 & 0 & & \langle \vec{\hat{w}}_p(i\tau), \vec{w}_p(i\tau) \rangle
\\
\end{array}
 \right].
\end{displaymath} 

The above procedure is the so--called QR decomposition of a matrix. In
practice, we proved by actually constructing the $ \textbf{Q}$ and $
\textbf{R}$ matrices via the Gram-Schmidt orthonormalization method, the
following theorem:
\begin{theorem}
  Let $\textbf{A}$ be an $n \times m$ $(n\ge m)$ matrix with linearly
  independent columns. Then $\textbf{A}$ can be uniquely factorized as
\begin{displaymath}
\textbf{A}=\textbf{Q} \cdot \textbf{R},
\end{displaymath} 
where $\textbf{Q}$ is an $n \times m$ matrix with orthogonal columns,
satisfying $\textbf{Q}^{\mathrm{T}} \textbf{Q} = \textbf{I}_m$ and
$\textbf{R}$ is an $m \times m$ invertible upper triangular matrix with
positive diagonal entries.
\label{Theorem:QR}
\end{theorem}

Although we presented the QR decomposition through the Gram-Schmidt
orthonormalization procedure this decomposition can also be achieved by
others, computationally more efficient techniques like for example the
Householder transformation \cite{GPL_90}\cite[\S 2.10]{NumRec}.

Observing that the quantities $\gamma_{ji}$, $j=1,2\ldots,p$, $i=1,2\ldots$,
needed for the evaluation of the LCEs through equation (\ref{eq:chi_p}) are
the diagonal elements of $\textbf{R}(i \tau)$ we can implement a variant of
the standard method for the computation on the LCEs, which is based on the QR
decomposition procedure \cite{ER_85,GPL_90,DV_95,DRV_97}. Similarly to the
procedure followed in Section \ref{Algorithm_all_LCEs}, in order to compute
the $p$ ($1<p\leq2N$) largest LCEs we follow the evolution of $p$ initially
orthonormal deviation vectors $\vec{\hat{w}}_j(0)=\vec{w}_j(0)$,
$j=1,2\ldots,p$, which can be considered as columns of a $2N \times p$ matrix
$\textbf{Q}(0)$. Every $t=\tau$ time units the matrix $\textbf{W}(i \tau)$,
$i=1,2,\ldots$, having as columns the deviation vectors
\begin{displaymath}
d_{\vec{x}((i-1)\tau)}\mbox{\cal{$\Phi$}}^{\tau}\, \vec{\hat{w}}_j((i-1)\tau)=
\vec{w}_j(i \tau), \,\,\,j=1,2,\ldots,p,
\end{displaymath} 
i.~e.~the columns of $\textbf{Q}((i-1)\tau)$ evolved in time interval $[
(i-1)\tau, i\tau]$ by the action of
$d_{\vec{x}((i-1)\tau)}\mbox{\cal{$\Phi$}}^{\tau}$, undergoes the QR
decomposition procedure
\begin{equation}
\textbf{W}(i \tau)=\textbf{Q}(i \tau) \cdot \textbf{R}(i \tau)
\label{eq:QR_matrixU}
\end{equation} 
and the new $\textbf{Q}(i \tau)$ is again evolved for the next time interval
$[ i\tau, (i+1) \tau]$, and so on and so forth. Then the LCEs are estimated
from the values of the diagonal elements of matrix $\textbf{R}(i \tau)$ as
\begin{equation}
\chi_p=\lim_{k \rightarrow \infty} \frac{1}{k \tau} \sum_{i=1}^k \ln
\textbf{R}_{pp}(i \tau).
\label{eq:chi_p_QR}
\end{equation} 
The corresponding algorithm is presented in pseudo-code in Table
\ref{tab:program_all_LCE_QR}.
\begin{table}
  \caption{\textbf{Discrete QR decomposition.} The algorithm for the
  computation of the $p$ largest LCEs $\chi_1, \chi_2,\ldots, \chi_p$
  according to the QR decomposition method.  The program computes the
  evolution of $X_1(t), X_2(t), \ldots, X_p(t)$ with respect to time $t$ up to
  a given upper value of time $t=T_M$ or until any of the these quantities
  becomes smaller than a low threshold value $X_{m}$.  }
\begin{tabular}{ll}
\hline
Input:  & 1. Hamilton equations of motion (\ref{eq:Hameq_gen}) and variational equations (\ref{eq:var}), or\\
        & \quad \quad equations of the map (\ref{eq:map_gen}) and of the tangent map (\ref{eq:w_map}).\\
        & 2. Number of desired LCEs $p$.\\
        & 3. Initial condition for the orbit $\vec{x}(0)$.\\
        & 4. Initial matrix $\textbf{Q}(0)$ having as columns \textit{orthonormal} deviation vectors \\
        & \quad \quad $\vec{w}_1(0)$, $\vec{w}_2(0)$, $\ldots$, $\vec{w}_p(0)$.\\
        & 5. Time interval $\tau$ between successive QR decompositions.\\
        & 6. Maximal time: $T_M$ and minimum allowed value of $X_1(t)$, \\
        & \quad \quad $X_2(t)$, $\ldots$, $X_p(t)$: $X_{m}$.\\
\hline
Step 1  & \textbf{Set} the stopping flag, $\textsl{SF} \gets 0$, and the counter, $k \gets 1$. \\
Step 2  & \textbf{While} $(\textsl{SF}=0)$\/ \textbf{Do}\\
        & \quad \quad \textbf{Evolve} the orbit and the matrix $\textbf{Q}((k-1) \tau)$ from time $t=(k-1)\tau$\\
        & \quad \quad  to $t=k \tau$, i.~e.~\textbf{Compute} $\vec{x}(k \tau)$ and $\textbf{W}(i \tau)$. \\
Step 3  & \quad \quad Perform the \textbf{QR decomposition} of $\textbf{W}(i \tau)$ according to (\ref{eq:QR_matrixU}):  \\            & \quad \quad \textbf{Compute} $\textbf{Q}(k \tau)$ and $\textbf{R}(k \tau)$. \\
        & \quad \quad \textbf{Compute} and \textbf{Store} current values of $X_j(k\tau)=\frac{1}{k \tau} \sum_{i=1}^k \ln \textbf{R}_{jj}(i \tau)$, \\ 
        & \quad \quad$j=1,2\ldots,p$.\\
Step 4  & \quad \quad \textbf{Set}  the counter $k \gets k+1$. \\ 
Step 5  & \quad \quad \textbf{If} $[ (k \tau > T_M)$ or (Any of $X_j((k-1)\tau) < X_{m}$, $j=1,2,\ldots, p)]$ \textbf{Then} \\
        & \quad \quad \quad \quad \textbf{Set}  $\textsl{SF} \gets 1$. \\
        & \quad \quad \textbf{End If}  \\
        & \textbf{End While}\\
Step 6  & \textbf{Report} the time evolution of $X_1(t), X_2(t), \ldots, X_p(t)$.\\
\hline
\end{tabular} 
\label{tab:program_all_LCE_QR}
\end{table}
From the above--presented analysis it becomes evident that the standard method
developed by Shimada and Nagashima \cite{SN_79} and Benettin et
al.~\cite{BGGS_80b} for the computation of the LCEs, is practically a QR
decomposition procedure performed by the Gram--Schmidt orthonormalization
method, although the authors of these papers formally do not refer to the
QR decomposition. We note that both the standard method and the QR
decomposition technique presented here can be used for the computation of part
$(p <2N)$ or of the whole $(p =2N)$ spectrum of LCEs.

As a final remark on the QR decomposition technique let us show the validity
of equation (\ref{eq:vols_eq}) by considering the QR decomposition of matrix
$\textbf{W}(i \tau)$ (\ref{eq:QR_matrixU}). According to equations
(\ref{eq:A1_norm_wedge}) and (\ref{eq:volume_def}) we have
\begin{eqnarray*}
\left\| \bigwedge_{j=1}^{p} \vec{w}_j(i \tau) \right\| & = &\sqrt{ \det \left(
 \textbf{W}^{\mathrm{T}}(i \tau) \cdot \textbf{W}(i \tau) \right) } \nonumber
 \\ & = &\sqrt{ \det \left( \textbf{R}^{\mathrm{T}}(i \tau) \cdot
 \textbf{Q}^{\mathrm{T}}(i \tau) \cdot \textbf{Q}(i \tau) \cdot \textbf{R}(i
 \tau)\right) } \nonumber \\ & = & \sqrt{ \det \textbf{R}^{\mathrm{T}}(i \tau)
 \det \textbf{R}(i \tau) } = \left| \det \textbf{R}(i \tau) \right| \nonumber
 \\ & = & \prod_{j=1}^p \gamma_{ji}= \prod_{j=1}^p\left\| \vec{u}_j(i \tau)
 \right\| =\left\| \bigwedge_{j=1}^{p} \vec{u}_j(i \tau) \right\|,
\end{eqnarray*} 
where the identities $\textbf{Q}^{\mathrm{T}} \textbf{Q} = \textbf{I}_p$ and $
\det \textbf{R}(i \tau)= \prod_{j=1}^p \gamma_{ji}$ have been used.

\subsection{Other methods for computing LCEs}
\label{Dis_Cont}

Over the years several methods have been proposed and applied for computing
the numerical values of the LCEs. The standard method we discussed so far, is
the first and probably the simplest method to address this problem. As we
showed in Section \ref{QR} the standard method, which requires successive
applications of the Gram-Schmidt orthonormalization procedure, is practically
equivalent to the QR decomposition technique.

The reorthonormalization of deviation vectors plays an indispensable role for
computing the LCEs and the corresponding methods can be distinguished in
discrete and continuous methods.  The \textit{discrete methods} iteratively
approximate the LCEs in a finite number of (discrete) time steps and therefore
apply to both continuous and discrete dynamical systems
\cite{GPL_90,DV_95,DRV_97}. The standard method and its QR decomposition
version, are discrete methods. A method is called \textit{continuous} when all
relevant quantities are obtained as solutions of certain ordinary differential
equations, which maintain orthonormality of deviation vectors
continuously. Therefore such methods can only be formulated for continuous
dynamical systems and not for maps. The use of continuous orthonormalization
for the numerical computation of LCEs was first proposed by Goldhirsch et
al.~\cite{GSO_87} and afterwards developed by several authors
\cite{GK_87,GPL_90,DV_95,DRV_97,CR_97,RHR_98,RS_00,LYOL_05,DE_06}.

Discrete and continuous methods are based on appropriate decomposition of
matrices performed usually by the QR decomposition or by the SVD
procedure. The discrete QR decomposition, which has already been presented in
Section \ref{QR} is the most frequently used method and has proved to be quite
efficient and reliable. The continuous QR decomposition, as well as methods
based on the SVD procedure are discussed in some detail at the end of the
current section.

Variants of these techniques have been also proposed by several authors. Let
us briefly refer to some of them. Rangarajan et al.~\cite{RHR_98} introduced a
method for the computation of part or of the whole spectrum of LCEs for
continuous dynamical systems, which does not require rescaling and
renormalization of vectors.  The key feature of their approach is the use of
explicit group theoretical representations of orthogonal matrices, which leads
to a set of coupled ordinary differential equations for the LCEs along with
the various angles parameterizing the orthogonal matrices involved in the
process. Ramasubramanian and Sriram \cite{RS_00} showed that the method is
competitive with the standard method and the continuous QR decomposition.

Carbonell et al.~\cite{CJB_02} proposed a method for the evaluation of the
whole spectrum of LCEs by approximating the differential equations describing
the evolution of an orbit of a continuous dynamical system and their
associated variational equations by two piecewise linear sets of ordinary
differential equations. Then an SVD or a QR decomposition--based method is
applied to these two new sets of equations, allowing us to obtain
approximations of the LCEs of the original system. An advantage of this method
is that it does not require the simultaneous integration of the two sets of
piecewise linear equations.

Lu et al.~\cite{LYOL_05} proposed a new continuous method for the computation
of few or of all LCEs, which is related to the QR decomposition
technique. According to their method one follows the evolution of orthogonal
vectors, similarly to the QR method, but does not require them to be
necessarily orthonormal. By relaxing the length requirement Lu et
al.~~\cite{LYOL_05} established a set of recursive differential equations for
the evolution of these vectors. Using symplectic Runge--Kutta integration
schemes for the evolution of these vectors they succeeded in preserving
automatically the orthogonality between any two successive
vectors. Normalization of vectors occurs whenever the magnitude of any vector
exceeds given lower or upper bounds.

Chen et al.~\cite{CDP_06} proposed a simple discrete QR algorithm for the
computation of the whole spectrum of LCEs of a continuous dynamical
system. Their method is based on a suitable approximation of the solution of
variational equations by assuming that the Jacobian matrix remains constant
over small integration time steps. Thus, the scheme requires the numerical
solution of the $2N$ equations of motion but not the solution of the $(2N)^2$
variational equations since their solution is approximated by an explicit
expression involving the computed orbit. This approach led to a
computationally fast evaluation of the LCEs for various multidimensional
dynamical systems studied in \cite{CDP_06}.

It is worth mentioning here a completely different approach, with respect to
the above--mentioned techniques, which was developed at the early 80's. In
particular, Fr\o{}yland proposed in \cite{F_83} an algorithm for the
computation of LCEs, which he claimed to be quite efficient in the case of
low--dimensional systems, and applied it to the Lorenz system
\cite{FA_84}. The basic idea behind this algorithm is the implementation of
appropriate differential equations describing the time evolution of volume
elements around the orbits of the dynamical system, instead of defining these
volumes through deviation vectors whose evolution is governed by the usual
variational equations (\ref{eq:var}).

Apart from the actual numerical computation of the values of the LCEs, methods
for the theoretical estimation of those values have been also developed. For
example, Li and Chen \cite{LC_04} provided a theorem for the estimation of
lower and upper bounds for the values of all LCEs in the case of discrete
maps. These results were also generalized for the case of continues dynamical
systems \cite{LX_04}. The validity of these estimates was demonstrated by a
comparison between the estimated bounds and the numerically computed spectrum
of LCEs of some specific dynamical systems \cite{LC_04,LX_04}.

Finally, let us refer to a powerful analytical method which allows one to
verify the existence of positive LCEs for a dynamical system, the so--called
\textit{cone technique}. The method was suggested by Wojtkowski \cite{W_85}
and has been extensively applied for the study of chaotic billiards
\cite{W_85,W_86,DL_91,M_94} and geodesic flows \cite{D_88,D_88b,BD_97}. A
concise description of the techniques can also be found in \cite{BP_06} and
\cite[\S 3.13]{ChernovM_2006}. Considering the space $\mathbb{R}^n$ a
\textit{cone} $\cal{C}_{\gamma}$, with $\gamma >0$, centered around
$\mathbb{R}^{n-k}$ is
\begin{equation}
\cal{C}_{\gamma}= \mbox{$\left\lbrace (\vec{u},\vec{v}) \in \mathbb{R}^k
\times \mathbb{R}^{n-k} : \| \vec{u} \| < \gamma \| \vec{v} \|\right\rbrace
\cup (\vec{0},\vec{0}) $}.
\label{eq:cone}
\end{equation} 
Note that $\left\lbrace\vec{0} \right\rbrace \times \mathbb{R}^{n-k} \subset
\cal{C}_{\gamma}$ for every $\gamma$. In the particular case of $n=3$, $k=2$,
$\cal{C}_{\gamma}$ corresponds to the usual 3--dimensional cone, while in the
case of the plane ($n=2$) a cone $\cal{C}_{\gamma}$ around an axis $L$ is the
set of vectors of $\mathbb{R}^2$ that make angle $\phi < \arctan \gamma$ with
the line $L$. In the case of Hamiltonian systems (and symplectic maps) a cone
can get the simple form $\delta \vec{q}\cdot \delta \vec{p} > 0$.  Finding an
invariant family of cones (\ref{eq:cone}) in $\cal{T}_{\vec{x}} \cal{S}$,
which are mapped strictly into themselves by
$d_{\vec{x}}\mbox{\cal{$\Phi$}}^t$, guarantees that the values of the $n-k$
largest LCEs are positive \cite{W_85,W_86}. We emphasize that the cone
technique is not used for the explicit numerical computation of the LCEs, but
for the analytical proof of the existence of positive LCEs, providing at the
same time some bounds for their actual values.

\subsubsection{Continuous QR decomposition methods}
\label{QR_cont}

The QR decomposition methods allow the computation of all or of the $p$ ($1<p
< 2N$) largest LCEs. Let us discuss in more detail the developed procedure
for both cases following mainly \cite{GPL_90,DV_95,LYOL_05}.

\paragraph{Computing the complete spectrum of LCEs}
\label{QR_cont_all}

The basic idea of the method is to avoid directly solving the differential
equation (\ref{eq:Y_ham}), by requiring $\textbf{Y}(t)=\textbf{Q}(t)
\textbf{R}(t)$ where $\textbf{Q}(t)$ is orthogonal and $\textbf{R}(t)$ is
upper triangular with positive diagonal elements, according to Theorem
\ref{Theorem:QR}. With this decomposition, one can write equation
(\ref{eq:Y_ham}) into the form
\begin{displaymath}
\textbf{Q}^{\mathrm{T}}\dot{\textbf{Q}}+\dot{\textbf{R}}\textbf{R}^{-1}=\textbf{Q}^{\mathrm{T}}\textbf{A}\textbf{Q},
\end{displaymath} 
where, for convenience, we dropped out the explicit dependence of the matrices
on time $t$, i.~e.~$\textbf{Q}(t) \equiv\textbf{Q}$. Since
$\textbf{Q}^{\mathrm{T}}\dot{\textbf{Q}}$ is skew and $
\dot{\textbf{R}}\textbf{R}^{-1}$ is upper triangular, one reads off the
differential equations
\begin{equation}
\dot{\textbf{Q}}=\textbf{Q}\textbf{S},
\label{eq:QRC_02}
\end{equation} 
where $\textbf{S}$ is the skew symmetric matrix 
\begin{displaymath}
\textbf{S}=\textbf{Q}^{\mathrm{T}}\dot{\textbf{Q}}
\end{displaymath} 
with elements
\begin{equation}
\textbf{S}_{ij}=\left\{\begin{array}{ccc}
(\textbf{Q}^{\mathrm{T}}\textbf{A}\textbf{Q})_{ij} & & i>j \\
0 & & i=j \\
-(\textbf{Q}^{\mathrm{T}}\textbf{A}\textbf{Q})_{ji} & & i<j
\end{array} \right. , \,\,\, i,j=1,2,\ldots,2N,
\label{eq:QRC_04}
\end{equation} 
and
\begin{equation}
\frac{\dot{\textbf{R}}_{pp}}{\textbf{R}_{pp}}=
(\textbf{Q}^{\mathrm{T}}\textbf{A}\textbf{Q})_{pp}, \,\,\, p,=1,2,\ldots,2N
\label{eq:QRC_05}
\end{equation} 
where $\textbf{R}_{pp}$ are the diagonal elements of $\textbf{R}$. As we have
already seen in equation (\ref{eq:chi_p_QR}) the LCEs are related to the
elements $\textbf{R}_{pp}$, through
\begin{displaymath}
\chi_p=\lim_{t \rightarrow \infty} \frac{1}{t}  \ln \textbf{R}_{pp}(t).
\end{displaymath} 
Thus, in order to compute the spectrum of LCEs only equations
(\ref{eq:QRC_02}) and (\ref{eq:QRC_05}) have to be solved simultaneously with
the equations of motion (\ref{eq:Hameq_gen}). In practice, the knowledge of
matrix $\textbf{R}$ is not necessary for the actual computation of the
LCEs. Noticing that
\begin{equation}
\frac{d}{dt}\left(\ln \textbf{R}_{pp} \right)=
\frac{\dot{\textbf{R}}_{pp}}{\textbf{R}_{pp}}=
(\textbf{Q}^{\mathrm{T}}\textbf{A}\textbf{Q})_{pp}=\vec{q}_p\cdot \textbf{A}
\vec{q}_p ,
\label{eq:QRC_07}
\end{equation} 
where $\vec{q}_p$ is the $p$th column vector of $\textbf{Q}$, we can compute
the LCEs using
\begin{displaymath}
\chi_p=\lim_{t \rightarrow \infty}\frac{1}{t} \int_0^t \vec{q}_p\cdot
\textbf{A} \vec{q}_p dt .
\end{displaymath} 
In practice, the LCEs can be estimated through a recursive formula. Let
\begin{displaymath}
X_p(k \tau)=\frac{1}{k \tau} \int_0^{k \tau} \vec{q}_p\cdot \textbf{A}
\vec{q}_p dt .
\end{displaymath} 
Then we have
\begin{eqnarray*}
X_p\left( (k+1) \tau\right) = \frac{1}{(k+1) \tau} \int_0^{(k+1) \tau}
\vec{q}_p\cdot \textbf{A} \vec{q}_p dt & &\nonumber \\ = \frac{1}{(k+1)\tau}
\int_0^{k \tau}\vec{q}_p\cdot \textbf{A} \vec{q}_p dt + \frac{1}{(k+1)\tau}
\int_{k \tau}^{(k+1) \tau}\vec{q}_p\cdot \textbf{A} \vec{q}_p dt & &.
\end{eqnarray*} 
Replacing the first integral with $k \tau X_p(k \tau)$ we get 
\begin{equation}
X_p\left( (k+1) \tau\right) = \frac{k}{k+1} X_p(k \tau)+\frac{1}{(k+1) \tau}
\int_{k \tau}^{(k+1) \tau} \vec{q}_p\cdot \textbf{A} \vec{q}_p dt ,
\label{eq:QRC_11}
\end{equation} 
and 
\begin{equation}
\chi_p=\lim_{k \rightarrow \infty} X_p(k \tau).
\label{eq:QRC_12}
\end{equation} 

The basic difference between the discrete QR decomposition method presented in
Section \ref{QR}, and the continuous QR method presented here, is that in the
first method the orthonormalization is performed numerically at discrete time
steps, while the latter method seeks to maintain the orthogonality via solving
differential equations that encode the orthogonality continously.

\paragraph{Computation of the $p>1$ largest LCEs}
\label{QR_cont_part}

If we want to compute the $p$ largest LCEs, with $1<p < 2N$, we change
equation (\ref{eq:Y_ham}) to
\begin{equation}
\dot{\textbf{Y}}(t)= \textbf{A}(t) \, \textbf{Y}(t)\,\,\,,\,\,\, \mbox{with}
\,\,\, \textbf{Y}(0)^{\mathrm{T}}\textbf{Y}(0)=\textbf{I}_{p}.
\label{eq:QRCp_01}
\end{equation}
where $\textbf{Y}(t)$ is in practice, the $2N \times p$ matrix having as
columns the $p$ deviation vectors $\vec{w}_1(t), \vec{w}_2(t), \ldots,
\vec{w}_p(t)$. Applying Theorem \ref{Theorem:QR} we get
$\textbf{Y}(t)=\textbf{Q}(t) \textbf{R}(t)$ where $\textbf{Q}(t)$ is
orthogonal so that the identity $\textbf{Q}^{\mathrm{T}}
\textbf{Q}=\textbf{I}$ holds but not the
$\textbf{Q}\textbf{Q}^{\mathrm{T}}=\textbf{I}$. Then from equation
(\ref{eq:QRCp_01}) we get
\begin{displaymath}
\dot{\textbf{R}}= \left( \textbf{Q}^{\mathrm{T}}\textbf{A} \textbf{Q}-
\textbf{S} \right) \textbf{R}
\end{displaymath}
where $\textbf{S}=\textbf{Q}^{\mathrm{T}}\dot{\textbf{Q}}$ is a $p \times p$
matrix whose elements are given by equation (\ref{eq:QRC_04}) for
$i,j=1,2,\ldots,p$.  Since $\textbf{R}$ is invertible, from the relations
\begin{displaymath}
\dot{\textbf{R}}\textbf{R}^{-1}= \textbf{Q}^{\mathrm{T}}\textbf{A} \textbf{Q}-
\textbf{S}
\end{displaymath}
and 
\begin{equation}
\dot{\textbf{Q}}=\textbf{A} \textbf{Q}-\textbf{Q}\dot{\textbf{R}}
\textbf{R}^{-1},
\label{eq:QRCp_02c}
\end{equation}
we obtain 
\begin{displaymath}
\dot{\textbf{Q}}= \left( \textbf{A}
-\textbf{Q}\textbf{Q}^{\mathrm{T}}\textbf{A}
+ \textbf{Q}\textbf{S}
\textbf{Q}^{\mathrm{T}} \right) \textbf{Q},
\end{displaymath}
or
\begin{equation}
\dot{\textbf{Q}}=  \textbf{H}( \textbf{Q},t)\textbf{Q},
\label{eq:QRCp_04}
\end{equation}
with 
\begin{displaymath}
\textbf{H}( \textbf{Q},t)=\textbf{A}
-\textbf{Q}\textbf{Q}^{\mathrm{T}}\textbf{A} + \textbf{Q}\textbf{S}
\textbf{Q}^{\mathrm{T}}.
\end{displaymath}

Notice that the matrix $\textbf{H}( \textbf{Q},t)$ in not necessarily
skew--symmetric, and the term $\textbf{Q}\textbf{Q}^{\mathrm{T}}$ is
responsible for lack of skew--symmetry in $\textbf{H}$. Of course for $p=2N$
equation (\ref{eq:QRCp_04}) reduces to equation
$\dot{\textbf{Q}}=\textbf{Q}\textbf{S}$ (\ref{eq:QRC_02}). The evolution of
the diagonal elements of $\textbf{R}$ are again governed by equation
(\ref{eq:QRC_05}), but for $ p< 2N$, and so the $p$ largest LCEs can be
computed again from equations (\ref{eq:QRC_07})--(\ref{eq:QRC_12}).
 
The main difference with respect to the case of the computation of the whole
spectrum is the numerical difficulties arising in solving equation
(\ref{eq:QRCp_04}), since $\textbf{H}$ is not skew--symmetric as was matrix
$\textbf{S}$ in equation (\ref{eq:QRC_02}). Due to this difference usual
numerical integration techniques fail to preserve the orthogonality of matrix
$\textbf{Q}$.

A central observation of \cite{DV_95} is that the matrix $\textbf{H}$ has a
weak skew--symmetry property. The matrix $\textbf{H}$ is called weak
skew--symmetric if
\begin{displaymath}
\textbf{Q}^{\mathrm{T}} \left( \textbf{H}( \textbf{Q},t) +
\textbf{H}^{\mathrm{T}}( \textbf{Q},t)\right) \textbf{Q}= \textbf{0} ,\,\,\,
\mbox{whenever} \,\,\,\textbf{Q}^{\mathrm{T}}\textbf{Q}=\textbf{I}_p.
\end{displaymath}
A matrix $\textbf{H}$ is said to be strongly skew--symmetric if it is
skew--symmetric, i.~e.~$\textbf{H}^{\mathrm{T}}=-\textbf{H}$. Christiansen and
Rugh \cite{CR_97} proposed a method according to which, the numerically
unstable equations (\ref{eq:QRCp_02c}) for the continuous orthonormalization
could be stabilized by the addition of an appropriate dissipation term. This
idea was also used in \cite{BR_01}, where it was shown that it is possible to
reformulate equation (\ref{eq:QRCp_04}) so that $\textbf{H}$ becomes strongly
skew--symmetric and thus, achieve a numerically stable algorithm for the
computation of few LCEs.

\subsubsection{Discrete and continuous methods based on the SVD procedure}
\label{SVD}

An alternative way of evaluating the LCEs is obtained by applying the SVD
procedure on the fundamental $2N \times 2N$ matrix $\textbf{Y}(t)$, which
defines the evolution of deviation vectors through equations (\ref{eq:w_ham})
and (\ref{eq:w0_map}) for continuous and discrete systems
respectively. According to the SVD algorithm a $2N \times p$ matrix $(p \leq
2N)$ $\textbf{B}$ can be written as the product of a $2N \times p$
column--orthogonal matrix $\textbf{U}$, a $p \times p$ diagonal matrix
$\textbf{F}$ with positive or zero elements $\sigma_i$, $i=1,\ldots, p$ (the
so--called {\it singular values}), and the transpose of a $p \times p$
orthogonal matrix $\textbf{V}$:
\begin{displaymath}
\textbf{B}=\textbf{U}\cdot \textbf{F} \cdot
\textbf{V}^{\mathrm{T}}. 
\end{displaymath}
We note that matrices $\textbf{U}$ and $\textbf{V}$ are orthogonal so
that:
\begin{equation}
\textbf{U}^{\mathrm{T}}\cdot \textbf{U}=\textbf{V}^{\mathrm{T}}\cdot
\textbf{V} = \mathrm{I}_p . 
\label{eq:UV}
\end{equation}
For a more detailed description of the SVD method, as well as an algorithm for
its implementation the reader is referred to \cite[Section 2.6]{NumRec} and
references therein.  The SVD is unique up to permutations of corresponding
columns, rows and diagonal elements of matrices $\textbf{U}$, $\textbf{V}$ and
$\textbf{F}$. Advanced numerical techniques for the computation of the
singular values of a product of many matrices can be found for example in
\cite{S_97,OS_00}.

So, for the purposes of our study let  
\begin{equation}
\textbf{Y}=\textbf{U}\cdot \textbf{F} \cdot
\textbf{V}^{\mathrm{T}}, 
\label{eq:SVD_Y}
\end{equation}
where we dropped out as before, the explicit dependence of the matrices on
time $t$. In those cases where all singular values are different, a unique
decomposition can be achieved by the additional request of a strictly
monotonically decreasing singular value spectrum, i.~e.~$\sigma_1(t) >
\sigma_2(t) > \cdots> \sigma_{2N}(t)$. Multiplying equation (\ref{eq:SVD_Y})
with the transpose
\begin{displaymath}
\textbf{Y}^{\mathrm{T}}=
\textbf{V}\cdot \textbf{F}^{\mathrm{T}}\cdot\textbf{U}^{\mathrm{T}}, 
\end{displaymath}
from the left we get
\begin{equation}
\textbf{Y}^{\mathrm{T}}\cdot\textbf{Y}= \textbf{V}\cdot
\textbf{F}^{\mathrm{T}} \cdot\textbf{U}^{\mathrm{T}}\cdot\textbf{U}
\cdot\textbf{F} \cdot \textbf{V}^{\mathrm{T}} = \textbf{V}
\cdot\mbox{diag}(\sigma_i^2(t)) \cdot \textbf{V}^{\mathrm{T}},
\label{eq:SVD_YYtr}
\end{equation}
where equation (\ref{eq:UV}) has been used. From equation (\ref{eq:SVD_YYtr})
we see that the eigenvalues of the diagonal matrix
$\mbox{diag}(\sigma_i^2(t))$, i.~e.~the squares of the singular values of
$\textbf{Y}(t)$, are equal to the eigenvalues of the symmetric matrix
$\textbf{Y}^{\mathrm{T}}\textbf{Y}$. Then from point 4 of the MET we conclude
that the LCEs are related to the singular values of $\textbf{Y}(t)$ through
\cite{GPL_90,S_97}
\begin{displaymath}
\chi_p=\lim_{t\rightarrow \infty} \frac{1}{t} \ln \sigma_i(t) , \,\,\,
p=1,2,\ldots,2N,
\end{displaymath} 
which implies that the LCEs can be evaluated as the limits for $t\rightarrow
\infty$ of the time rate of the logarithms of the singular values.

Theoretical aspects of the SVD technique, as well as a detailed study of its
ability to approximate the spectrum of LCEs can be found in
\cite{OS_00,DV_02,DE_06}. Continuous \cite{GK_87,GPL_90,DL_06} and discrete
\cite{S_97} versions of the SVD algorithm have been applied for the
computation of few or of all  LCEs, although this approach is not widely
used. A basic problem of these methods is that they fail to compute the
spectrum of LCEs if it is degenerate, i.~e.~when two or more LCEs are equal or
very close to each other, due to the appearance of ill--conditioned matrices.

\section{Chaos detection techniques}
\label{other_chaos_methods}

A simple, qualitative way of studying the dynamics of a Hamiltonian system is
by plotting the successive intersections of its orbits with a Poincar\'{e}
surface of section (PSS) (e.~g.~\cite{HH_64}
\cite[p.~17--20]{LichtenbergL_1992}). Similarly, in the case of symplectic maps
one simply plots the phase space of the system. This qualitative method has
been extensively applied to 2d maps and to 2D Hamiltonians, since in these
systems the PSS is a 2--dimensional plane. In such systems one can visually
distinguish between regular and chaotic orbits since the points of a regular
orbit lie on a torus and form a smooth closed curve, while the points of a
chaotic orbit appear randomly scattered. In 3D Hamiltonian systems (or 4d
symplectic maps) however, the PSS (or the phase space) is 4--dimensional and
the behavior of the orbits cannot be easily visualized. Things become even
more difficult and deceiving for multidimensional systems.  One way to
overcome this problem is to project the PSS (or the phase space) to spaces
with lower dimensions (see e.g.\ \cite{VBK_96,VIB_97,PP_08}) although these
projections are often very complicated and difficult to interpret. Thus, we
need fast and accurate numerical tools to give us information about the
regular or chaotic character of orbits, mainly when the dynamical system has
many degrees of freedom.

The most commonly employed method for distinguishing between regular and
chaotic behavior is the evaluation of the mLCE $\chi_1$, because if $\chi_1>0$
the orbit is chaotic.  The main problem of using the value of $\chi_1$ as an
indicator of chaoticity is that, in practice, the numerical computation may
take a huge amount of time, in particular for orbits which stick to regular
ones for a long time before showing their chaotic behavior. Since $\chi_1$ is
defined as the limit for $t \rightarrow \infty$ of the quantity $X_1(t)$
(\ref{eq:X_1_t}), the time needed for $X_1(t)$ to converge to its limiting
value is not known a priori and may become extremely long. Nevertheless, we
should keep in mind that the mLCE gives us more information than just
characterizing an orbit as regular or chaotic, since it also quantifies the
notion of chaoticity by providing a characteristic time scale for the studied
dynamical system, namely the Lyapunov time (\ref{eq:lyap_time}).

In order to address the problem of the fast and reliable determination of the
regular or chaotic nature of orbits, several methods have been developed over
the years with varying degrees of success.  These methods can be divided in
two major categories: Some are based on the study of the evolution of
deviation vectors from a given orbit, like the computation of $\chi_1$, while
others rely on the analysis of the particular orbit itself.

Among other chaoticity detectors, belonging to the same category with the
evaluation of the mLCE, are the fast Lyapunov indicator (FLI)
\cite{FLG_97,FGL_97,FL_00,LF_01,FLFF_02,GLF_02} and its variants
\cite{B_05,B_06}, the smaller alignment index (SALI)
\cite{S_01,SABV_03,SABV_04} and its generalization, the so--called generalized
alignment index (GALI) \cite{SBA_07,SBA_08}, the mean exponential growth of
nearby orbits (MEGNO) \cite{CS_00,CGS_03}, the relative Lyapunov indicator
(RLI) \cite{SEE_00,SESF_04}, the average power law exponent (APLE)
\cite{LVE_08}, as well as methods based on the study of spectra of quantities
related to the deviation vectors like the stretching numbers
\cite{FFL_93,LFD_93,VC_94,VVT_00}, the helicity angles (the angles of
deviation vectors with a fixed direction) \cite{CV_96}, the twist angles (the
differences of two successive helicity angles) \cite{CV_97}, or the study of
the differences between such spectra \cite{LF_98,VCE_98}.

In the category of methods based on the analysis of a time series constructed
by the coordinates of the orbit under study, one may list the frequency map
analysis of Laskar \cite{L_90,LFC_92,L_93,L_99}, the `0--1' test
\cite{GM_04,GM_05}, the method of the low frequency spectral analysis
\cite{VI_92,KV_04}, the `patterns method' \cite{S_05,S_06}, the recurrence
plots technique \cite{ZPRTK_07,ZTRK_07} and the information entropy index
\cite{NCW_96}. One could also refer to several ideas presented by various
authors that could be used in order to distinguish between chaoticity and
regularity, like the differences appearing for regular and chaotic orbits in
the time evolutions of their correlation dimension \cite{F_00}, in the time
averages of kinetic energies related to the virial theorem \cite{H_05} and in
the statistical properties of the series of time intervals between successive
intersections of orbits with a PSS \cite{KV_08}.

A systematic and detailed comparative study of the efficiency and reliability
of the various chaos detection techniques has not been done yet, although
comparisons between some of the existing methods have been performed
sporadically in studies of particular dynamical systems
\cite{S_01,SABV_04,S_06a,S_06b,K_07,LVE_08,BBB_08}.

Let us now focus our attention on the behavior of the FLI and of the GALI and
on their connection to the LCEs. The FLI was introduced as an indicator of
chaos in \cite{FLG_97,FGL_97} and after some minor modifications in its
definition, it was used for the distinction between resonant and not resonant
regular motion \cite{FL_00,FLFF_02}. The FLI is defined as
\begin{displaymath}
\mbox{FLI}(t)=\sup_t\,\, \ln \| \vec{w}(t) \|,
\end{displaymath} 
where $\vec{w}(t)$ is a deviation vector from the studied orbit at point
$\vec{x}(t)$, which initially had unit norm, i.~e.~$\| \vec{w}(0) \|=1$. In
practice, $\mbox{FLI}(t)$ registers the maximum length that an initially
unitary deviation vector attains from the beginning of its evolution up to the
current time $t$. Using the notation appearing in equation
(\ref{eq:compute_x1}), the FLI can be computed as
\begin{displaymath}
\mbox{FLI}(k \tau)=\sup_k\,\,\ \sum_{i=1}^k \ln
\frac{D_i}{D_0}=\sup_k\,\,\sum_{i=1}^k \ln \alpha_i,
\end{displaymath} 
with the initial norm $D_0$ of the deviation vector being $D_0=1$.

According to equation (\ref{eq:approx_norm}) the norm of $\vec{w}(t)$
increases linearly in time in the case of regular orbits. On the other hand,
in the case of chaotic orbits the norm of any deviation vector exhibits an
exponential increase in time, with an exponent which approximates $\chi_1$ for
$t \rightarrow \infty$. Thus, the norm of a deviation vector reaches rapidly
completely different values for regular and chaotic orbits, which actually
differ by many orders of magnitude. This behavior allows FLI to discriminate
between regular orbits, for which FLI has relatively small values, and chaotic
orbits, for which FLI gets very large values.

The main difference of FLI with respect to the evaluation of the mLCE by
equation (\ref{eq:compute_x1}) is that FLI registers the \textit{current}
value of the norm of the deviation vector and does not try to compute the 
 limit value, for $t\rightarrow \infty$, of the mean of 
stretching numbers as $\chi_1$ does. By dropping
the time average requirement of the stretching numbers, FLI succeeds in
determining the nature of orbits faster than the computation of the mLCE.

The generalized alignment index of order $p$ (GALI$_p$) is determined through
the evolution of $2 \leq p \leq 2N$ initially linearly independent deviation
vectors $\vec{w}_i(0)$, $i=1,2,\ldots,p$ and so it is more related to the
computation of many LCEs than to the computation of the mLCE. The evolved
deviation vectors $\vec{w}_i(t)$ are normalized from time to time in order to
avoid overflow problems, but their directions are left intact. Then, according
to \cite{SBA_07} GALI$_p$ is defined to be the volume of the
$p$--parallelogram having as edges the $p$ unitary deviation vectors
$\hat{\vec{w}}_i(t)$, $i=1,2,\ldots,p$
\begin{equation}
\mbox{GALI}_p(t)=\| \hat{\vec{w}}_1(t)\wedge \hat{\vec{w}}_2(t)\wedge \cdots
\wedge\hat{\vec{w}}_p(t) \|. 
\label{eq:GALI}
\end{equation}
In \cite{SBA_07} the value of GALI$_p$ is computed according to equation
(\ref{eq:A1_norm_wedge}), while in \cite{AB_06,SBA_08} a more efficient
numerical technique based on the SVD algorithm is applied. From the definition
of GALI$_p$ it becomes evident that if at least two of the deviation vectors
become linearly dependent, the wedge product in (\ref{eq:GALI}) becomes zero
and the GALI$_p$ vanishes.

In the case of a chaotic orbit all deviation vectors tend to become linearly
dependent, aligning in the direction  which corresponds to
the mLCE and GALI$_{p}$ tends to zero exponentially following the law
\cite{SBA_07}:
\begin{displaymath}
\mbox{GALI}_p(t) \sim e^{-\left[ (\chi_1-\chi_2) + (\chi_1-\chi_3)+
\cdots+ (\chi_1-\chi_p)\right]t},
\end{displaymath}
where $\chi_1,\chi_2, \ldots, \chi_p$ are the $p$ largest LCEs. On the other
hand, in the case of regular motion all deviation vectors tend to fall on the
$N$--dimensional tangent space of the torus on which the motion lies. Thus, if
we start with $p\leq N$ general deviation vectors they will remain linearly
independent on the $N$--dimensional tangent space of the torus, since there is
no particular reason for them to become linearly dependent. As a consequence
GALI$_p$ remains practically constant for $p\leq N$. On the other hand,
GALI$_p$ tends to zero for $p>N$, since some deviation vectors will eventually
become linearly dependent, following a particular power law which depends on
the dimensionality $N$ of the torus and the number $p$ of deviation
vectors. So, the generic behavior of GALI$_p$ for regular orbits lying on
$N$--dimensional tori is given by \cite{SBA_07}
\begin{equation}
\mbox{GALI}_p(t) \sim \left\{ \begin{array}{ll} \mbox{constant} & \mbox{if
$2\leq p \leq N$} \\ \frac{1}{t^{2(p-N)}} & \mbox{if $N< p \leq 2N$} \\
\end{array}\right. .
\end{equation}

The different behavior of GALI$_p$ for regular orbits, where it remains
different from zero or tends to zero following a power law, and for chaotic
orbits, where it tends exponentially to zero, makes GALI$_p$ an ideal
indicator of chaoticity independent of the dimensions of the system
\cite{SBA_07,SBA_08,BMC_08}. GALI$_p$ is a generalization of the SALI method
\cite{S_01,SABV_03,SABV_04} which is related to the evolution of only two
deviation vectors. Actually $\mbox{GALI}_2 \propto \mbox{SALI}$. However,
GALI$_p$ provides significantly more detailed information on the local
dynamics, and allows for a faster and clearer distinction between order and
chaos. It was shown recently \cite{CB_07,SBA_08} that GALI$_p$ can also be
used for the determination of the dimensionality of the torus on which regular
motion occurs.
 
As we discussed in Section \ref{allLCEs_Benettin} the alignment of all
deviation vectors to the direction corresponding to the mLCE is a basic
problem for the computation of many LCEs, which is overcome by successive
orthonormalizations of the set of deviation vectors. The GALIs on the other
hand, exploit exactly this `problem' in order to determine rapidly and with
certainty the regular or chaotic nature of orbits.

It was shown in Section \ref{Def_LCE} that the values of all LCEs (and
therefore the value of the mLCE) do not depend on the particular used norm. On
the other hand, the quantitative results of all chaos detection techniques
based on quantities related to the dynamics of the tangent space on a finite
time, depend on the used norm, or on the coordinates of the studied
system. For example, the actual values of the finite time mLCE $X_1(t)$
(\ref{eq:X_1_t}) will be different for different norms, or for different
coordinates, although its limiting value for $t\rightarrow \infty$, i.~e.~the
mLCE $\chi_1$, will be always the same. Other chaos detection methods, like
the FLI and the GALI, which depend on the current values of some norm--related
quantities and not on their limiting values for $t\rightarrow \infty$, will
attain different values for different norms and/or coordinate
systems. Although the values of these indices will be different, one could
expect that their qualitative behavior would be independent of the chosen norm
and the used coordinates, since these indices depend on the geometrical
properties of the deviation vectors. For example, the GALI quantifies the
linear dependence or independence of deviation vectors, a property which
obviously does not depend on the particular used norm or coordinates. Indeed,
some arguments explaining the independence of the behavior of the GALI method
on the chosen coordinates can be found in \cite{SBA_07}. Nevertheless, a
systematic study focused on the influence of the used norm on the qualitative
behavior of the various chaos indicators has not been performed yet, although
it would be of great interest.

\section{LCEs of dissipative systems and time series}
\label{LCEs_dis_time_series}

The presentation of the LCEs in this report was mainly done in connection to
conservative dynamical systems, i.~e.~autonomous Hamiltonian flows and
symplectic maps. The restriction to conservative systems is not necessary
since the theory of LCEs, as well as the techniques for their evaluation are
valid for general dynamical systems like for example dissipative ones. In
addition, within what is called time series analysis (see
e.g.~\cite{Kantz_1997}) it is of great interest to measure LCEs in order to
understand the underlying dynamics that produces any time series of
experimental data. For the completeness of our presentation we devote the last
section of our report to a concise survey of results concerning the LCEs of
dissipative systems and time series.

\subsection{Dissipative systems}
\label{LCEs_dis}

In contrast to Hamiltonian systems and symplectic maps for which the
conservation of the phase space volume is a fundamental constraint of the
motion, a dissipative system is characterized by a decrease of the phase space
volume with increasing time. This leads to the contraction of motion on a
surface of lower dimensionality than the original phase space, which is called
\textit{attractor}. Thus any dissipative dynamical system will have at least
one negative LCE, the sum of all its LCEs (which actually measures the
contraction rate of the phase space volume through equation (\ref{eq:MET_05}))
is negative and after some initial transient time the motion occurs on an
attractor.

Any continuous $n$--dimensional dissipative dynamical system without a
stationary point (which is often called a \textit{fixed point}) has at least
one LCE equal to zero \cite{H_83} as we have already discussed in Section
\ref{Spectrum_properties}. For regular motion the attractor of dissipative
flows represents a fixed point having all its LCEs negative, or a quasiperiodic
orbit lying on a $p$--dimensional torus $(p<n)$ having $p$ zero LCEs while the
rest $n-p$ exponents are negative. For dissipative flows in three or more
dimensions there can also exist attractors having a very complicated
geometrical structure which are called `strange'.

\textit{Strange attractors} have one or more positive LCEs implying that the
motion on them is chaotic. The exponential expansion indicated by a positive
LCE is incompatible with motion on a bounded attractor unless some sort of
folding process merges separated orbits. Each positive exponent corresponds to
a direction in which the system experiences the repeated stretching and
folding that decorrelates nearby orbits on the attractor. A simple geometrical
construction of a hypothetical strange attractor where orbits are bounded
despite the fact that nearby orbits diverge exponentially can be found in
\cite[Sect.~1.5]{LichtenbergL_1992}.

The numerical methods for the evaluation of the mLCE, of the $p$ ($1<p<n$)
largest LCEs and of the whole spectrum of them, presented in Sections
\ref{maximal} and \ref{allLCEs}, can be applied also to dissipative
systems. Actually, many of these techniques were initially used in studies of
dissipative models \cite{NS_77,SN_79,FA_84,GPL_90}.  For a detailed
description of the dynamical features of dissipative systems, as well as of
the behavior of LCEs for such systems the reader is referred, for example, to
\cite{O_81,ER_85} \cite[Sect.~1.5, Chapt.~7 and 8]{LichtenbergL_1992} and
references therein.

\subsection{Computing LCEs from a time series}
\label{LCEs_time_series}

A basic task in real physical experiments is the understanding of the
dynamical properties of the studied system by the analysis of some observed
time series of data. The knowledge of the LCEs of the system is one important
step towards the fulfillment of this goal.  Usually, we have no knowledge of
the nonlinear equations that govern the time evolution of the system which
produces the experimental data. This lack of information makes the computation
of the spectrum of LCEs of the system a hard and challenging task.

The methods developed for the determination of the LCEs from a scalar time
series have as starting point the technique of \textit{phase space
reconstruction} with \textit{delay coordinates} \cite{PCFS_80,T_81,RSS_83}
\cite[Chapt.~3 and 9]{Kantz_1997}. This technique is used for recreating a
$d$--dimensional phase space to capture the behavior of the dynamical system
which produces the observed scalar time series.

Assume that we have $N_D$ measurements of a dynamical quantity $x$ taken at
times $t_n=t_0+n \tau$, i.~e.~$x(n)\equiv x(t_0+n \tau)$, $n=0,1,2,\ldots,
N_D-1$. Then we produce $N_d=N_D-(d-1)T$ $d$--dimensional vectors
$\vec{x}(t_n)$ from the $x$'s as
\begin{displaymath}
\vec{x}(t_n)= \left[ \begin{array}{cccc}
x(n) & x(n+T) &  \ldots & x(n+(d-1)T)
\end{array} \right] ^{\mathrm{T}},
\end{displaymath} 
where $T$ is the (integer) delay time. With this procedure we construct $N_d$
points in a $d$--dimensional phase space, which can be treated as successive
points of a hypothetical orbit. We assume that the evolution of $\vec{x}(t_n)$
to $\vec{x}(t_{n+1})$ is given by some map and we seek to evaluate the LCEs of
this orbit.

The first algorithm to compute LCEs for a time series was introduced by Wolf
et al.~\cite{WSSV_85}. According to their method (which is also referred as
the \textit{direct method}), in order to compute the mLCE we first locate the
nearest neighbor (in the Euclidean sense) $\vec{x}(t_k)$, to the initial point
$\vec{x}(t_0)$ and define the corresponding deviation vector $\vec{w}(t_0)=
\vec{x}(t_0)-\vec{x}(t_k)$ and its length $L(t_0)=\|\vec{w}(t_0)\|$. The
points $\vec{x}(t_0)$ and $\vec{x}(t_k)$ are considered as initial conditions
of two nearby orbits and are followed in time. Then the mLCE is evaluated by
the method discussed in Section
\ref{maximal_algorithm}, which approximates deviation vectors by differences
of nearby orbits. So, at some later time $t_{m_1}$ (which is fixed a priori or
determined by some predefined threshold violation of the vector's length) the
evolved deviation vector $\vec{w}'(t_{m_1})=
\vec{x}(t_{m_1})-\vec{x}(t_{k+{m_1}})$ is normalized and its length
$L'(t_{m_1})=\|\vec{w}'(t_{m_1})\|$ is registered. The `normalization' of the
evolved deviation vector is done by looking for a new data point, say
$\vec{x}(t_l)$, whose distance $L(t_{m_1})=\|\vec{x}(t_{m_1})-\vec{x}(t_l) \|$
from the studied orbit is small and the corresponding deviation vector
$\vec{w}(t_{m_1})=\vec{x}(t_{m_1})-\vec{x}(l)$ has the same direction with
$\vec{w}'(t_{m_1})$. Of course with finite amount of data, one cannot hope to
find a replacement point $\vec{x}(l)$ which falls exactly on the direction of
$\vec{w}'(t_{m_1})$ but chooses a point that comes as close as
possible. Assuming that such point is found the procedure is repeated and an
estimation $X_1(t_{m_n})$ of the mLCE $\chi_1$ is obtained by an equation
analogous to equation (\ref{eq:X_1_t_sum1}):
\begin{displaymath}
X_1(t_{m_n})=\frac{1}{t_{m_n}-t_0} \sum_{i=1}^n \ln
\frac{L_1'(t_{m_i})}{L(t_{m_{i-1}})},
\end{displaymath} 
with $m_0=0$.  A Fortran code of this algorithm with fixed time steps between
replacements of deviation vectors is given in \cite{WSSV_85}.

Generalizing this technique by evolving simultaneously $p>1$ deviation
vectors, i.~e.~following the evolution of the orbit under study, as well as of
$p$ nearby orbits, we can, in principle, evaluate the $p$--mLCE $\chi_1^{(p)}$
of the system, which is equal to the sum of the $p$ largest 1--LCEs (see
equation (\ref{eq:limit_pmLCE})). Then the values of $\chi_i$ $i=1,2,\ldots,p$
can be computed from equation (\ref{eq:comp_lce_1}). This procedure
corresponds to a variant of the standard method for computing the LCEs,
presented in \cite{SN_79} and discussed in Section \ref{allLCEs_Benettin}, in
that deviation vectors are defined as differences of neighboring orbits. The
implementation of this approach requires the repeated replacement of the
deviation vectors, i.~e.~the replacement of the $p$ points close to the
evolved orbit under consideration, when the lengths of the vectors exceed some
threshold value. This replacement should be done in a way that the volume of
the corresponding $p$--parallelogram is small, and in particular smaller than
the replaced volume, and the new $p$ vectors point more or less to the same
direction like the old ones. This procedure is explained in detail in
\cite{WSSV_85} for the particular case of the computation of
$\chi_1^{(2)}=\chi_1+\chi_2$, where a triplet of points is involved.

It is clear that in order to achieve a good replacement of the evolved $p$
vectors, which will lead to a reliable estimation of the LCEs, the numerical
data have to satisfy many conditions. Usually this is not feasible due to the
limited number of data points. So the direct method of \cite{WSSV_85} does not
yield very precise results for the LCEs. Another limitation of the method,
which was pointed out in Wolf et al.~\cite{WSSV_85}, is that it should not be
used for finding negative LCEs which correspond to shrinking directions, due
to a cut off in small distances implied mainly by the level of noise of the
experimental data. An additional disadvantage of the direct method is that
many parameters which influence the estimated values of the LCEs like the
embedding dimension $d$, the delay time $T$, the tolerances in direction
angles during vector replacements and the evolution times between
replacements, have to be tuned properly in order to obtain reliable results.

A different approach for the computation of the whole spectrum of LCEs is
based on the numerical determination of matrix $\textbf{Y}_n$, $n=1,2,\ldots$,
of equation (\ref{eq:w0_map}), which defines the evolution of deviation
vectors in the reconstructed phase space. This method was introduced in
\cite{SS_85} and was studied in more detail in \cite{ER_85,EOKRC_86} (see also
\cite[Chapt.~11]{Kantz_1997}). According to this approach, often called the
\textit{tangent space method}, matrix $\textbf{Y}_n$ is evaluated for each
point of the studied orbit through local linear fits of the data. In
particular, for every point $\vec{x}(t_n)$ of the orbit we find all its
neighboring points, i.~e.~points whose distance from $\vec{x}(t_n)$ is less
than a predefined small value $\epsilon$. Each of these point define a
deviation vector. Then we find the next iteration of all these points and see
how these vectors evolve. Keeping only the evolved vectors having length less
than $\epsilon$ we evaluate matrix $\textbf{Y}_n$ through a
least--square--error algorithm. By repeating this procedure for the whole
length of the studied orbit we are able to evaluate at each point of the orbit
matrix $\textbf{Y}_n$ which defines the evolution of deviation vectors over
one time step. Then by applying the QR decomposition version of the standard
method, which was presented in Section \ref{QR}, we estimate the values of the
LCEs. The corresponding algorithm is included in the TISEAN software package
of nonlinear time series analysis methods developed by Hegger et
al.~\cite{HKS_99}. It is also worth mentioning that Brown et al.~\cite{BBA_91}
improved the tangent space method by using higher order polynomials for the
local fit.

If, on the other hand, we are interested only in the evaluation of the mLCE of
a time series we can apply the algorithm proposed by Rosenstein et
al.~\cite{RCD_93} and Kantz \cite{K_94}. The method is based on the
statistical study of the evolution of distances of neighboring orbits. This
approach is in the same spirit of Wolf et al.~\cite{WSSV_85} although being
simpler since it compares distances and not directions. A basic difference with
the direct method is that for each point of the reference orbit not one, but
several neighboring orbits are evaluated leading to improved estimates of the
mLCE with smaller statistical fluctuations even in the case of small data
sets. This algorithm is also included in the TISEAN package \cite{HKS_99},
while its Fortran and C codes can be found in \cite[Appendix B]{Kantz_1997}.

\subsection*{Acknowledgments}
The author is grateful to the referee (A.~Giorgilli) whose constructive
remarks and perceptive suggestions helped him improve significantly the
content and the clarity of the paper.  Comments from Ch.~Antonopoulos,
H.~Christodoulidi, S.~Flach, H.~Kantz, D.~Krimer, T.~Manos and R.~Pinto are
deeply appreciated. The author would also like to thank G.~Del Magno for the
careful reading of the manuscript, for several suggestions and for drawing his
attention to the cone technique. This work was supported by the Marie Curie
Intra--European Fellowship No MEIF--CT--2006--025678.

\appendix
\section*{Appendix}

\section{Exterior algebra and wedge product: Some basic notions}
\label{Wedge}

We present here some basic results of the exterior algebra theory along with
an introduction to the theory of wedge products following \cite{AB_02} and
textbooks such as \cite{Spivak_1965,Greub_1978,Spivak_1999}. We also provide
some simple illustrative examples of these results.

Let us consider an $M$--dimensional vector space $V$ over the field of real
numbers $\mathbb{R}$. The \textit{exterior algebra} of $V$ is denoted by
$\Lambda(V)$ and its multiplication, known as the \textit{wedge product} or
the \textit{exterior product}, is written as $\wedge$. The wedge product is
associative:
\[
(\vec{u} \wedge \vec{v}) \wedge \vec{w}= \vec{u} \wedge (\vec{v} \wedge
\vec{w}) 
\]
for $\vec{u}, \vec{v}, \vec{w} \in V$ and bilinear
\begin{eqnarray*}
 (c_1 \vec{u} + c_2 \vec{v}) \wedge \vec{w}= c_1(\vec{u}\wedge \vec{w}) +
c_2(\vec{v}\wedge \vec{w}), & & \nonumber \\ \vec{w} \wedge (c_1 \vec{u} + c_2
\vec{v}) = c_1(\vec{w}\wedge \vec{u}) + c_2(\vec{w}\wedge \vec{v}), & &
\end{eqnarray*}
for $\vec{u}, \vec{v}, \vec{w} \in V$ and $c_1, c_2 \in \mathbb{R}$. The wedge
product is also alternating on $V$
\[
\vec{u}\wedge \vec{u}=\vec{0}
\]
for all vectors $\vec{u}\in V$. Thus we have that
\[
\vec{u}\wedge \vec{v}=- \vec{v}\wedge\vec{u} 
\]
for all vectors $\vec{u},\vec{v} \in V $ and
\begin{equation}
\vec{u}_1\wedge \vec{u}_2\wedge \cdots \wedge\vec{u}_k=\vec{0}
\label{eq:w_e}
\end{equation}
whenever $\vec{u}_1, \vec{u}_2,\ldots ,\vec{u}_k\in V$ are linearly
dependent. Elements of the form $\vec{u}_1\wedge \vec{u}_2\wedge \cdots
\wedge\vec{u}_k$ with $\vec{u}_1, \vec{u}_2,\ldots ,\vec{u}_k\in V$ are called
\textit{$k$--vectors}. The subspace of $\Lambda(V)$ generated by all
$k$--vectors is called the \textit{$k$--th exterior power of} $V$ and denoted
by $\Lambda^k(V)$.

Let $\{\vec{\hat{e}}_1,\vec{\hat{e}}_2,\ldots,\vec{\hat{e}}_M\}$ be an
orthonormal basis of $V$, i.~e.~$\vec{\hat{e}}_i$, $i=1,2,\ldots,M$
are linearly independent vectors of unit magnitude and
\[
\vec{\hat{e}}_i \cdot \vec{\hat{e}}_j = \delta_{ij}
\]
where `$\,\cdot\,$' denotes the inner product in $V$ and
\[
\delta_{ij}=\left\{ \begin{array}{ll}
1&\mbox{for} \,\,\, i=j\\
0&\mbox{for} \,\,\, i\neq j\\ \end{array} \right. .
\]
It can be easily seen that the set
\begin{equation}
\{\vec{\hat{e}}_{i_1} \wedge \vec{\hat{e}}_{i_2} \wedge \cdots \wedge
\vec{\hat{e}}_{i_k}\,\, | \,\, 1 \leq i_1 < i_2 < \cdots < i_k \leq M \}
\label{eq:w_basis}
\end{equation}
is a basis of $\Lambda^k(V)$ since any wedge product of the form
$\vec{u}_1\wedge \vec{u}_2\wedge \cdots \wedge\vec{u}_k$ can be written as a
linear combination of the $k$--vectors of equation (\ref{eq:w_basis}). This is
true because every vector $\vec{u}_i$, $i=1,2,\ldots,k$ can be written as a
linear combination of the basis vectors $\vec{\hat{e}}_i$, $i=1,2,\ldots,M$
and using the bilinearity of the wedge product this can be expanded to a
linear combination of wedge products of those basis vectors. Any wedge product
in which the same basis vector appears more than once is zero, while any wedge
product in which the basis vectors do not appear in the proper order can be
reordered, changing the sign whenever two basis vectors change places. 
The dimension $d_k$ of $\Lambda^k(V)$ is equal to the binomial coefficient
\[
d_k=\mbox{dim}\Lambda^k(V)=\left(\begin{array}{c} M \\k \end{array} \right)=
\frac{M!}{k!(M-k)!}.
\]
\label{eq:w_binom}

Ordering the elements of basis (\ref{eq:w_basis}) of $\Lambda^k(V)$ according
to the standard \textit{lexicographical order}
\begin{equation}
\vec{\omega}_i=\vec{\hat{e}}_{i_1} \wedge \vec{\hat{e}}_{i_2} \wedge \cdots
\wedge \vec{\hat{e}}_{i_k} , \,\,\, 1 \leq i_1 < i_2 < \cdots < i_k \leq M
 , \,\,\, i=1,2,\cdots,d_k,
\label{eq:omega}
\end{equation}
any $k$--vector $\vec{\bar{u}}\in \Lambda^k(V)$ can be represented as
\begin{equation}
\vec{\bar{u}}=\sum_{i=1}^{d_k} \bar{u}_i \vec{\omega}_i \,\, ,\,\, \bar{u}_i
\in \mathbb{R}.
\label{eq:u_sum}
\end{equation}
A $k$--vector which can be written as the wedge product of $k$ linear
independent vectors of $V$ is called \textit{decomposable}. Of course, if the
$k$ vectors are linearly dependent we get the zero $k$--vector
(\ref{eq:w_e}). Note that not all $k$--vectors are decomposable. For example
the $2$--vector $\vec{\bar{u}}=\vec{e}_1 \wedge \vec{e}_2 + \vec{e}_3 \wedge
\vec{e}_4 \in \Lambda^2 (\mathbb{R}^4) $ is not decomposable as it cannot be
written as $ \vec{u}_1 \wedge \vec{u}_2$ with $\vec{u}_1, \vec{u}_2 \in
\mathbb{R}^4$.

Let us consider a decomposable $k$--vector $\vec{\bar{u}}=\vec{u}_1\wedge
\vec{u}_2\wedge \cdots \wedge\vec{u}_k$. Then the coefficients $\bar{u}_i$ in
(\ref{eq:u_sum}) are the minors of matrix $\textbf{U}$ having as columns the
coordinates of vectors $\vec{u}_i$, $i=1,2,\ldots,k$ with respect to the
orthonormal basis $\vec{\hat{e}}_i$, $i=1,2,\ldots,M$. In matrix form we have
\begin{equation}
\left[
\begin{array}{cccc}
 \vec{u}_1 & \vec{u}_2 & \cdots & \vec{u}_k \end{array} \right] = \left[
\begin{array}{cccc}
 \vec{\hat{e}}_1 & \vec{\hat{e}}_2 & \cdots & \vec{\hat{e}}_M
\end{array} \right] \cdot
\left[
\begin{array}{cccc}
u_{11} & u_{12} & \cdots & u_{1k} \\ u_{21} & u_{22} & \cdots & u_{2k} \\
\vdots & \vdots & & \vdots \\ u_{M1} & u_{M2} & \cdots & u_{Mk} \end{array}
\right] = \left[
\begin{array}{cccc}
 \vec{\hat{e}}_1 & \vec{\hat{e}}_2 & \cdots &
\vec{\hat{e}}_M
\end{array} \right] \cdot \textbf{U} \,\,\, 
\label{eq:w_matrix}
\end{equation}
where $u_{ij}$, $i=1,2,\ldots,M$, $j=1,2,\ldots,k$ are real
numbers. Then, the wedge product $\vec{u}_1\wedge \vec{u}_2\wedge
\cdots \wedge\vec{u}_k$ is written as
\begin{equation}
\begin{array}{c}
\displaystyle \vec{\bar{u}}= \vec{u_1}\wedge \vec{u}_2\wedge \cdots
\wedge\vec{u}_k = \sum_{i=1}^{d_k} \bar{u}_i \vec{\omega}_i= \\ \displaystyle
\sum_{1 \leq i_1 < i_2 < \cdots < i_k \leq M}  \left|
\begin{array}{cccc}
u_{i_1 1} & u_{i_1 2} & \cdots & u_{i_1 k} \\ u_{i_2 1} & u_{i_2 2} & \cdots &
u_{i_2 k} \\ \vdots & \vdots & & \vdots \\ u_{i_k 1} & u_{i_k 2} & \cdots &
u_{i_k k} \end{array} \right| \vec{\hat{e}}_{i_1}\wedge \vec{\hat{e}}_{i_2}
\wedge \cdots \wedge \vec{\hat{e}}_{i_k},
\end{array}
\label{eq:w_wedge}
\end{equation} 
where the sum is performed over all possible combinations of $k$ indices out
of the $M$ total indices and $|\,\,|$ denotes the determinant. So, the
coefficient of a particular $k$--vector $\vec{\hat{e}}_{i_1}\wedge
\vec{\hat{e}}_{i_2} \wedge \cdots \wedge \vec{\hat{e}}_{i_k}$ is the
determinant of the $k\times k$ submatrix of the $M \times k$ matrix of
coefficients appearing in equation (\ref{eq:w_matrix}) formed by its $i_1$,
$i_2$, $\ldots$, $i_k$ rows.

The inner product on $V$ induces an \textit{inner product} on each vector
space $\Lambda^k(V)$ as follows: Considering two decomposable $k$--vectors
\begin{displaymath}
\vec{\bar{u}}= \vec{u_1}\wedge \vec{u}_2\wedge \cdots \wedge\vec{u}_k \,\,\,
\mbox{and} \,\,\, \vec{\bar{v}}= \vec{v_1}\wedge \vec{v}_2\wedge \cdots
\wedge\vec{v}_k,
\end{displaymath}
with $\vec{u}_i, \vec{v}_j \in V$, $i,j=1,2,\ldots, k$, the inner product of
$\vec{\bar{u}}$, $\vec{\bar{v}}$ $\in \Lambda^k(V)$ is defined by
\begin{equation}
\langle \vec{\bar{u}},\vec{\bar{v}} \rangle_k \stackrel{\rm def}{=} 
\left|
\begin{array}{cccc}
\vec{u}_1 \cdot \vec{v}_1 & \vec{u}_1 \cdot \vec{v}_2 & \cdots & \vec{u}_1
\cdot \vec{v}_k \\ \vec{u}_2 \cdot \vec{v}_1 & \vec{u}_2 \cdot \vec{v}_2 &
\cdots & \vec{u}_2 \cdot \vec{v}_k \\ \vdots & \vdots & & \vdots \\ \vec{u}_k
\cdot \vec{v}_1 & \vec{u}_k \cdot \vec{v}_2 & \cdots & \vec{u}_k \cdot
\vec{v}_k \end{array} \right|= \left|
\textbf{U}^{\mathrm{T}}\cdot \textbf{V} \right|
\label{eq:A1_inner}
\end{equation}
where $\textbf{U}$, $\textbf{V}$ are matrices having as columns the
coefficients of vectors $\vec{u}_i$, $\vec{v}_i$, $i=1,2,\ldots,k$ with
respect to the orthonormal
$\{\vec{\hat{e}}_1,\vec{\hat{e}}_2,\ldots,\vec{\hat{e}}_M\}$ (see equation
(\ref{eq:w_matrix})). Since every element of $\Lambda^k(V)$ is a sum of
decomposable element, this definition extends by bilinearity to any
$k$--vector. Obviously for the basis (\ref{eq:omega}) of $\Lambda^k(V)$ we
have
\[
\langle \vec{\omega}_i,\vec{\omega}_j \rangle_k = \delta_{ij}\,\,, \,\,
i,j=1,2,\ldots, d_k,
\]
implying that the basis is orthonormal.  Inner product (\ref{eq:A1_inner})
defines a \textit{norm} $\|\,\,\|$ for $k$--vectors by
\[
\| \vec{\bar{u}} \|= \sqrt{ \langle \vec{\bar{u}},\vec{\bar{u}} \rangle_k}=
\sqrt{ \left| \textbf{U}^{\mathrm{T}}\cdot
\textbf{U} \right|}.
\]
Thus, the norm of a decomposable $k$--vector (\ref{eq:w_wedge}) is given by
\begin{equation}
\begin{array}{c}
\displaystyle \| \vec{\bar{u}} \|= \| \vec{u_1}\wedge \vec{u}_2\wedge \cdots
\wedge\vec{u}_k \| = \sqrt{ \left| \textbf{U}^{\mathrm{T}}\cdot
\textbf{U}\right| }= \left( \sum_{i=1}^{d_k} \bar{u}_i^2
\right)^{1/2}= \\ \displaystyle \left\lbrace \sum_{1 \leq i_1 < i_2 < \cdots <
i_k \leq M}  \left|
\begin{array}{cccc}
u_{i_1 1} & u_{i_1 2} & \cdots & u_{i_1 k} \\ u_{i_2 1} & u_{i_2 2} & \cdots &
u_{i_2 k} \\ \vdots & \vdots & & \vdots \\ u_{i_k 1} & u_{i_k 2} & \cdots &
u_{i_k k} \end{array} \right|^2 \right\rbrace^{1/2},
\end{array}
\label{eq:A1_norm_wedge}
\end{equation} 
and it measures the volume $\mbox{vol}(P_k)$ of the $k$--parallelogram $P_k$
having as edges the $k$ vectors $\vec{u_1}, \vec{u}_2, \cdots, \vec{u}_k$,
since this volume is defined as (see e.~g.~\cite[p.~472]{Hubbard_1999})
\begin{equation}
\mbox{vol}(P_k)=\sqrt{\left| \textbf{U}^{\mathrm{T}}\cdot
\textbf{U}\right|  }\,\,.
\label{eq:volume_def}
\end{equation}

The use of a different orthonormal basis does not change the numerical value
of $\mbox{vol}(P_k)$. This can be easily seen as follows: Let
$\vec{\hat{f}}_i$, $i=1,2,\cdots,M$ be a different orthonormal basis of $V$
related to basis $\vec{\hat{e}}_i$ through
\[
\left[
\begin{array}{cccc}
 \vec{\hat{e}}_1 & \vec{\hat{e}}_2 & \cdots & \vec{\hat{e}}_M
\end{array} \right]  = \left[
\begin{array}{cccc}
 \vec{\hat{f}}_1 & \vec{\hat{f}}_2 & \cdots & \vec{\hat{f}}_M
\end{array} \right] \cdot  \textbf{A}
\]
where $\textbf{A}$ is an \textit{orthogonal} $M\times M$ matrix,
i.~e.~$\textbf{A}^{-1}=\textbf{A}^{\mathrm{T}}$. From equation
(\ref{eq:w_matrix}) we get
\[
\left[
\begin{array}{cccc}
 \vec{u}_1 & \vec{u}_2 & \cdots & \vec{u}_k \end{array} \right]= \left[
\begin{array}{cccc}
 \vec{\hat{f}}_1 & \vec{\hat{f}}_2 & \cdots & \vec{\hat{f}}_M
\end{array} \right] \cdot  \textbf{A} \cdot \textbf{U},
\]
whence the volume $\mbox{vol}'(P_k)$ with respect to the new basis
$\vec{\hat{f}}_i$, $i=1,2,\cdots,M$ is given by
\[
\mbox{vol}'(P_k)=\sqrt{\left| \left( \textbf{A} \cdot \textbf{U}\right)
^{\mathrm{T}} \cdot\textbf{A} \cdot\textbf{U}\right| }= \sqrt{\left|
\textbf{U}^{\mathrm{T}}\cdot \textbf{A}^{-1}\cdot \textbf{A}
\cdot\textbf{U}\right| } =\sqrt{\left| \textbf{U}^{\mathrm{T}}\cdot
\textbf{U}\right| }=\mbox{vol}(P_k),
\]
where the orthogonality of $\textbf{A}$ was used. This result is not
surprising since an orthogonal matrix corresponds to a rotation that leaves
unchanged the norms of vectors and the angles between them.

Finally we note that the equality
\[
\left| \textbf{U}^{\mathrm{T}}
\textbf{U}\right| = \sum_{1 \leq i_1 < i_2 < \cdots < i_k
\leq M} \left|
\begin{array}{cccc}
u_{i_1 1} & u_{i_1 2} & \cdots & u_{i_1 k} \\ u_{i_2 1} & u_{i_2 2} & \cdots &
u_{i_2 k} \\ \vdots & \vdots & & \vdots \\ u_{i_k 1} & u_{i_k 2} & \cdots &
u_{i_k k} \end{array} \right|^2
\]
appearing in equation (\ref{eq:A1_norm_wedge}) is the so--called
\textit{Lagrange identity} (e.~g.~\cite[p.~108]{Greub_1978},
\cite[p.~103]{Bourbaki_1958}).

\subsection{An illustrative example}
\label{wedge_example}

In order to illustrate the content of the previous section we consider here a
specific example. Let $V$ be the vector space of $M=4$--dimensional real
vectors, i.~e.~$V=\mathbb{R}^4$ and
\begin{equation}
\vec{\hat{e}}_1=(1,0,0,0)\,\,, \,\, \vec{\hat{e}}_2=(0,1,0,0)\,\,,
\,\,\vec{\hat{e}}_3=(0,0,1,0)\,\,, \,\,\vec{\hat{e}}_4=(0,0,0,1)\,\,,
\label{eq:W2_basis}
\end{equation} 
the usual orthonormal basis of $\mathbb{R}^4$. Then the lexicographically
ordered orthonormal basis (\ref{eq:omega}) of the $d_2=6$--dimensional vector
space $\Lambda^2 (\mathbb{R}^4)$ is
\begin{equation}
\begin{array}{ccc}
\vec{\omega}_1=\vec{\hat{e}}_1 \wedge \vec{\hat{e}}_2 \,\,, \,\,&
\vec{\omega}_2=\vec{\hat{e}}_1 \wedge \vec{\hat{e}}_3 \,\,,\,\, &
\vec{\omega}_3=\vec{\hat{e}}_1 \wedge \vec{\hat{e}}_4 \,\,,\,\, \\
\vec{\omega}_4=\vec{\hat{e}}_2 \wedge \vec{\hat{e}}_3 \,\,, \,\,&
\vec{\omega}_5=\vec{\hat{e}}_2 \wedge \vec{\hat{e}}_4 \,\,,\,\, &
\vec{\omega}_6=\vec{\hat{e}}_3 \wedge \vec{\hat{e}}_4 \,\,.
\end{array}
\label{eq:L2_basis}
\end{equation} 
The $\Lambda^3 (\mathbb{R}^3)$ vector space has dimension $d_3=4$ and
the set
\[
\begin{array}{cc}
\vec{y}_1=\vec{\hat{e}}_1 \wedge \vec{\hat{e}}_2 \wedge \vec{\hat{e}}_3\,\,,
\,\,& \vec{y}_2=\vec{\hat{e}}_1 \wedge \vec{\hat{e}}_2 \wedge
\vec{\hat{e}}_4\,\,, \,\, \\ \vec{y}_3=\vec{\hat{e}}_1 \wedge \vec{\hat{e}}_3
\wedge \vec{\hat{e}}_4\,\,, \,\,& \vec{y}_4=\vec{\hat{e}}_2 \wedge
\vec{\hat{e}}_3 \wedge \vec{\hat{e}}_4\,\,, \,\,
\end{array}
\]
as an orthonormal basis, while the $d_4=1$--dimensional vector space
$\Lambda^4 (\mathbb{R}^4)$ is spanned by vector
\[
\vec{x}_1=\vec{\hat{e}}_1 \wedge \vec{\hat{e}}_2 \wedge \vec{\hat{e}}_3 \wedge
\vec{\hat{e}}_4.
\]

Let us now consider $4$ linearly independent vectors $\vec{u_1}$, $\vec{u}_2$,
$\vec{u}_3$, $\vec{u}_4$ of $\mathbb{R}^4$ and the matrix
\[
\textbf{U}=[u_{ij}]= [
\begin{array}{cccc}
\vec{u_1} & \vec{u}_2 &\vec{u}_3 &\vec{u}_4 
\end{array}
]=\left[
\begin{array}{cccc}
u_{1 1} & u_{1 2} & u_{1 3} & u_{1 4} \\ u_{2 1} & u_{2 2} & u_{2 3} & u_{2 4}
\\ u_{3 1} & u_{3 2} & u_{3 3} & u_{3 4} \\ u_{4 1} & u_{4 2} & u_{4 3} & u_{4
4} \end{array} \right]\,\,\, , \,\,\, i,j=1,2,3,4,
\]
having as columns the coordinates of these vectors with respect to the
basis (\ref{eq:W2_basis}) of $\mathbb{R}^4$.

Considering basis (\ref{eq:L2_basis}) of $\Lambda^2 (\mathbb{R}^4)$
the 2--vector $\vec{u}_1 \wedge \vec{u}_2$ is given by
\[
\begin{array}{c}
\vec{u_1}\wedge \vec{u}_2 =
\left| \begin{array}{cc}
u_{11} & u_{12} \\ u_{21} & u_{22}
\end{array} \right| \vec{\omega}_1 +
\left| \begin{array}{cc}
u_{11}& u_{12} \\ u_{31}& u_{32}
\end{array} \right| \vec{\omega}_2 + 
\left| \begin{array}{cc} u_{11}& u_{12} \\ u_{41} &u_{42}
\end{array} \right| \vec{\omega}_3 + \\
  \left| \begin{array}{cc}
u_{21} &u_{22} \\ u_{31} &u_{32}
\end{array} \right| \vec{\omega}_4 +
  \left| \begin{array}{cc}
u_{21}& u_{22} \\ u_{41} &u_{42}
\end{array} \right| \vec{\omega}_5 +
  \left| \begin{array}{cc}
u_{31} &u_{32} \\ u_{41} &u_{42}
\end{array} \right| \vec{\omega}_6
\end{array}
\]
according to equation (\ref{eq:w_wedge}). For the norm of this vector we get
from equations (\ref{eq:A1_inner}) and (\ref{eq:A1_norm_wedge}):
\[
\begin{array}{c}
\| \vec{u_1}\wedge \vec{u}_2 \|^2=  
\left|
\begin{array}{cc}
\| \vec{u}_1 \|^2  & \vec{u}_1 \cdot \vec{u}_2 \\ 
\vec{u}_2 \cdot \vec{u}_1 & \| \vec{u}_2 \|^2 \end{array} \right|= 
  \left| \begin{array}{cc}
u_{11} &u_{12} \\ u_{21} &u_{22}
\end{array} \right|^2   + 
\left| \begin{array}{cc}
u_{11} &u_{12} \\ u_{31} &u_{32}
\end{array} \right|^2  + \\
  \left| \begin{array}{cc}
u_{11}& u_{12} \\ u_{41}& u_{42}
\end{array} \right|^2  + 
  \left| \begin{array}{cc}
u_{21} &u_{22} \\ u_{31} &u_{32}
\end{array} \right|^2  +
  \left| \begin{array}{cc}
u_{21}& u_{22} \\ u_{41}& u_{42}
\end{array} \right|^2  +
  \left| \begin{array}{cc}
u_{31} &u_{32} \\ u_{41} &u_{42}
\end{array} \right|^2, 
\end{array}
\]
where $\| \,\, \|$ is used also for denoting the usual Euclidian norm of a
vector.

In a similar way we conclude that the norm of the 3--vector produced by
$\vec{u}_1$, $\vec{u}_2$, $\vec{u}_3$
\[
\begin{array}{c}
\vec{u}_1\wedge \vec{u}_2 \wedge \vec{u}_3=   \left| \begin{array}{ccc}
u_{11} & u_{12} & u_{13}\\ u_{21} & u_{22} & u_{23} \\u_{31} & u_{32} & u_{33}
\end{array} \right| \vec{y}_1 +
  \left| \begin{array}{ccc} u_{11} & u_{12} & u_{13}\\ u_{21} & u_{22} &
u_{23} \\u_{41} & u_{42} & u_{43}
\end{array} \right| \vec{y}_2 + \\
  \left| \begin{array}{ccc}
u_{11} & u_{12} & u_{13}\\ u_{31} & u_{32} & u_{33} \\u_{41} & u_{42} & u_{43}
\end{array} \right| \vec{y}_3 + 
  \left| \begin{array}{ccc}
u_{21} & u_{22} & u_{23}\\ u_{31} & u_{32} & u_{33} \\u_{41} & u_{42} & u_{43}
\end{array} \right| \vec{y}_4 
\end{array}
\]
is
\[
\begin{array}{c}
\| \vec{u_1}\wedge \vec{u}_2 \wedge \vec{u}_3\|^2=  
\left|
\begin{array}{ccc}
\| \vec{u}_1 \|^2 & \vec{u}_1 \cdot \vec{u}_2 & \vec{u}_1 \cdot \vec{u}_3\\
\vec{u}_2 \cdot \vec{u}_1 & \| \vec{u}_2 \|^2 & \vec{u}_2 \cdot \vec{u}_3\\
\vec{u}_3 \cdot \vec{u}_1 &\vec{u}_3 \cdot \vec{u}_2 & \| \vec{u}_3 \|^2
\end{array} 
\right|= \\
\left| \begin{array}{ccc}
u_{11} & u_{12} & u_{13}\\ u_{21} & u_{22} & u_{23} \\u_{31} & u_{32} & u_{33}
\end{array} \right|^2 + 
\left| \begin{array}{ccc}
u_{11} & u_{12} & u_{13}\\ u_{21} & u_{22} & u_{23} \\u_{41} & u_{42} & u_{43}
\end{array} \right|^2+
\left| \begin{array}{ccc}
u_{11} & u_{12} & u_{13}\\ u_{31} & u_{32} & u_{33} \\u_{41} & u_{42} & u_{43}
\end{array} \right|^2  + 
\left| \begin{array}{ccc}
u_{21} & u_{22} & u_{23}\\ u_{31} & u_{32} & u_{33} \\u_{41} & u_{42} & u_{43}
\end{array}  \right|^2  ,
\end{array}
\]
while the norm of the 4--vector produced by $\vec{u}_1$,
$\vec{u}_2$, $\vec{u}_3$, $\vec{u}_4$
\[
\vec{u_1}\wedge \vec{u}_2 \wedge \vec{u}_3 \wedge \vec{u}_4= \left|  
\textbf{U}\right| \vec{x}_1
\]
is given by
\[
\begin{array}{c}
\| \vec{u}_1\wedge \vec{u}_2 \wedge \vec{u}_3\wedge \vec{u}_4\|^2=  
\left|
\begin{array}{cccc}
\| \vec{u}_1 \|^2 & \vec{u}_1 \cdot \vec{u}_2 & \vec{u}_1 \cdot \vec{u}_3&
\vec{u}_1 \cdot \vec{u}_4\\ \vec{u}_2 \cdot \vec{u}_1 & \| \vec{u}_2 \|^2 &
\vec{u}_2 \cdot \vec{u}_3& \vec{u}_2 \cdot \vec{u}_4 \\ \vec{u}_3 \cdot
\vec{u}_1 &\vec{u}_3 \cdot \vec{u}_2 & \|\vec{u}_3 \|^2 &\vec{u}_3 \cdot
\vec{u}_4 \\ \vec{u}_4 \cdot \vec{u}_1 &\vec{u}_4 \cdot \vec{u}_2 &\vec{u}_4
\cdot \vec{u}_3 & \|\vec{u}_4 \|^2
\end{array} 
\right|= 
\left|  \textbf{U}\right|^2
\end{array}.
\]


\end{document}